\documentclass[a4paper,fleqn]{cas-sc}
\usepackage{graphicx,bm}
\usepackage{amssymb}
\usepackage[numbers,sort&compress]{natbib}
\usepackage{subfig,float}
\usepackage{ulem,verbatim}
\usepackage[figurename=Fig.]{caption}
\usepackage{titletoc}
\definecolor{prcolor}{RGB}{0,128,172}
\hypersetup{colorlinks=true, linkcolor=prcolor, citecolor=prcolor, urlcolor=prcolor}
\def\tsc#1{\csdef{#1}{\textsc{\lowercase{#1}}\xspace}}
\tsc{WGM}
\tsc{QE}
\tsc{EP}
\tsc{PMS}
\tsc{BEC}
\tsc{DE}

\newcommand{\ket}[1]{\left| {#1} \right\rangle}

\ExplSyntaxOn
\keys_set:nn { stm / mktitle } {nologo}
\ExplSyntaxOff
\begin{document}
\let\WriteBookmarks\relax
\def\floatpagepagefraction{1}
\def\textpagefraction{.001}
\shorttitle{Xue-Jia Yu, Limei Xu, and Hai-Qing Lin}
\shortauthors{Xue-Jia Yu et~al.}

\title [mode = title]{Topological Physics in Quantum Critical Systems}                      
\author[a,b,c]{Xue-Jia Yu}[orcid=https://orcid.org/0000-0002-1935-1463]
\ead{xuejiayu815@gmai.com}
\address[a]{Eastern Institute of Technology, Ningbo 315200, China}
\address[b]{Department of Physics, Fuzhou University, Fuzhou 350116, Fujian, China}
\address[c]{Fujian Key Laboratory of Quantum Information and Quantum Optics, College of Physics and Information Engineering,Fuzhou University, Fuzhou, Fujian 350108, China}

\author[d,e,f]{Limei Xu}[orcid=https://orcid.org/0000-0001-9368-8796]
\address[d]{International Center for Quantum Materials, School of Physics, Peking University, Beijing, 100871, China}
\ead{limei.xu@pku.edu.cn}
\address[e]{Interdisciplinary Institute of Light-Element Quantum Materials and Research Center for Light-Element Advanced Materials, Peking University, Beijing, 100871, China}
\address[f]{Collaborative Innovation Center of Quantum Matter, Beijing, 100871, China}

\author[g]{Hai-Qing Lin}[orcid=https://orcid.org/0000-0001-6497-8030]
\ead{hqlin@zju.edu.cn}
\address[g]{Institute for Advanced Study in Physics and School of Physics, Zhejiang University, Hangzhou 310058, China}

\begin{abstract}
Topology forms a cornerstone in modern condensed matter and statistical physics, offering a new framework to classify the phases and phase transitions beyond the traditional Landau paradigm. However, it is widely believed that topological properties are destroyed when the bulk energy gap closes, making it highly nontrivial to consider topology in gapless quantum critical systems. 
To address these challenges, recent advancements have sought to generalize the notion of topology to systems without a bulk energy gap, including quantum critical points and critical phases, collectively referred to as gapless symmetry-protected topological states. 
Extending topology to gapless quantum critical systems challenges the traditional belief in condensed matter physics that topological edge states are typically tied to the presence of a bulk energy gap. 
Furthermore, it suggests that topology plays a crucial role in classifying quantum phase transitions even if they belong to the same universality class, fundamentally enriching the textbook understanding of phase transitions. Given its importance, here we give a pedagogical review of the current progress of topological physics in quantum critical systems. 
We introduce the topological properties of quantum critical points and generalize them to stable critical phases, both for noninteracting and interacting systems. 
Additionally, we discuss further generalizations and future directions, including higher dimensions, nonequilibrium phase transitions, and realizations in modern experiments.
\end{abstract}

\begin{keywords}
\sep Phase transition 
\sep Quantum criticality 
\sep Gapless system 
\sep Deconfined quantum criticality 
\sep Non-equilibrium phase transition 
\sep Gapped phase of matter  
\sep Topological edge state 
\sep Symmetry-protected topological phase 
\sep \textcolor{red}{Topologically protected Dirac cone} 
\sep Symmetry-enriched quantum criticality 
\sep Gapless symmetry-protected topological phase 
\sep Intrinsically gapless symmetry-protected topological phase  
\sep Topological holography  
\end{keywords}

\maketitle

\tableofcontents
              
\section{Introduction}
\label{section1}

The Landau symmetry-breaking theory (paradigm) has long been regarded as the cornerstone for characterizing quantum phases and phase transitions~\cite{sachdev1999quantum,sachdev2023quantum}. 
However, the development of topological physics over the past few decades has introduced a new paradigm for classifying quantum phases and phase transitions in modern condensed matter and statistical physics~\cite{Qi2011rmp,Hasan2010rmp,Wen2017rmp,senthil2015symmetry,Chiu2016rmp,Haldane2017rmp}. Traditionally, the study of topological phases in condensed matter physics has been limited to \emph{gapped} phases, \emph{with the prevailing belief that nontrivial topological properties must be destroyed when the bulk energy gap closes.} 
Moreover, it is commonly believed that \emph{the universality class of phase transitions---one of the most fundamental concepts in statistical physics---is solely determined by critical exponents.} 
These two long-standing beliefs have shaped our understanding of the topological phases of matter and quantum phase transitions.

However, recently, the discovery of nontrivial topology in gapless quantum critical systems~\cite{Verresen2018prl,Jones2019JSP,verresen2020topologyedgestatessurvive,Kumar2021SR,Tam2022prx,Balabanov2021PRR,Rahul2021JPSJ,Balabanov2022PRB,Tarighi2022PRB,Wangke2022scipost,Jones2022JSP,Rahul2022SR,Ortega2023PRB,Kumar2023prb,Kumar_2023,Jones2023prl,florescalderon2023topologicalquantumcriticalitymultiplicative,Zhang2024pra,Choi2024prb,zhou2024topologicaledgestatesfloquet,cardoso2024gaplessfloquettopology,zhong2025quantumentanglementfermionicgapless,Scaffidi2017prx,Parker2018prb,Parker2019PRL,Verresen2021prx,Duque2021prb,Yu2022prl,niu2021emergentgaplesstopologicalluttinger,Fraxanet2022prl,Smith2022PRR,Ma2022scipost,Thorngren2022prb,Ye2022SciPost,Hidaka2022prb,chang2022absencefriedeloscillationsentanglement,Liu2021prb,Umberto2021scipost,verresen2021quotientsymmetryprotectedtopological,Nathanan2023scipost_a,Nathanan2023scipost_b,Mondal2023prb,wu2023impurityscreeningdefects11d,Prembabu2024prb,Yu2024prl,Yu2024prb,Zhong2024pra,Wei2024JHEP,verresen2024higgscondensatessymmetryprotectedtopological,thorngren2023higgscondensatessymmetryprotectedtopological,Yanghong2023prb,Li2024scipost,Su2024prb,Li2022prb,Li2023prb,wang2023stabilityfinestructuresymmetryenriched,li2023intrinsicallypurelygaplesssptnoninvertibleduality,Wen2023prb,wen2023classification11dgaplesssymmetry,huang2023topologicalholographyquantumcriticality,Yang2025cp,song2025boundaryphasetransitionstwodimensional,wen2024topologicalholographyfermions,huang2024fermionicquantumcriticalitylens,wen2025stringcondensationtopologicalholography,myersonjain2024pristinepseudogappedboundariesdeconfined,ando2024gauginglatticegappedgaplesstopological,ando2024gaugetheorymixedstate,bhardwaj2024hassediagramsgaplessspt,antinucci2025symtft31dgaplessspts,zhou2024interactioninducedphasetransitionstopological,li2024noninvertiblesymmetryenrichedquantumcritical,yu2025gaplesssymmetryprotectedtopologicalstates,tan2025exploringnontrivialtopologyquantum,yang2025deconfinedcriticalityintrinsicallygapless,rey2025incommensurategaplessferromagnetismconnecting}, including both phase transition points and stable critical phases described by conformal field theory (CFT), has challenged the conventional beliefs outlined above. 
This breakthrough has attracted growing attention in modern condensed matter and statistical physics communities. 
Unlike the extensively studied topological phase transitions in previous literature, which typically involve either transitions driven by topological defects or transitions between gapped phases characterized by different topological invariants, this review introduces a fundamentally different form of gapless topology: the emergence of nontrivial topology intrinsic to quantum critical systems themselves, now known as gapless symmetry-protected topological (SPT) states~\cite{Keselman2015prb,Scaffidi2017prx,Verresen2021prx,Yu2022prl,Yu2024prl}. These states can be explored in both free fermion systems and strongly interacting many-body systems, providing a new paradigm for understanding topological phases and classifying quantum phase transitions. 
Specifically, the discovery of nontrivial topology in quantum critical systems not only \emph{overturns the common belief that topological edge states must be protected by a bulk energy gap but also opens new avenues for classifying phase transitions that share the same critical exponents.} This fundamentally enriches the textbook understanding of topological phases and phase transitions. 
This review aims to start with the fundamental concepts of phase transitions and progressively introduce the role of topology in quantum critical systems.

Before delving into the details, we first provide an intuitive physical picture to understand the coexistence of topological edge states and gapless bulk fluctuations~\cite{Verresen2021prx,Thorngren2022prb}. 
Consider a conformally invariant quantum critical chain of finite size $L$. 
In such a system, the bulk finite-size gap typically scales as $1/L$. To ensure the stability of boundary modes---preventing them from coupling to each other through the gapless bulk in the thermodynamic limit---the energy splitting of edge modes should decay \emph{faster} than the bulk finite-size gap $1/L$. This requirement can be satisfied if the energy splitting of edge modes scales either exponentially or algebraically as $1/L^a$ with $a > 1$. The former corresponds to critical systems with a gapped sector, while the latter describes those without one (a more detailed discussion is provided in Sec.~\ref{section3}). This picture naturally includes the gapped case: when a finite bulk energy gap is present, edge modes are exponentially localized near the boundary and must decay faster than the bulk energy gap (which remains constant in this case). This prevents edge modes at the two boundaries from mixing, preserving their stability in gapped topological phases. This framework is essential for defining topological edge modes in various quantum critical systems. In the following subsections, we begin by introducing the basic concepts of quantum phase transitions and topological phases of matter.

\subsection{Phase transition and quantum criticality}
\begin{figure*}[h]
    \centering
\includegraphics[width=0.8\linewidth]{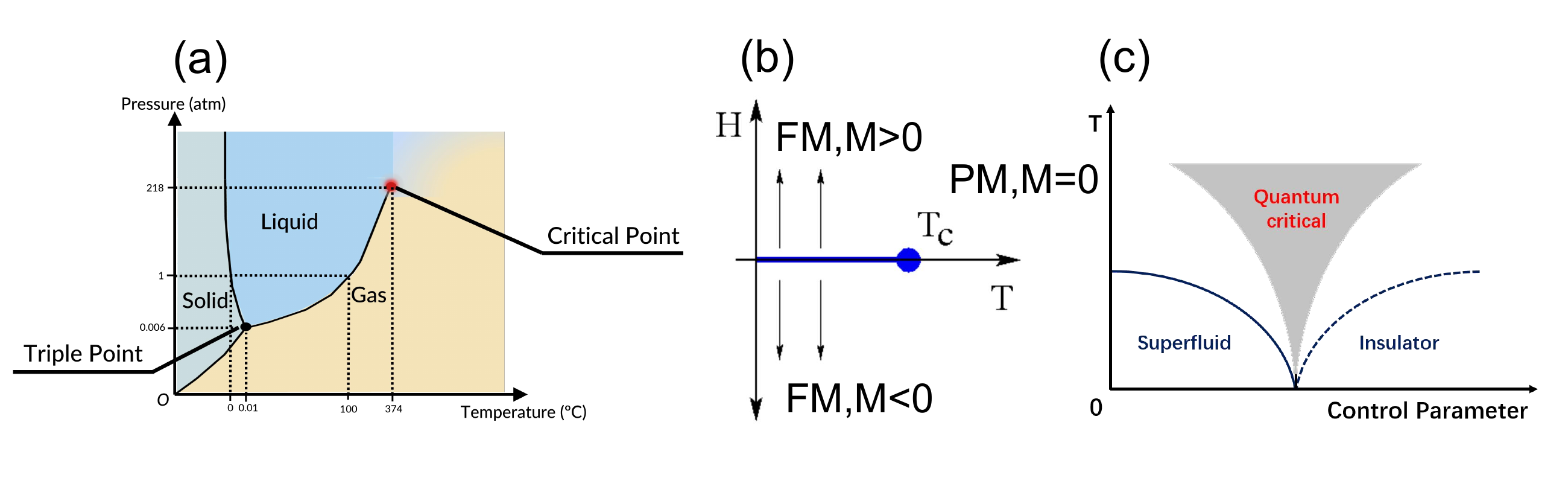}
    \caption{(a) Typical example of a classical phase transition: liquid-gas transition. The black solid line indicates the first-order phase transition, while the red dot marks the continuous critical point. The intersection of the three phases is known as the triple point. (b) Typical example of a classical phase transition: magnetic transition. The blue solid line represents the first-order phase transition line. The ferromagnetic phase (FM) is characterized by magnetization $M>0$ or $M<0$, indicating alignment along the direction of the external magnetic field, respectively. The paramagnetic phase (PM) exhibits no spontaneous magnetization. The critical temperature $T_c$ marks the continuous phase transition point, also referred to as the Curie point. (c) Typical example of a quantum phase transition: superfluid-insulator transition. At zero temperature, quantum phase transitions can be driven by tuning non-thermal parameters, such as the interaction strength or chemical potential. This tuning induces a transition between the superfluid phase, which spontaneously breaks $U(1)$ symmetry, and the Mott insulating phase, which preserves it. The figures are adapted from Ref.~\cite{landau2013statistical,sachdev1999quantum}.}
    \label{fig1}
\end{figure*}

As a typical example of gapless quantum critical systems, we first introduce the basic concept of quantum phase transitions. In both condensed matter and statistical physics, researchers are often interested in emergent phenomena that arise from the collective behavior of a large number of particles in the thermodynamic limit. 
Such systems are referred to as many-body systems. A phase, whether it is a classical phase observed in everyday life or a less common quantum phase, represents a region of material that is chemically uniform, physically distinct, and often mechanically separable. 
Nature exhibits a wide variety of phases of matter, ranging from familiar examples such as water, ice, and magnets to more exotic forms of quantum matter, including superconductors and superfluids~\cite{sachdev2023quantum}. 
Despite the complexity of real materials, their essential physics can often be captured using simplified effective model Hamiltonians~\cite{landau2013statistical,anderson1972more,kaul2013bridging}. A classic example is the Ising model, which provides a minimal description of the magnetic properties of real materials. Thermal or quantum fluctuations can drive the system into different phases of matter and induce transitions between them. 
The point at which a transition occurs is known as a classical or quantum critical point~\cite{sachdev1999quantum,Sondhi1997rmp}. 
Typical examples of these phase transitions include the liquid-gas transition, the magnetic ordering transition in the magnet, and the superfluid-insulator transition in the ultracold atom, as illustrated in Fig.~\ref{fig1} (a-c). 
According to Ehrenfest's classification principle~\cite{landau2013statistical}, phase transitions are broadly categorized as either first-order or continuous transitions. In this review, we primarily focus on gapless quantum critical systems that typically emerge at continuous phase transition points. 
The continuous transitions between different phases can exhibit universal behavior in the low-energy, long-wavelength limit. 
This gives rise to the concept of \emph{universality class}, one of the most fundamental concepts in modern physics~\cite{Vojta_2003}. Specifically, it asserts that while the microscopic Hamiltonians of various systems may be different, their critical points can exhibit the same low-energy physics. 
Thus, a central task in condensed matter and statistical physics is the classification of phases and phase transitions, establishing a unified framework for understanding the behavior of various systems and the fundamental principles that govern their transitions.

We begin by introducing classical phase transitions driven by thermal fluctuations to illustrate the universal properties of critical points. At the critical temperature, the correlation length $\xi$ diverges, indicating that the system lacks a characteristic length scale.
This divergence implies scale invariance and, in many cases, conformal invariance~\cite{francesco2012conformal,ginsparg1988appliedconformalfieldtheory,cardy1996scaling}. 
The lattice spacing $a$ in a microscopic lattice model becomes negligible compared to $\xi$. 
As a result, the critical behavior of the system can be described by a universal, effective continuum field theory that abstracts away the microscopic details of the underlying lattice model. This provides a qualitative explanation for the existence of universality classes, even when starting from completely different microscopic models. Furthermore, conformally invariant critical points can be described by CFT. In particular, for classical models in two spatial dimensions or quantum models in $(1+1)$-dimensional spacetime, the conformal data from CFT---such as the central charge, scaling dimensions (which are directly related to critical exponents) and operator product expansion coefficient---fully determine the universality classes of phase transitions~\cite{cardy1996scaling}. 
The universal behavior near a critical point leads to a set of critical exponents, ${\alpha, \beta, \gamma, \delta, \nu, \eta}$, which define scaling relations for various physical observables~\cite{kardar2007statistical_a,kardar2007statistical_b}. Taking the Ising model as an example, these relationships can be expressed as:
\begin{equation}
\label{scalinglaw}
\begin{split}
\begin{aligned}
& C \propto t^{-\alpha}, \quad \chi \propto t^{-\gamma}, \\ 
& \xi \propto t^{-\nu}, \quad M \propto (-t)^{\beta}, \\ 
& M \propto h^{1/\delta}, \quad \langle \sigma(r) \sigma(0) \rangle \propto \frac{1}{r^{d-2+\eta}},
\end{aligned}
\end{split}
\end{equation}
where $C$, $\chi$, $\xi$, $M$, $h$, and $\sigma$ represent the specific heat, magnetic susceptibility, correlation length, magnetization, external magnetic field, and Ising spin, respectively.
We define $t = (T - T_c)/T_c$ as the reduced temperature, and $r$ as the distance between the two different spins.
These critical exponents satisfy four universal scaling laws~\cite{kardar2007statistical_a,kardar2007statistical_b}. 
Consequently, traditional equilibrium statistical physics textbooks assert that to classify the universality class of a critical point, one must determine at least two critical exponents or the conformal data of the underlying CFT. 
In the 1960s and 1970s, Landau, Ginzburg, Wilson, and Fisher developed a comprehensive theory of \emph{equilibrium} phase transitions and critical phenomena~\cite{landau2013statistical}. 
This monumental achievement, now known as the Landau-Ginzburg-Wilson-Fisher (LGWF) theory, or the LGWF paradigm, is one of the cornerstones in modern physics. 
The central concept of the LGWF theory is \emph{spontaneous symmetry breaking}, which provides a framework for classifying different phases and phase transitions. 
In equilibrium systems, a phase may not preserve the symmetries of the underlying microscopic lattice Hamiltonian, indicating that the symmetry is spontaneously broken. 
The classification of phases then corresponds to identifying different patterns of symmetry breaking, which are quantified by an order parameter. 
The transition from a symmetry-preserving phase to a symmetry-breaking phase cannot occur smoothly.
It must pass through a singularity, i.e., a phase transition point, where the Landau order parameter changes from zero to a finite value as the system crosses the transition point. 
We can compute the universal critical exponents using the renormalization group technique, with tools such as the $\epsilon$-expansion. 
Consequently, the LGWF symmetry-breaking theory has long been regarded as the paradigm for phase transitions in condensed matter and statistical physics.

\begin{figure}[h]
    \centering
\includegraphics[width=0.65\linewidth]{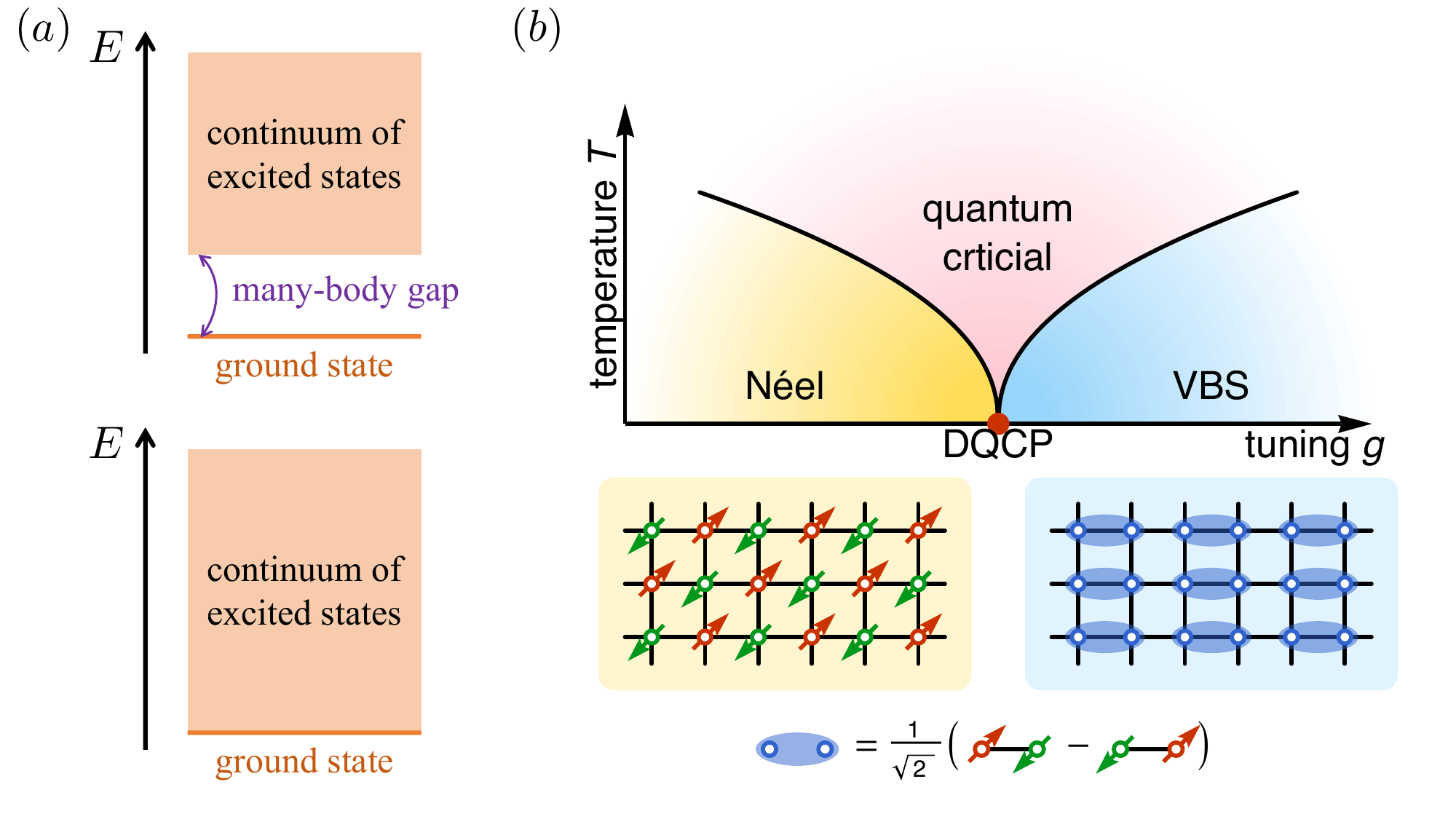}
    \caption{(a) The schematic energy spectrum of gapped (upper row) and gapless (lower row) system (phase). (b) The Néel-VBS deconfined quantum critical point (DQCP) is depicted in the schematic finite-temperature phase diagram. The configurations of the Néel antiferromagnetic phase and the VBS order are illustrated at the bottom of the phase diagram. The figures are adapted from Ref.~\cite{Oshikawa2000PRL,Ma2019prl}.}
    \label{fig2}
\end{figure}

In the context of quantum systems, we focus on the nontrivial universal behavior of the \emph{ground state}, as quantum fluctuations dominate over thermal fluctuations at zero temperature. 
Thus, a quantum phase transition refers to a transition between distinct quantum ground states, typically occurring at zero temperature. 
Here, we emphasize two key concepts that will be discussed in the following sections: gapped and gapless systems (or phases). 
As illustrated in Fig.~\ref{fig2} (a), a gapped system is characterized by a finite energy gap, $\Delta = E_1 - E_0$, between the first excited state $E_1$ and the ground state $E_0$ in the thermodynamic limit; in contrast, a gapless system is defined by the vanishing of $\Delta$ in the thermodynamic limit. 
Generally, the phases of matter on either side of a quantum phase transition are both gapped. However, as the control parameter approaches the critical point, the characteristic energy scale of fluctuations above the ground state vanishes, rendering the system gapless at the transition point.
Specifically, in a continuous quantum phase transition, as the control parameter varies, the energy gap $\Delta$ gradually closes and then reopens on the other side of the transition. 
Consequently, the quantum critical point represents a \emph{special type of gapless quantum critical system} that exists only at transition points in parameter space. A more detailed discussion of this will be presented in later sections. 
In this review, we do not consider first-order phase transitions, as they typically retain a finite energy gap at the transition point.

To describe the universal behavior at continuous quantum phase transitions, we adapt the LGWF theory, originally developed for classical phase transitions, by invoking the quantum-classical correspondence: a quantum system in $d+1$ spacetime dimensions (i.e., $d$ spatial dimensions and one imaginary time dimension) can be mapped to a classical system in $d+1$ spatial dimensions~\cite{sachdev1999quantum}. This correspondence enables the application of field-theoretic methods for classical phase transitions to quantum systems.
Within this framework, the effective Lagrangian of the quantum system typically takes the form of a classical field theory with an extra dimension. However, although the quantum-classical correspondence is a powerful concept, it does not imply that all quantum phase transitions are equivalent to classical phase transitions in one higher dimension. 
In particular, while many quantum critical points---referred to as conventional quantum critical points---can be analyzed through this correspondence, others, known as unconventional quantum critical points, require distinct theoretical treatment~\cite{xu2012unconventional}. Conventional quantum critical points 
typically occur at the transition between an ordered phase with spontaneous symmetry breaking and a disordered phase that is gapped and nondegenerate. 
The disordered phase is ``featureless'', meaning that it is adiabatically connected to a fully gapped product state without nontrivial correlations or entanglement. 
In contrast, unconventional quantum critical points violate the quantum-classical correspondence, and their quantum nature is essential to understanding the critical behavior. 
Moreover, there are several reasons why additional theoretical frameworks for quantum phase transitions are necessary~\cite{sachdev1999quantum}:
\begin{itemize}
\item[$\bullet$] The most intriguing properties of quantum critical points often pertain to their real-time dynamics, particularly at long times. However, analytic continuation from imaginary-time correlations to real-time correlations is an ill-defined problem~\cite{SHAO20231}, necessitating a theory directly addressing real-time dynamics.  
\item[$\bullet$] Quantum critical systems introduce a fundamentally new timescale characterizing dynamic properties, which is absent in classical phase transitions.  
\item[$\bullet$] If the low-energy effective field theory describing the quantum critical point includes topological terms, such as a Berry phase, the critical behavior is intrinsically quantum, with no classical counterpart. 
\end{itemize}

Let us briefly comment on the effects of finite temperature on quantum critical points. 
Depending on the relative magnitudes of the thermal equilibration time $\tau_{\text{eq}}$ (or the energy gap $\Delta$), and the thermal energy $k_B T$, the finite-temperature phase diagram can be divided into three regions, as illustrated in Fig.~\ref{fig1} (c). Of particular interest is the quantum critical region, which occupies a specific sector region of the phase diagram. In this region, the universal critical scaling behaviors at finite temperature are inherited from the quantum critical point at $T=0$, providing a promising opportunity to experimentally observe quantum critical scaling.

Before the 1980s, the LGWF symmetry-breaking theory was regarded as a central framework for understanding phase transitions in condensed matter and statistical physics.  
However, the experimental discoveries of high-temperature superconductors~\cite{muller1987discovery,Lee2006rmp} and the quantum Hall effect~\cite{Klitzing1980prl,Klitzing1986rmp} challenged this perspective. 
It has since become clear that the LGWF theory does not provide a complete description of all phase transitions. 
A typical example is the confinement-deconfinement transition in the three-dimensional $\mathbb{Z}_2$ gauge theory, which occurs without symmetry breaking and was first proposed by Franz Wegner~\cite{wegner1971duality}. 
Another well-known example is the topological phase transition in the two-dimensional classical XY model, also known as the Kosterlitz-Thouless transition~\cite{Kosterlitz_1973,Kosterlitz2017rmp}. This transition involves a change from a low-temperature quasi-long-range ordered phase (where vortices are bound) to a high-temperature short-range ordered phase (where vortices condense). 
Notably, throughout the transition, there is no long-range order and, consequently, no spontaneous symmetry breaking, in stark contrast to the LGWF paradigm. Instead, topological defects play a crucial role in driving the transition. 
Another prominent example of an unconventional quantum phase transition is deconfined quantum criticality~\cite{senthil2004deconfined,senthil2023deconfinedquantumcriticalpoints}, a continuous phase transition between two ordered phases that break incompatible symmetries and are also associated with topological defects~\cite{senthil2004deconfined,Senthil2004prb}. 
In this context, the topological defect of the group $G_1$ carries the nontrivial quantum number of the other incompatible group $G_2$, and therefore the condensed topological defect will not only recover $G_1$ symmetry but result in $G_2$ spontaneous symmetry breaking phase. 
This is a manifestation of the mixed anomaly between group $G_1$ and $G_2$, which is encoded in the Wess-Zemino-Witten (WZW) topological term in the low-energy effective field theory at the quantum critical point and does not have any classical counterparts. 
In standard LGWF theory, a continuous phase transition between the two ordered phases is possible only if one of the two groups $G_1, G_2$, is a subgroup of the other. In this sense, we say deconfined criticality is a novel quantum phase transition beyond the LGWF paradigm, also known as non-Landau phase transition~\cite{senthil2023deconfinedquantumcriticalpoints,Bi2020prr}. 
A prototypical example of deconfined quantum criticality is the transition between a N\'eel order and a valence bond solid (VBS) order in quantum magnets, as illustrated in Fig.~\ref{fig2} (b). 
The critical theory of such a transition is 
described in terms of a continuum field theory with emergent gauge fields coupled to matter fields carrying fractional quantum numbers of the microscopic global symmetry. 
These fractionalized degrees of freedom and associated deconfined gauge fields only emerge at the critical point, and are hence dubbed `deconfined quantum critical points'. The study of deconfined criticality has advanced significantly in the past two decades through field theory~\cite{Levin2004prb,Senthil2006prb,Grover2008prl,Melitski2008prb,Nahum2011prl,Swingle2012prb,Wang2017prx,You2018prx,Bi2019prx,Bi2020prr,Ma2020prb,Nahum2020prb}, numerical simulations~\cite{Sandvik2007prl,Melko2008prl,Kaul2008prb,Sandvik2010prl,Kaul2012prl,Block2013prl,Nahum2015prx,Nahum2015prl,shao2016quantum,Qin2017prx,Ma2018prb,Sun2018prl,Ma2019prl,Serna2019prb,zhao2019symmetry,li2019deconfinedquantumcriticalityemergent,liu2019superconductivity,Jiang2019prb,Roberts2019prb,Huang2019prb,Sandvik_2020,Zhao2020prl,Wang2022SciPost,Zhao2022prl,Liu2023prl,Yang2023prb,song2025evolution,DEmidio2024prl,Deng2024prl,zhang2025diracspinliquidunnecessary,Zhou2024prx,Chen2024prl,takahashi2024so5multicriticalitytwodimensionalquantum,chen2024emergentconformalsymmetrymulticritical}, and experiments~\cite{Guo2020prl,Lee2023prl,guo2023deconfinedquantumcriticalpoint,cui2023proximate,song2024unconventional}, with detailed discussions that can be found in review articles~\cite{senthil2023deconfinedquantumcriticalpoints}. 
A particularly notable feature of deconfined criticality is its dual descriptions. For example, the non-compact $\mathbb{CP}^1$ gauge theory is dual to the nonlinear sigma model with a WZW term or a QED$_3$ theory with two flavors of Dirac fermions ($N_f=2$), forming a duality web that describes the same fixed point~\cite{SEIBERG2016395,SENTHIL20191}. 
This duality property directly enforces the emergent $SO(5)$ symmetry at the critical point in the infrared (IR) limit. 
Moreover, the unconventionality of deconfined criticality lies in the fact that the WZW topological term in its low-energy description has no classical counterpart. This highlights the fundamental role of topology in understanding unconventional quantum phase transitions.

\subsection{Topological phase of matter}
In addition to describing phase transitions, Landau paradigm also suggests that quantum phases of matter can be classified based on spontaneous symmetry breaking~\cite{landau2013statistical}. 
The emergence of different states of matter corresponds to the breaking of distinct symmetries, often accompanied by the opening of energy gaps (or mass generation in the particle physics literature~\cite{Nambu2009rmp}). 
In contrast, a quantum phase without symmetry breaking is referred to as a symmetric or disordered phase. 
According to Landau theory, all symmetric phases should belong to the same class, namely featureless product states. 
However, advancements over the past several decades~\cite{Haldane1983prl,HALDANE1983464,Haldane1988prl,wen1990topological} have revealed that topology can further classify these gapped symmetric phases, resulting in topologically distinct classes. Specifically, certain gapped symmetric phases, though lacking symmetry breaking, may still belong to different phases due to their nontrivial topology and cannot be adiabatically connected without undergoing a phase transition. 
The discovery of topological phases of matter beyond the LGWF paradigm has opened new avenues for investigating and classifying gapped quantum phases, generating significant interest over the past two decades. 
In this section, we briefly introduce three representative types of gapped topological phases: band topology in free fermion systems, SPT phases in strongly interacting many-body systems, and intrinsic topological order with fractionalized excitations. 
More detailed discussions on these topological phases can be found in several influential review articles~\cite{Qi2011rmp,Hasan2010rmp,Wen2017rmp,senthil2015symmetry,Chiu2016rmp,Haldane2017rmp}.

\begin{figure}[h]
    \centering
\includegraphics[width=0.45\linewidth]{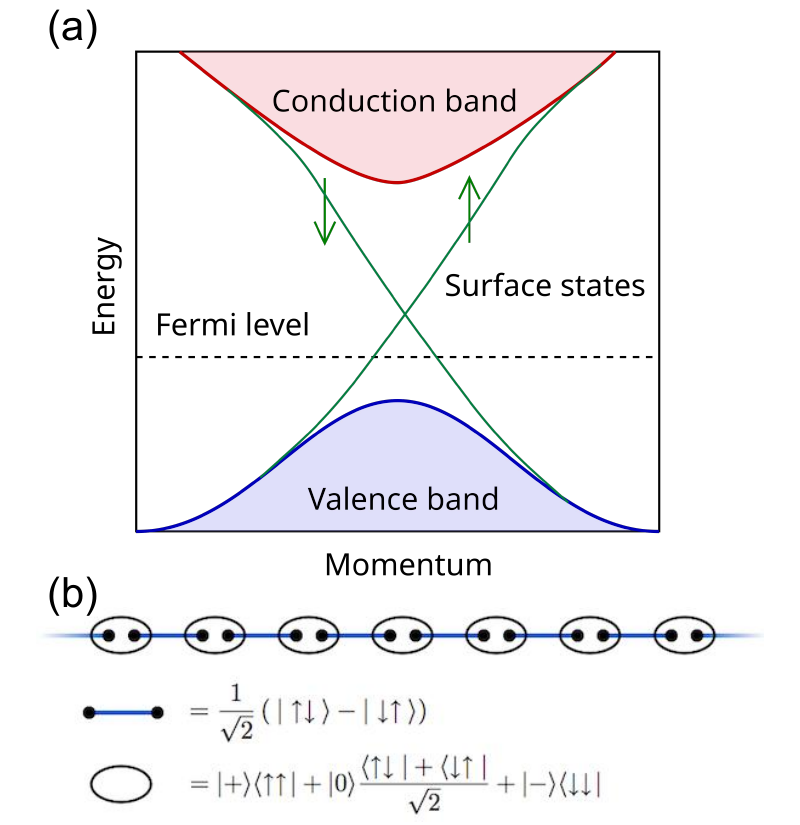}
    \caption{(a) Schematic band structure of a topological insulator. (b) Topological edge modes in the SPT phase (Haldane phase) of a one-dimensional spin-1 Heisenberg chain with open boundary conditions. The figures are adapted from Ref.~\cite{Moore2010Nature,hasan2011three,BRUNE2013125}.}
    \label{fig3}
\end{figure}

We begin by introducing the concept of band topology, which includes well-known examples such as topological insulators and superconductors. 
The classification of these quantum matters based on the topology of the band structures gained prominence following the first discovery of topological insulating states driven by spin-orbit couplings~\cite{Kane2005prla,Kane2005prlb,Bernevig2006prl,bernevig2006quantum}. The topology of a material's band structure becomes nontrivial when the conduction and valence bands invert due to strong spin-orbit coupling. 
In materials with a topologically nontrivial band structure---known as topological insulators---the bulk remains insulating, while the surface hosts metallic states, as illustrated in Fig.~\ref{fig3} (a). 
These surface states are characterized by two-dimensional Dirac points, named because the low-energy excitations near these points in momentum space can be described by a Dirac Hamiltonian for massless Dirac fermions~\cite{shen2012topological}. 
Around these points, the energy dispersion is linear in momentum, and spin is locked to momentum. Importantly, these surface states are topologically protected, meaning that they remain stable against disorder as long as the topology of the bulk band structure remains unchanged. 
More generally, the solvability of free-fermion systems has enabled the development of a unified theory of topological insulators. 
Furthermore, different topological phases are characterized by distinct topological invariants, typically defined in momentum or parameter space as integrals of the Berry curvature~\cite{Xiao2010rmp}. These nontrivial bulk topological invariants enforce the presence of gapless edge modes near the boundaries, establishing the well-known principle of bulk-boundary correspondence~\cite{bernevig2013topological}. 
Several theoretical lattice models have played a crucial role in illustrating band topology, including the Su-Schrieffer-Heeger model in one dimension~\cite{Su1979prl}, the Haldane model~\cite{Haldane1988prl}, and the Kane-Mele model in two dimensions~\cite{Kane2005prla,Kane2005prlb}. Despite the richness of possible topological phases, the classification of band topology in free-fermion tight-binding models is fortunately determined by time-reversal, particle-hole, and chiral symmetries, resulting in the well-known tenfold way classification table~\cite{Altland1997prb,Kitaev2009AIP,Ryu_2010}. 
As a result, topological insulators are characterized by a nontrivial topological invariant in the bulk and gapless topologically protected edge modes at the boundary. 
In addition, many topological materials have been experimentally discovered based on theoretical predictions and first-principle calculations~\cite{Yan_2012,xiao2021first}. The rapid development of this field in recent years includes the identification of numerous novel topological phases in the past few years, including high-order topological phases~\cite{schindler2018higher}, non-Hermitian topological phases~\cite{Gong2018prx}, etc. Detailed discussions of these novel phases can be found in review articles~\cite{xie2021higher,okuma2023non,ashida2020non}. 
Beyond topological insulators, superconducting pairing offers an alternative mechanism for opening energy gaps, resulting in band structures analogous to band insulators. This naturally extends the notion of band topology to superconductors, giving rise to the concept of topological superconductors~\cite{Kitaev_2001,Fu2008prl}. 
A particularly intriguing feature of these systems is the emergence of Majorana zero modes at topological defects, which are of fundamental interest for fault-tolerant topological quantum computing~\cite{Nayak2008rmp}. 
A detailed discussion of topological superconductors can be found in review articles~\cite{Sato_2017}, as this topic is not the focus of this section.

Following the theoretical prediction and experimental discovery of non-interacting topological phases, increasing attention has been given to understanding similar topological phenomena in strongly interacting many-body systems. To explore the interplay between strong correlations and topological properties, we focus on gapped phases with nontrivial topological characteristics that are protected by global symmetries, commonly referred to as SPT phases~\cite{Gu2009prb,chen2012symmetry,Chen2013prb,Pollmann2012prb,chen2014symmetry}. It is worth noting that topological insulators and superconductors, which are also protected by global symmetries, are sometimes classified as SPT phases in the literature. However, to avoid potential confusion, we use the term SPT phases exclusively for symmetry-protected interacting topological phases, while referring to topological insulators and superconductors collectively as band topology in this review. Under periodic boundary conditions, the bulk properties of SPT phases are indistinguishable from those of trivial paramagnetic phases. However, under open boundary conditions, SPT phases exhibit topologically protected edge states near the boundaries, as illustrated in Fig.~\ref{fig3} (b). 
Although a local order parameter cannot be used to probe SPT phases, their topological nature can be detected using a nonlocal string order parameter~\cite{Nijs1989prb}. Thus, SPT phases can be regarded as topological paramagnetic phases. 
More importantly, in the past two decades, it has been widely believed that the stability of topologically protected edge modes in the SPT phase is ensured by the bulk energy gap since it prevents the mixing of different boundary edge states and protects them from instability. 
The existence of this gap has made the study of SPT phases relatively tractable, leading to significant advancements over the past decade, including the classification of bosonic and fermionic SPT phases in various dimensions~\cite{Wen2014prb,Cheng2018prb,Lan2017prb,Wang2020prx,Bi2015prb,Else2014prb,Wen2015nsr}, the construction of lattice model to realize SPT phase numerically, and experimental realization of SPT phase in solid materials and quantum simulation platforms, see review articles~\cite{senthil2015symmetry,Wen2017rmp}. 
From the perspective of quantum entanglement, SPT phases are classified as short-range entangled phases, which can be constructed using finite-depth unitary~\cite{Gu2009prb,Chen2013prb}. In contrast, long-range entangled phases, also known as intrinsically topological orders, exhibit fractionalized anyonic excitations and are closely tied to topological quantum computing~\cite{Nayak2008rmp}. 
Examples of such phases include the fractional quantum Hall liquids (experimentally observed~\cite{Tsui1982prl,Stormer1999rmp}) and gapped spin liquid phases (important progress has been made in quantum simulator experiments~\cite{semeghini2021probing,satzinger2021realizing,Zhou2017rmp,Savary_2017}). Despite topological ordered phases are of fundamental interest and hold significant practical importance, this review focuses exclusively on symmetry-protected topologies and does not address intrinsically topological orders. 
For further details on the latter, we refer readers to Ref.~\cite{wen2004quantum,Wen2017rmp,Wen2015nsr,Yu2024prlemergence}.

\subsection{Gapless quantum critical systems}
The main focus of this review is on gapless quantum many-particle systems. 
Specifically, quantum many-particle systems that feature excitations at arbitrarily low energies are pervasive across various energy scales and are of significant interest to both the condensed matter and statistical physics communities. 
Such systems appear either at continuous phase transition points or in stable phases collectively referred to as gapless phases. Unlike gapped phases, gapless phases exhibit distinct characteristics, such as 
power-law decaying correlations~\cite{cardy1996scaling}, and the complex entanglement induced by quantum fluctuations on all length scales~\cite{LAFLORENCIE20161}. 
These systems are typically described by quantum field theories 
and cannot simply be understood as perturbative corrections to free-particle theories. 
Prominent examples of gapless phases currently under study include one-dimensional Luttinger liquids~\cite{giamarchi2003quantum,Haldane_1981,Kane1992prl}, Fermi liquids (metals)~\cite{landau2013statistical}, continuous phase transition points~\cite{sachdev1999quantum}, superfluids~\cite{Leggett1999rmp}, gapless quantum spin liquids~\cite{Gingras_2014,Zhou2017rmp,Savary_2017,Liu2022prx,LIU20221034}, non-Fermi liquids~\cite{VARMA2002267,lee2018recent,Stewart2001rmp}, strange metals~\cite{greene2020strange,phillips2022stranger}, topological (semi) metals~\cite{Armitage2018rmp,vafek2014dirac,burkov2018weyl}, as well as fractionalized Fermi liquids, which have recently been proposed in connection with cuprate physics~\cite{Senthil2003prl,Zhang2020prr,Maine2023PNAS}. 
Theoretical understanding of gapless phases often relies on tools like bosonization and CFT. 
Additionally, for 1+1 dimensions, these phases can also be effectively simulated using the density matrix renormalization group (DMRG) due to their logarithmic entanglement scaling~\cite{White1992prl,schollwock2011density}. 
However, in higher dimensions, theoretical tools and numerical algorithms for gapless quantum matter remain limited, as quantum Monte Carlo methods could suffer from the sign problem and tensor network-based methods may face challenges due to increased entanglement and less efficient contraction schemes. 
Despite these challenges, recent advances in analytical and numerical techniques have led to significant progress, including:
\begin{itemize}
\item[$\bullet$] The Sachdev-Ye-Kitaev (SYK) model~\cite{Sachdev1993PRL,Chowdhury2022rmp}, provides a solvable many-body framework for non-Fermi liquids and quantum matter without quasiparticles at finite density.
\item[$\bullet$] New insights into constraints imposed by anomalies on gapless states~\cite{Shi2022SciPost,Ning2020prr,Ye2022SciPost}. 
\item[$\bullet$] Non-perturbative approaches like the conformal bootstrap for strongly interacting field theories~\cite{Poland2019rmp}.
\item[$\bullet$] Numerical studies (e.g. quantum Monte Carlo, density matrix renormalization group algorithm) of quantum phase transitions beyond Landau paradigm, particularly those involving itinerant fermions~\cite{Liu2018prb,Liu2019PNAS,Xu_2019}. 
\item[$\bullet$] A series of dualities proposed for 2+1-dimensional quantum critical points~\cite{SEIBERG2016395,SENTHIL20191}.
\end{itemize}
In this review, we primarily focus on more tractable gapless systems described by CFT, which we refer to as gapless quantum critical systems. These include conformal critical points and stable critical phases.

Experimentally, quantum simulation platforms face significant challenges in simulating gapless quantum matter due to the difficulty in accessing critical ground states. 
Nevertheless, progress has been made in exploring novel gapless phases in condensed matter materials, such as quantum spin liquid Mott insulators~\cite{Kundu2020prl,Isono2014prl}, Moir\'e van der Waals heterostructures~\cite{castellanos2022van}, heavy fermion metals~\cite{Stewart1984rmp}, and the normal state of cuprate superconductors~\cite{Sachdev2003rmp}. 

Despite these advances, understanding gapless quantum matter---particularly in higher dimensions---remains challenging. 
On the other hand, the development of the topological phase of matter has revolutionized our understanding of phases and phase transitions. 
This naturally raises the question: Can topologically protected edge states remain robust even in the gapless critical systems? 
Furthermore, Can quantum critical points themselves exhibit nontrivial topological properties and serve as the basis for a new classification scheme purely from a topological perspective? 
In this review, we survey recent progress in understanding these intriguing possibilities with a particular focus on gapless symmetry-protected topological phases (gSPT) in both non-interacting~\cite{Verresen2018prl,Jones2019JSP,verresen2020topologyedgestatessurvive,Kumar2021SR,Tam2022prx,Balabanov2021PRR,Rahul2021JPSJ,Balabanov2022PRB,Tarighi2022PRB,Wangke2022scipost,Jones2022JSP,Rahul2022SR,Ortega2023PRB,Kumar2023prb,Kumar_2023,Jones2023prl,florescalderon2023topologicalquantumcriticalitymultiplicative,Zhang2024pra,Choi2024prb,zhou2024topologicaledgestatesfloquet,cardoso2024gaplessfloquettopology,zhong2025quantumentanglementfermionicgapless} and interacting systems~\cite{Scaffidi2017prx,Parker2018prb,Parker2019PRL,Verresen2021prx,Duque2021prb,Yu2022prl,niu2021emergentgaplesstopologicalluttinger,Fraxanet2022prl,Smith2022PRR,Ma2022scipost,Thorngren2022prb,Ye2022SciPost,Hidaka2022prb,chang2022absencefriedeloscillationsentanglement,Liu2021prb,Umberto2021scipost,verresen2021quotientsymmetryprotectedtopological,Nathanan2023scipost_a,Nathanan2023scipost_b,Mondal2023prb,wu2023impurityscreeningdefects11d,Prembabu2024prb,Yu2024prl,Yu2024prb,Zhong2024pra,Wei2024JHEP,verresen2024higgscondensatessymmetryprotectedtopological,thorngren2023higgscondensatessymmetryprotectedtopological,Yanghong2023prb,Li2024scipost,Su2024prb,Li2022prb,Li2023prb,wang2023stabilityfinestructuresymmetryenriched,li2023intrinsicallypurelygaplesssptnoninvertibleduality,Wen2023prb,wen2023classification11dgaplesssymmetry,huang2023topologicalholographyquantumcriticality,Yang2025cp,song2025boundaryphasetransitionstwodimensional,wen2024topologicalholographyfermions,huang2024fermionicquantumcriticalitylens,wen2025stringcondensationtopologicalholography,myersonjain2024pristinepseudogappedboundariesdeconfined,ando2024gauginglatticegappedgaplesstopological,ando2024gaugetheorymixedstate,bhardwaj2024hassediagramsgaplessspt,antinucci2025symtft31dgaplessspts,zhou2024interactioninducedphasetransitionstopological,li2024noninvertiblesymmetryenrichedquantumcritical,yu2025gaplesssymmetryprotectedtopologicalstates,tan2025exploringnontrivialtopologyquantum,yang2025deconfinedcriticalityintrinsicallygapless,rey2025incommensurategaplessferromagnetismconnecting}. 
The topological classification of the critical quantum matter, especially for quantum critical points, could provide a new paradigm for phase transitions in statistical physics and topology in condensed matter physics.

\color{red}
For convenience, we provide a conceptual summary, which recurs throughout the main text and clarifies the distinctions and relationships between these concepts.
\begin{itemize}

\item[$\bullet$] \textbf{Topological nontrivial quantum critical point:} A continuous quantum critical point can host stable topological edge states, which may either be conformally invariant or lack conformal symmetry with a dynamical exponent $z \neq 1$.

\item[$\bullet$] \textbf{Symmetry-enriched quantum critical point:} A conformally invariant quantum critical point can split into distinct types of critical points that feature nontrivial topological edge states when additional global symmetries are imposed. This phenomenon, also known as symmetry-enriched CFT, represents a special class of topologically nontrivial quantum critical points.

\item[$\bullet$] \textbf{The gSPT phase:} A gapless quantum phase (not just a discrete point) that supports symmetry-protected topological edge modes. In this paper, the term gSPT refers only to the simplest case, namely one that has a gapped counterpart and contains additional gapped degrees of freedom (see Sec.~\ref{section3}). In a broader sense, topologically nontrivial quantum critical points and symmetry-enriched CFTs can be viewed as a special type of gSPT, which occurs only at a single point rather than over an extended phase.

\item[$\bullet$] \textbf{Intrinsically gSPT phase:} A gapless symmetry-protected topological phase without a gapped counterpart under the same symmetries and dimensionality arises as a consequence of an emergent anomaly at low energies. Such a gapless topological phase exhibits exponentially localized edge modes due to the presence of additional gapped degrees of freedom in the bulk.

\item[$\bullet$] \textbf{Purely gSPT phase:} A gapless symmetry-protected topological phase without any additional gapped degrees of freedom in the bulk exhibits algebraically localized edge states, which are usually absent in gapped systems. Furthermore, a purely gapless SPT is non-intrinsic in the sense that it has a gapped counterpart under the same symmetry and dimensionality.

\item[$\bullet$] \textbf{Intrinsically purely gSPT phase:} A gapless symmetry-protected topological phase without a gapped counterpart and additional gapped degree of freedom in the bulk. Unfortunately, to the best of our knowledge, no lattice realization of such exotic gapless phases has been found so far, making it a worthwhile direction to explore.

\item[$\bullet$] \textbf{Bosonic gSPT phase:} A gapless symmetry-protected topological phase is realized in interacting bosonic many-body systems, such as quantum spin and boson systems.

\item[$\bullet$] \textbf{Fermionic gSPT phase:} A gapless symmetry-protected topological phase can be realized in fermionic systems, either as a non-interacting free fermion or as an interacting fermionic system.

\item[$\bullet$] \textbf{Symmetry-flux operator:} This is a nonlocal operator that can be regarded as a string operator at the critical point, where the nontrivial symmetry charge of such a string operator implies the existence of stable topological edge states in certain symmetry-enriched CFTs.

\item[$\bullet$] \textbf{Conformal boundary condition:} A boundary condition that preserves conformal symmetry in the infrared limit and constitutes a fixed point under the boundary renormalization group flow. Conformal boundary conditions are uniquely characterized by the Affleck-Ludwig boundary $g$ function and can be used to determine the universality classes of surface criticality.

\item[$\bullet$] \textbf{Emergent anomaly:} An emergent anomaly can occur whenever the (nonanomalous) microscopic symmetry is not faithfully represented on the gapless modes. This emergent anomaly enforces that the topological phase exists only in gapless systems, leading to the notion of igSPT. In this context, symmetry extension plays a crucial role in the lattice realization. Specifically, the total symmetry is $\Gamma$, fitting into the extension $1 \rightarrow H \rightarrow \Gamma \rightarrow G \rightarrow 1$. One starts from a $G$-symmetric gapless system or CFT with a quantum anomaly, and further stacks an $H$-symmetric SPT on top of $G$ defects. Due to the nontrivial symmetry extension, the induced gapped sector carries an opposite quantum anomaly that cancels the anomaly in the gapless sector, resulting in an anomaly-free model.

\end{itemize}

\color{black}

The rest of the paper is organized as follows: In Sec.~\ref{section2}, we discuss the topological physics of analytically more tractable quantum critical systems---non-interacting free fermion systems---leading to the concept of fermionic gSPT states. In Sec.~\ref{section3}, we extend the discussion to more challenging interacting critical spin systems, including both critical points and stable critical phases, collectively referred to as bosonic gSPT states. 
Finally, in Sec.~\ref{section4}, we conclude with a discussion of related developments and potential future directions.

\section{Topological physics in free fermion quantum critical systems}
\label{section2}

\subsection{Topology and edge modes at quantum critical point }
\label{section2.1}
To begin, we provide a comprehensive understanding of topological physics in quantum critical systems by exploring exactly solvable free-fermion models. Specifically, we explain how topology can protect exponentially localized zero-energy edge modes at critical points in free-fermion chains with chiral symmetry, belonging to the BDI or AIII symmetry class.

\begin{figure}[h]
    \centering
\includegraphics[width=0.55\linewidth]{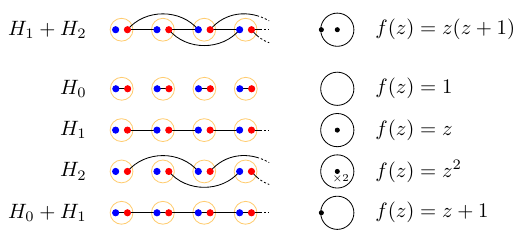}
    \caption{The schematic diagram illustrates the use of the $\alpha$-chain to construct various fermionic gapped and gapless topological phases. Each fermionic site is decomposed into Majorana modes: $\gamma$ (blue) and $\tilde{\gamma}$ (red), with bonds representing hopping terms of varying range in the Hamiltonian. 
    $H_0 + H_1$ corresponds to the standard topologically trivial critical Majorana chain, while $H_1 + H_2$ represents the topologically nontrivial critical Majorana chain with edge modes. 
    The associated complex function $f(z)$ (discussed in the main text) and its zeros in the complex plane are also depicted. The figures are adapted from Ref.~\cite{Verresen2018prl}.}
    \label{fig4}
\end{figure}

There has been significant progress in studying gapless phases exhibiting edge modes~\cite{Kestner2011prb,Cheng2011prb,Fidkowski2011prb,Sau2011prb,grover2012quantumcriticalitytopologicalinsulators,Kraus2013prl,Keselman2015prb,Baum2015PRB,Iemini2015prl,Lang2015prb,Montorsi2017prb,Ruhman2017prb,Scaffidi2017prx,JIANG2018753,Zhang2018prl,Ji2020PRR} in the past few years. 
However, these edge modes are typically exponentially localized and originate from gapped degrees of freedom. The first systematic exploration of topological edge states in critical systems was pioneered by Verresen et al.~\cite{Verresen2018prl,Verresen2017prb}, who introduced a free Majorana fermion chain with varying hopping ranges, referred to as the $\alpha$-chain, as depicted in Fig.~\ref{fig4}. 
In this model, the fermionic operator $c_n,c_n^\dagger$ at each site can be decomposed into two Majorana modes: $\gamma_n = c_n^\dagger + c_n$ and $\tilde \gamma_n = i(c_n^\dagger - c_n)$, represented by the red and blue circles in Fig.~\ref{fig4}. 
The model respects both particle-hole ($\mathcal{C}$) and time-reversal symmetry ($\mathcal{T}$), placing it in the BDI symmetry class (or equivalently, in the AIII symmetry class for complex fermion representation) according to the AZ ten-fold classification scheme~\cite{Altland1997prb,Kitaev2009AIP,Ryu_2010}. 
In this section, we first introduce the physical picture of the non-interacting $\alpha$ chain before discussing the general theory. 
The $\alpha$-chain is defined as:
\begin{equation} 
H_\alpha = \frac{i}{2} \sum_n \tilde \gamma_n \gamma_{n+\alpha} \qquad (\alpha \in \mathbb Z) \; . 
\end{equation} 
A schematic representation of the distinct gapped and gapless phases is illustrated in Fig.~\ref{fig4}. For $\alpha = 1$, the model reduces to the familiar Kitaev chain, which hosts a decoupled Majorana mode near the boundaries~\cite{Kitaev_2001}. The Hamiltonian $ H_\alpha $ exhibits $ |\alpha| $ Majorana zero modes per edge, effectively corresponding to a stack of $|\alpha|$ Kitaev chains. 
To diagonalize this Hamiltonian, we express the $\alpha$-chain $H_\alpha = \frac{i}{2} \sum \tilde \gamma_n \gamma_{n+\alpha}$ in terms of the Fourier modes of the complex fermion $c_k$, which yields:
\begin{equation} 
H_\alpha = - \frac{1}{2}\sum_k \left( c_k^\dagger \quad c_{-k} \right) \; H_{\alpha,k} \; \left(\begin{array}{c} c_k \\ c_{-k}^\dagger \end{array}\right) \qquad \textrm{ where } H_{\alpha,k} = \cos(-k\alpha) \; \sigma^z + \sin(-k \alpha ) \; \sigma^y \; . 
\end{equation} 
For the full Hamiltonian $H = \sum_\alpha t_\alpha \; H_\alpha$ we obtain:
\begin{equation} 
H = - \frac{1}{2}\sum_k \left( c_k^\dagger \quad c_{-k} \right) \; H_k \; \left(\begin{array}{c} c_k \\ c_{-k}^\dagger \end{array}\right) \qquad \textrm{ where } H_k = \varepsilon_k \; \left( \cos(-\varphi_k) \; \sigma^z + \sin(-\varphi_k) \; \sigma^y \right) \; . 
\end{equation} 
Here, $\varphi_k$ represent Bogoliubov rotation angle, and $\varepsilon_k$ denotes the single-particle spectrum. The Hamiltonian $ H_k $ can be interpreted as a two-dimensional vector, which can be aligned with the $ \sigma^z $-axis by rotating it by $ -\varphi_k $ around the $ \sigma^x $-axis. This rotation is implemented using $ U(\vartheta) = \exp(-i \vartheta \sigma^x / 2) $, yielding $H_k = \varepsilon_k \; U(\varphi_k) \; \sigma^z \; U(-\varphi_k)$. Hence, 
\begin{align} 
H &= - \frac{1}{2}\sum_k \varepsilon_k \; \left( d_k^\dagger \quad d_{-k} \right) \; \sigma^z \; \left(\begin{array}{c} d_k \\ d_{-k}^\dagger \end{array}\right) = - \sum_k \varepsilon_k \; d_k^\dagger d_k + \frac{1}{2} \sum_k \varepsilon_k \; .
\end{align}
Now, we consider a critical point between the phases described by the fixed-point Hamiltonians $ H_{\alpha} $, specifically $ H_0 + H_1 $ and $ H_1 + H_2 $. 
As illustrated in Fig.~\ref{fig4}, it is unambiguously demonstrated that although the critical points in the fermionic chains $ H_0 + H_1 $ and $ H_1 + H_2 $ belong to the same universality class---both described by a CFT with central charge $ c = 1/2 $---they cannot be smoothly connected without undergoing a new phase transition. 
This is due to the presence of a decoupled Majorana mode near the boundaries in $ H_1 + H_2 $, which is absent in $ H_0 + H_1 $. This distinction brings out the notion of nontrivial topology at the quantum critical point, also referred to as a fermionic gSPT phase, since the emergence of nontrivial gapless topology occurs in a free-fermion critical system.

\begin{figure}[h]
    \centering
\includegraphics[width=0.65\linewidth]{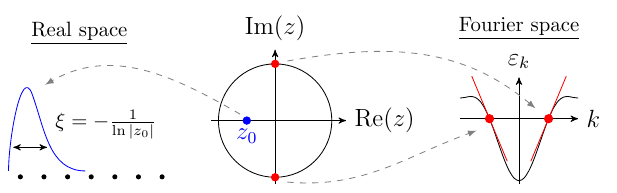}
    \caption{The middle figure displays the zeros within the unit disk (blue) and on the unit circle (red). The former represents an edge mode (per edge) with a localization length of $\xi = -\frac{1}{\ln |z_0|}$, while the latter corresponds to massless Majorana fields in the low-energy limit, each with a central charge of $c=1/2$. The figures are adapted from Ref.~\cite{Verresen2018prl}.}
    \label{fig5}
\end{figure}

A natural question arises: How can we topologically classify the critical $\alpha$-chains within the BDI class? More specifically, what is the topological invariant in this case? To address this, Verresen et al.~\cite{Verresen2018prl} first proposed a topological invariant based on a complex function that counts its zeros and poles inside the unit circle. Specifically, the set $\{H_\alpha\}_{\alpha \in \mathbb{Z}}$ forms a basis for any translationally invariant Hamiltonian in this class:  
\begin{equation}
H_\textrm{BDI} = \frac{i}{2}\sum_{\alpha=-\infty}^{+\infty} t_\alpha \; \left( \sum_{n \in \textrm{sites}} \tilde \gamma_n \gamma_{n+\alpha} \right) = \sum_{\alpha} t_\alpha H_\alpha \; ,\label{ham:BDI} 
\end{equation} 
where we assume $t_\alpha$ is nonzero only for a finite number of $\alpha$. 
The Hamiltonian $H_\textrm{BDI}$ is thus fully determined by the set of coefficients $\{t_\alpha\}$ or equivalently by its Fourier transform, $f(k) := \sum_\alpha t_\alpha e^{ik\alpha}$.  
In this representation, $H_\textrm{BDI}$ can be efficiently diagonalized. If $f(k) = \varepsilon_k e^{i\varphi_k}$ (where $\varepsilon_k, \varphi_k \in \mathbb{R}$), the topological invariant for gapped phases is the winding number of $f(k)$ around the origin. Since $\varepsilon_k \neq 0$, the phase $e^{i\varphi_k}$ defines a well-behaved mapping from $S^1$ (the Brillouin zone) to $S^1$ (the complex unit circle). However, this invariant fails when the system becomes gapless. 
To overcome this limitation, Verresen et al.~\cite{Verresen2018prl,Jones2019JSP} extended the framework using analytic continuation: the function $f(k)$ can be interpreted as an analytic function restricted to the unit circle in the complex plane. 
Denoting $z = e^{ik}$, we rewrite $f(k)$ as $f(z) = \sum_{\alpha=-\infty}^{\infty} t_\alpha \; z^\alpha $. 
If $f(z)$ has no zeros on the unit circle (i.e., the system is gapped), Cauchy's principle states that the winding number equals the difference between the number of zeros ($N_z$) and the order of the pole ($N_p$) within the unit disk. For gapless systems, where zeros lie \emph{on} the unit circle, the standard winding number definition breaks down. However, the quantity $N_z - N_p$ remains well-defined and can thus be regarded as a topological invariant in gapless systems. Specifically, a (non-degenerate) zero $e^{ik_0}$ of $f(z)$ implies that $\varepsilon_k \sim k - k_0$, contributing a massless Majorana fermion with central charge $c = \frac{1}{2}$ to the critical field theory (see Fig.~\ref{fig5}). 
With these theoretical insights, several fundamental questions naturally arise:
\begin{itemize}
\item[$\bullet$] How do we characterize topological edge states at critical points?
\item[$\bullet$] Given two gapped 
topological phases, what is the universality class of the critical point between them? 
\end{itemize}

To address these questions, 
if two Hamiltonians share the same CFT description, one might expect them to belong to the same universality class. 
However, we have seen that $H_1 + H_2$ and $H_0 + H_1$ exhibit different topological invariants ($\omega = 1$ and $\omega = 0$, respectively), despite having the same CFT description. 
This topological distinction can be understood as follows: Any translation-invariant Hamiltonian $H_\textrm{BDI}$ can be associated with a complex function $f(z)$, whose zeros and poles characterize its topological properties, as illustrated in Fig.~\ref{fig5}. 
Let ${z_i}, i=1,...,N_z$ denote the set of distinct zeros of $f(z)$ with corresponding multiplicities ${m_i}$, and let $N_p$ represent the order of the pole at the origin. Then, the topological invariant for both gapped and gapless states is given by $\omega = N_z - N_p$. For gapped phases, there are no zeros on the unit circle of $f(z)$, resulting in $c=0$, since $c$ is half the number of zeros on the unit circle. 
In this context, $\omega$ reduces to the winding number, which typically characterizes nontrivial gapped topological phases in the BDI or AIII symmetry classes~\cite{Hasan2010rmp,Qi2011rmp}. However, in gapless phases, zeros exist on the unit circle of $f(z)$, with their multiplicity defining the dynamical exponent. 
Furthermore, the zeros within the unit disk correspond to a localized edge mode near the boundary, even without a bulk energy gap. For example, the two critical Hamiltonians indeed have a zero on the unit circle, and the edge modes are determined by the zeros strictly within the unit disk, as shown in Fig.~\ref{fig4}. 
Consequently, we can directly infer that both critical chains have a central charge of $c=1/2$ since each has exactly one zero on the unit circle. 
However, the critical chain $H_1 + H_2$ has a nontrivial topological invariant $\omega=1$ because it possesses one zero within the unit disk, leading to topologically protected edge modes. 
Based on these observations, we summarize two key theorems in Ref.~\cite{Verresen2018prl} that answer the questions posed above:

\begin{itemize}
\item[$\bullet$] \textbf{Theorem 1}: For free-fermion quantum critical systems belonging to the BDI class: (1) if the topological invariant $\omega> 0$, then each boundary has $\omega$ Majorana zero modes, (2) the modes have localization length $\xi_i = -\frac{1}{\ln|z_i|}$ where $\{ z_i \}$ are the $\omega$ largest zeros of $f(z)$ within the unit disk.
\item[$\bullet$] \textbf{Theorem 2}: As before, $2c$ should be understood as counting the number of zeros on the unit circle. 
If these zeros are non-degenerate, the bulk is a CFT with central charge $c$, which is generically equal to $\frac{|\omega_1 - \omega_2|}{2}$.
\end{itemize}

From a lattice simulation perspective, we want to identify lattice observables that correspond to the continuum CFT description discussed earlier. 
In the BDI or AIII symmetry class, both gapped and gapless phases are characterized by a topological invariant $\omega$ and a central charge $c$. Reference~\cite{Jones2019JSP} establishes a lattice-continuum correspondence that bridges this gap. 
Specifically, in gapped topological phases, non-local string order parameters allow us to extract the winding number $\omega$. However, in gapless phases, the scaling dimensions of these string operators serve as order parameters, encoding both $\omega$ and $c$, thereby providing a lattice-accessible observable for these crucial quantities.

\begin{figure}[h]
    \centering
\includegraphics[width=0.65\linewidth]{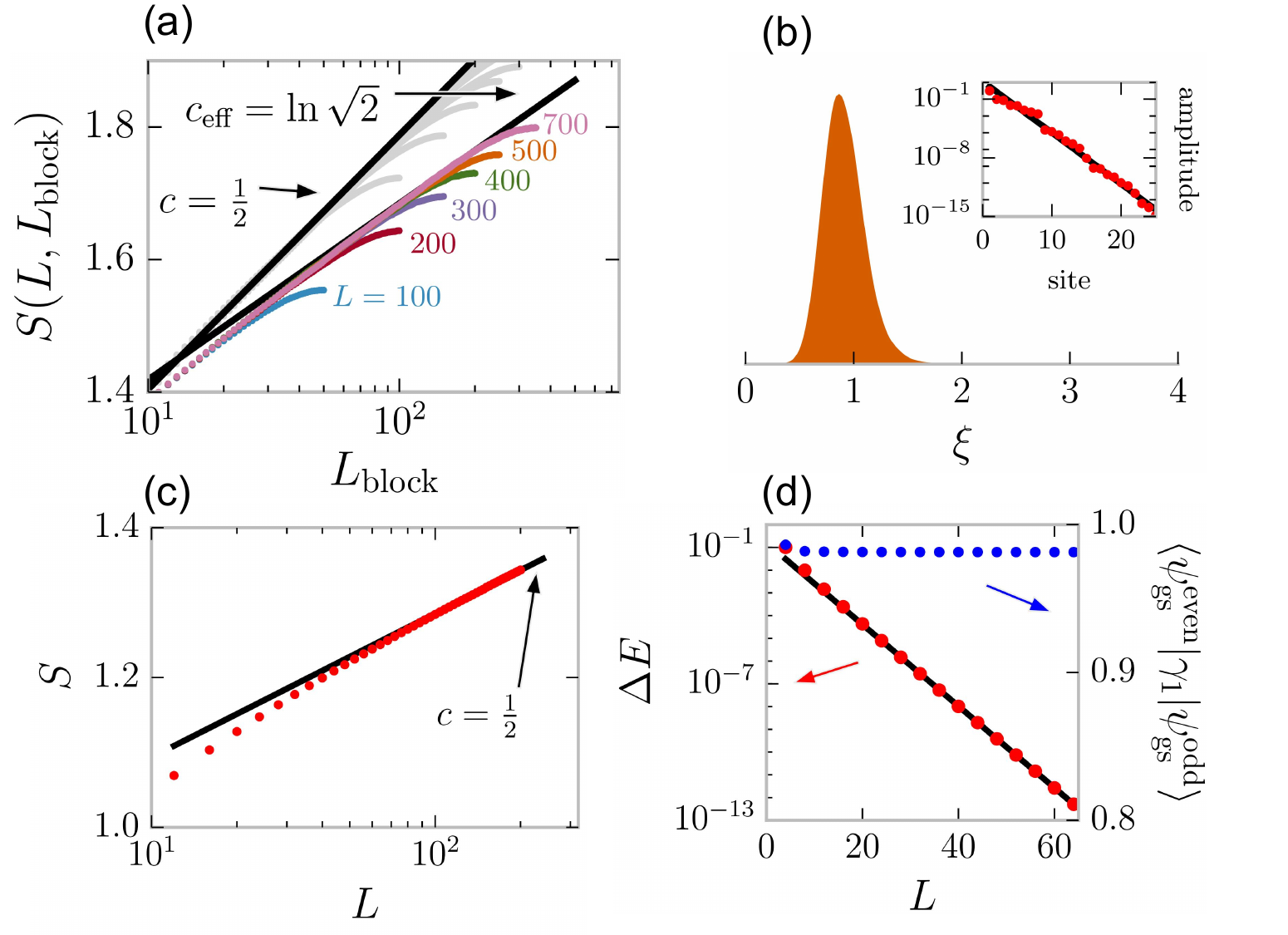}
    \caption{(a) Entanglement scaling (averaged over $10^5$ states) across different system sizes $L$ suggests an infinite randomness fixed point with $c_\textrm{eff} = \ln \sqrt{2}$ (black lines are for reference; gray represents the clean case). (b) Distribution of edge mode localization lengths over disorder realizations (inset: edge mode for a single realization). (c) The bulk is described by the $c = \frac{1}{2}$ Majorana CFT (black line serves as a guide). (d) The energy splitting between fermionic parity sectors is exponentially small. The figures are adapted from Ref.~\cite{Verresen2018prl}.}
    \label{fig6}
\end{figure}

Additionally, the nontrivial topological edge modes of the critical Majorana $\alpha$ chain can remain stable in the presence of disorder and interactions. 
To demonstrate this, we first examine the effects of disorder~\cite{Verresen2018prl,Monthus_2018,Duque2021prb}. 
Specifically, consider the Hamiltonian $H = \frac{i}{2} \sum_{\alpha=0}^3 \sum_n t_{\alpha}^{(n)} \tilde{\gamma}_n \gamma_{n+\alpha}$. 
For the clean model, $ t_1^{(n)} = t_2^{(n)} = 1 $ and $ t_0^{(n)} = t_3^{(n)} = a $, where $ -1 < a < \frac{1}{3} $, the system is critical with $ c = \frac{1}{2} $ and $ \omega = 1 $. This setup reduces to $ H_1 + H_2 $ when $ a = 0 $. To introduce a strong disorder, one can independently draw $ t_1^{(n)} $ and $ t_2^{(n)} $ from a uniform distribution over $[0, 1]$ and $ t_0^{(n)} $ and $ t_3^{(n)} $ from $[-0.5, 0]$. 
As confirmed in the reference~\cite{Verresen2018prl}, the system flows to the infinite random fixed point, with an effective central charge $ c_\textrm{eff} = \ln \sqrt{2} $. Furthermore, the average entanglement entropy $ S(L, L_{\textrm{block}}) $ scales asymptotically as $ S \sim \frac{c_\textrm{eff}}{3} \ln L_\textrm{block} $ (for $ 1 \ll L_\textrm{block} \ll \frac{L}{2} $), as shown in Fig.~\ref{fig6} (a). 
Under open boundary conditions, we observe one exponentially localized Majorana edge mode per edge. Figure~\ref{fig6} (b) shows the distribution of localization lengths across disorder realizations, with the inset showing a specific example where the amplitude of the edge mode is exponential decay with lattice distance.

Next, we consider the effects of interactions, which are known to reduce the gapped classification from $ \mathbb{Z} $ to $ \mathbb{Z}_8 $~\cite{Fidkowski2010prb}. 
Consider the interacting Hamiltonian~\cite{Verresen2018prl}:  \begin{equation} \label{ham:interaction}
H = H_1 + H_2 + U \sum_{n=1}^L \gamma_{n} \gamma_{n+1} \gamma_{n+2} \gamma_{n+3} + (\gamma \leftrightarrow \tilde \gamma)\, .
\end{equation}
The critical point does not shift for $ U \neq 0 $ because \eqref{ham:interaction} is self-dual under the transformations $ \gamma_n \to \gamma_{3-n} $ and $ \tilde{\gamma}_n \to \tilde{\gamma}_{-n} $. 
Using the DMRG method, the authors of reference~\cite{Verresen2018prl} performed finite-size scaling under open boundary conditions for $ U = 0.3 $. 
In Fig.~\ref{fig6} (c), the quantum criticality of the interacting Hamiltonian is confirmed through the scaling of the entanglement entropy of a bipartition into two halves of length $ L/2 $, consistent with the CFT prediction $ S \sim \frac{c}{6} \ln L $~\cite{ginsparg1988appliedconformalfieldtheory,francesco2012conformal}. Figure~\ref{fig6} (d) illustrates the ground state degeneracy under open boundary conditions. 
These states differ only near the edges, since $ \langle \psi_\textrm{gs}^\textrm{even} | \gamma_1 | \psi_\textrm{gs}^\textrm{odd} \rangle $ remains finite as $ L \to \infty $, where $\psi_\textrm{gs}^{\textrm{odd/even}}$ is the ground state in odd/even fermi parity subspace. 
Consequently, the topological edge mode at the critical point is protected by symmetry and does not arise from fine-tuning, as its stability against both disorder and interaction effects.

\subsection{The general theory for topological edge modes in free fermion quantum critical systems}
\label{section2.2}
After illustrating the concept of topological edge modes at criticality for the BDI class, we now generalize the existence of topological invariants and localized edge modes at phase transitions across \emph{all} AZ symmetry classes in \emph{various dimensions}. More importantly, we seek to review the progress that has been made in uncovering the underlying mechanism and developing intuitive physical pictures to explain why topological edge modes emerge at these transition points. 

To address these fundamental and intriguing questions, Verresen~\cite{verresen2020topologyedgestatessurvive} proposed a new mechanism aimed at establishing a general theory of topological physics in free-fermion critical systems. 
Specifically, Verresen demonstrated that topological invariants and localized edge modes can persist at phase transitions between the gapped phases with \emph{different nonzero topological invariants} in arbitrary dimensions. For example, while phase transitions in Chern insulators between $\mathcal{C} = 0 \leftrightarrow \mathcal{C} = 1$ and $\mathcal{C} = 1 \leftrightarrow \mathcal{C} = 2$ are both exhibited by a single Dirac cone, however, Verresen generalized the definition of $\mathcal{C}$~\cite{verresen2020topologyedgestatessurvive} to the critical systems with fractional values. 
In this framework, the former transition corresponds to $\mathcal{C} = \frac{1}{2}$ and the latter to $\mathcal{C} = \frac{3}{2}$, making them topologically distinct Dirac cones that are necessarily separated by a phase transition of phase transitions.
Moreover, at the $\mathcal{C} = \frac{3}{2}$ critical point, an exponentially localized chiral edge mode emerges. 
This phenomenon is not fine-tuned in the sense that the topological edge mode remains localized when tuning from a $\mathcal{C} = 1$ Chern insulator to a $\mathcal{C} = 2$ Chern insulator. 

\begin{figure*}[h]
    \centering
\includegraphics[width=1.0\linewidth]{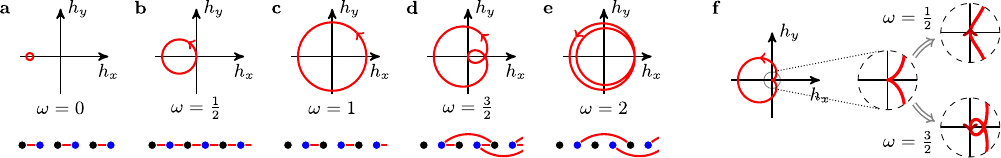}
    \caption{(a)-(e) Gapped and gapless noninteracting quantum phases are characterized by integer and half-integer topological invariants $\omega$, respectively. 
    The Bloch Hamiltonian $\mathcal{H}(\vec{k}) = \vec{h}(\vec{k)} \cdot \vec{\sigma}$ (top) and the corresponding real-space Majorana hopping configuration (bottom) are illustrated. The critical point with $\omega = \frac{3}{2}$ is topologically nontrivial, leading to the emergence of a localized edge mode (see (d)). (f), The mother theory can be perturbed into the five phases shown in (a)-(e). A perturbation linear in $\vec{k}$ (near $\vec{k} \approx 0$) flows to $\omega = \frac{1}{2}$ or $\omega = \frac{3}{2}$, representing a topological transition between two gapless phases. The figures come from the reference.~\cite{verresen2020topologyedgestatessurvive}}
    \label{fig7}
\end{figure*}

We briefly outline the general theory of the topological critical free-fermion systems, using the familiar one-dimensional BDI or AIII symmetry class model as an illustrative example. 
This framework naturally generalizes to arbitrary dimensions and symmetry classes, as discussed in detail in Ref.~\cite{verresen2020topologyedgestatessurvive}. 
We begin by introducing topological invariants and then discuss the associated edge modes. 
To define and understand topological invariants at criticality, we consider one-dimensional free-fermion topological insulators protected by sublattice symmetry (AIII class) or topological superconductors protected by spinless time-reversal symmetry (BDI class)~\cite{Altland1997prb}. 
Both classes share a unified description through the single-particle Hamiltonian $\mathcal{H}(\vec{k}) = \vec{h}(\vec{k})\cdot \vec{\sigma}$, where symmetry enforces $h_z(k) = 0$ in some basis, and $\vec{\sigma}=(\sigma_x,\sigma_y,\sigma_z)$ represents the Pauli matrices. 
Consequently, $\vec{h}(\vec{k})$ can be visualized as a closed loop in the $x$-$y$ plane, with its winding number around the origin serving as the topological invariant for gapped phases, as illustrated in Fig.~\ref{fig7} (a), (c), and (e) for windings $\omega = 0, 1, 2$. 
However, as shown in Fig.~\ref{fig7} (b) and (d), two such loops can appear similar near $\vec{k} \approx 0$, where $\vec{h}(\vec{k})$ is linear. This suggests that the low-energy theory corresponds to a relativistic fermion and represents the transition points between the gapped phases. 
Additionally, Fig.~\ref{fig7} (d) shows that at large momentum (high-energy sector), the transition between $\omega = 1$ and $\omega = 2$ also encloses the origin. 
To incorporate topological information from this high-energy sector, Verresen introduced an invariant~\cite{verresen2020topologyedgestatessurvive} by calculating the winding number while excluding an infinitesimal region around the transition point: $\omega = \frac{1}{2\pi} \lim_{\varepsilon \to 0} \int_{|\vec{h}| > \varepsilon} \mathrm{d} \left( \textrm{arg}(\vec{h})\right) \in \frac{1}{2} \mathbb{Z}.\label{eq:defomega} $ This definition ensures that the invariant is equal to the average of the topological invariants of the neighboring gapped phases. 
For the two critical points shown in Fig.~\ref{fig7} (b) and (d), the corresponding topological invariants are $\omega = \frac{1}{2}$ and $\frac{3}{2}$, respectively. To gain intuition about why half-integer values arise, we note that a line passing through the origin spans a $180^\circ$ angle. 
As long as $\partial_k \vec{h}(0) \neq 0$, $\omega$ remains invariant, making it a robust characteristic of the low-energy universality class. The only way to interpolate between these two critical points is by tuning through a multicritical point that is no longer described by a single relativistic fermion. A possible ``topological phase transition of phase transitions'' is depicted in Fig.~\ref{fig7} (f), where the dispersion becomes quadratic ($|\vec{h}| \sim k^2$), resulting in a nonconformal multicritical point. 
This demonstrates that although transitions may locally share the same universality class, they can be globally distinguished by their topological properties. Furthermore, to illustrate the above theory in two dimensions, the Chern number can also be extended to cases where the low-energy theory is local (i.e., $\vec{h}(\vec{k})$ is smooth wherever it vanishes):
\begin{equation}
\mathcal C = \frac{1}{4\pi} \lim_{\varepsilon \to 0} \iint_{|\vec{h} (\vec{k})|>\varepsilon} \vec{\hat h} \cdot \big( \partial_{k_x} \vec{\hat h} \times \partial_{k_y} \vec{\hat h} \big)
\mathrm d^2 \vec{k} \quad \in \frac{1}{2} \mathbb Z, \label{eq:defchern}
\end{equation}
where $\vec{\hat{h}} = \frac{\vec{h}}{|\vec{h}|}$. The transition between $\mathcal{C} = n$ and $\mathcal{C} = n+1$ corresponds to a Dirac cone with $\mathcal{C} = n + \frac{1}{2}$. This aligns with the parity anomaly in $(2+1)$-dimensional QED, which necessitates a Chern-Simons term with a half-integer prefactor~\cite{Haldane1988prl,Redlich1984prl,ALVAREZGAUME1985288,Seiberg2016PTEP}. This framework also allows for topologically distinct Dirac cones separated by multicritical points. For instance, a transition between $\mathcal{C} = \frac{1}{2}$ and $\mathcal{C} = \frac{3}{2}$ could involve the Dirac cone becoming quadratic. 

Having established bulk topological invariants, we now turn to one of their most intriguing consequences: topologically protected edge states. 
In gapped topological phases, this phenomenon can be understood as a spatial interface between a topological phase and a trivial ``vacuum''. 
At the boundary between two topologically distinct gapped regions, the interface must be gapless, leading to the emergence of a degenerate zero-energy mode. This concept extends to the critical case, where exponentially localized edge modes appear despite the absence of a gap. For simplicity, we begin by reviewing edge modes in the one-dimensional setting.

\begin{figure}[h]
    \centering
\includegraphics[width=0.65\linewidth]{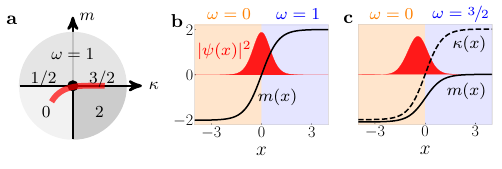}
    \caption{(a) The schematic phase diagram near the multi-critical point in Fig.~\ref{fig7}(f), showing the values of the topological invariant.
    The critical line with $\omega = \frac{3}{2}$ is not perturbatively connected to the trivial phase ($\omega = 0$). 
    Any path connecting the two (e.g., the red line) must pass through a transition point. 
    (b) A spatial interface between the trivial and topological gapped phases. This interface hosts an exponentially localized zero-energy mode (red) resulting from the band inversion mechanism. (c) A spatial interface between a topologically nontrivial critical point and the vacuum. Here, a localized edge mode arises again, but this time due to the kinetic inversion mechanism. The figures are adapted from Ref.~\cite{verresen2020topologyedgestatessurvive}.}
    \label{fig8}
\end{figure}

The claim is that the transition points between gapped phases with $\omega = n$ and $\omega = n+1$ supports $n$ exponentially localized edge modes, as illustrated in Fig.~\ref{fig7} (b) and (d) for $n = 1$. Importantly, the phase $\omega = \frac{3}{2}$ cannot be continuously deformed into the trivial phase $\omega = 0$ without undergoing a phase transition. A representative phase diagram is shown in Fig.~\ref{fig8} (a), where the red path interpolating between $\omega = \frac{3}{2}$ and $\omega = 0$ necessarily passes through a multicritical point. 
Even if one can avoid this direct path by traversing the neighboring gapped regions where $\omega = 1$ or $\omega = 2$, a phase transition to $\omega = 0$ remains unavoidable. 
This unavoidable singularity between $\omega = 0$ and $\omega = \frac{3}{2}$ guarantees the presence of an edge mode at their spatial interface, analogous to the case of gapped topological phases~\cite{bernevig2013topological}.

We first focus on topological edge states in gapped phases. To this end, we establish a spatial interface between the $\omega = 0$ and $\omega = 1$ topological phases, which requires a model that realizes both phases. This can be achieved by shifting the loop in Fig.~\ref{fig7} (b) to the left or right, described by $ h(k) = h_x + i h_y = (e^{ik} - 1) + m $, which is trivial for $ m < 0 $ and topologically nontrivial for $ m > 0 $. To examine the universal low-energy physics, we expand near the gap-closing point at $ \vec{k} \approx 0 $, yielding $ h(k) \approx ik + m $. Transforming to real space, we obtain $ D_x := h(-i\partial_x) = \partial_x + m $, leading to the Hamiltonian:
\begin{equation}
\mathcal H(x) = \left( \begin{array}{cc} 0 & D_x^\dagger \\ D_x & 0\end{array} \right)
\textrm{ has }
\omega = \left\{ \begin{array}{ll}
0 & \textrm{if } m <0,\\
1 & \textrm{if } m >0.
\end{array} \right. \label{eq:gapped}
\end{equation}
Next, we consider a mass profile $ m(x)$ that interpolates between the $ \omega = 0 $ and $ \omega = 1 $ gapped phases, as illustrated in Fig.~\ref{fig8} (b). Jackiw and Rebbi~\cite{Jackiw1976prd} showed that such a configuration supports a localized zero-energy mode. Specifically, the zero-energy condition $ D_x \psi(x) = 0 $ leads to the solution: $ D_x \psi(x) = 0 \Rightarrow \psi(x) \propto e^{-\int_0^x m(x')}\label{eq:jackiwrebbi}$, which demonstrates that when the sign of the mass term reverses, an exponentially localized edge state emerges at the interface, with an inverse decay length $ \sim 1/|m(\pm\infty)|$. This phenomenon of mass (or band) inversion underlies the bulk-boundary correspondence in topological insulators and superconductors~\cite{bernevig2013topological,shen2012topological}.

The generalization to the gapless case is conceptually straightforward but leads to surprising results. As before, to establish an interface between $\omega = 0$ and $\omega = \frac{3}{2}$, we require a model that realizes both phases. A natural candidate for this is the multicritical point separating $\omega = \frac{1}{2}$ and $\omega = \frac{3}{2}$, where $h(k) \approx k^2$. By incorporating linear and constant perturbations, we obtain $h(k) = k^2 - i\kappa k + m$, which in real space takes the form:  
\begin{equation}
D_x := h(-i\partial_x) = -\partial_x^2 - \kappa(x) \partial_x + m(x). \label{eq:criticalinterface}
\end{equation}
In particular, setting $m = 0$ corresponds to the transition depicted in Fig.~\ref{fig7}(f), while taking $\kappa \to -\infty$ (with $m/\kappa$ finite) recovers the gapped case. Here, $\kappa$ and $m$ are referred to as the kinetic and mass terms, respectively.  Now, we consider a spatial interface between the topological critical phase $\omega = \frac{3}{2}$ and the trivial vacuum, as shown in Fig.~\ref{fig8} (a,c). In this case, it is the \emph{kinetic} term, rather than the mass term, that changes sign across the transition. This guarantees an exponentially localized solution to the zero-energy equation $D_x \psi(x) = 0$ near the interface. For $x \to +\infty$ (where $m \to 0$ and $\kappa$ approaches a constant, as shown in Fig.~\ref{fig8}(c)), we obtain:  
\begin{equation}
D_x = -\partial^2 - \kappa \partial = -\partial(\partial + \kappa).
\end{equation}
Thus, one solution remains constant while the other decays:  
\begin{equation}
\psi(x) \sim \left\{ \begin{array}{ll}
\exp\left( - \kappa x \right) & \textrm{as } x \to + \infty, \\
\exp \left( - \frac{\kappa x}{2} \left[ 1 \pm \sqrt{ 1 + \frac{4m}{\kappa^2} } \right]  \right) & \textrm{as } x \to - \infty. \\
\end{array} \right. \label{eq:psicrit}
\end{equation}
This exponential localization requires $\kappa > 0$ asymptotically on the right and $\kappa < 0$ on the left. 
This novel phenomenon, termed \emph{kinetic inversion}, underscores the stability of the edge mode against arbitrary deformations of $D_x$ (e.g., higher-order derivative corrections), which can be rigorously established via the index theorem~\cite{Atiyah1968aom,Fukaya2017prd} for gapless systems~\cite{verresen2020topologyedgestatessurvive}.  
In the limit $ x \to +\infty $, Eq.~\eqref{eq:criticalinterface} simplifies to $ D_x \propto -\partial_x - \frac{1}{\kappa} \partial_x^2 $. 
Hence, the localization length of the edge mode, given by $ \sim 1/\kappa $, represents an irrelevant perturbation in the sense of renormalization group compared to the linear kinetic term. 
This perspective is quite general, as even in the gapped case, the localization length $ \sim 1/m $ serves as the prefactor of the kinetic term that is more irrelevant relative to the mass term. 

Finally, we note that the stability of topological edge modes at criticality with interaction can be understood through anomaly considerations. 
The prefactors of topological response terms can remain quantized even in a gapless bulk, as exemplified by the parity anomaly of the Dirac cone~\cite{verresen2020topologyedgestatessurvive,Verresen2021prx}. 
Furthermore, edge modes can remain stable in the presence of interactions, a phenomenon referred to as symmetry-enriched criticality~\cite{Verresen2021prx} and gSPT phases~\cite{Scaffidi2017prx}, which will be reviewed in the Sec.~\ref{section3}.  
A key qualitative effect of interactions is that they cause the edge mode to become algebraically localized. For instance, at $\omega = \frac{3}{2}$, a universal finite-size energy splitting scales as $\sim 1/L^{14}$---a contribution that dominates over exponential terms. 
Determining the scaling power for algebraically localized edge modes requires a case-by-case analysis, and extending this framework to higher-dimensional systems remains an intriguing direction for future research, which we will briefly discuss in the last few sections.

\subsection{Topological semimetal as a special class of fermionic gSPT phases}
\label{section2.3}
This review focuses on gapless topological phases, which have traditionally referred to topological semimetals and have been extensively studied in the condensed matter community, attracting significant theoretical and experimental interest over the past decade~\cite{Wan2011prb,Burkov2016NM,Lv2021rmp,yan2017topological,hasan2017discovery,Armitage2018rmp,vafek2014dirac}. 
In these systems, gaplessness arises from discrete points in momentum space, such as Weyl and Dirac points, which are protected by certain symmetries. However, in this section, we highlight the key differences between topological semimetals and gSPT states---the primary focus of this review---showing that topological semimetals can be understood as a specific subclass of fermionic gSPT phases. 

\begin{figure}[h]
    \centering
\includegraphics[width=0.65\linewidth]{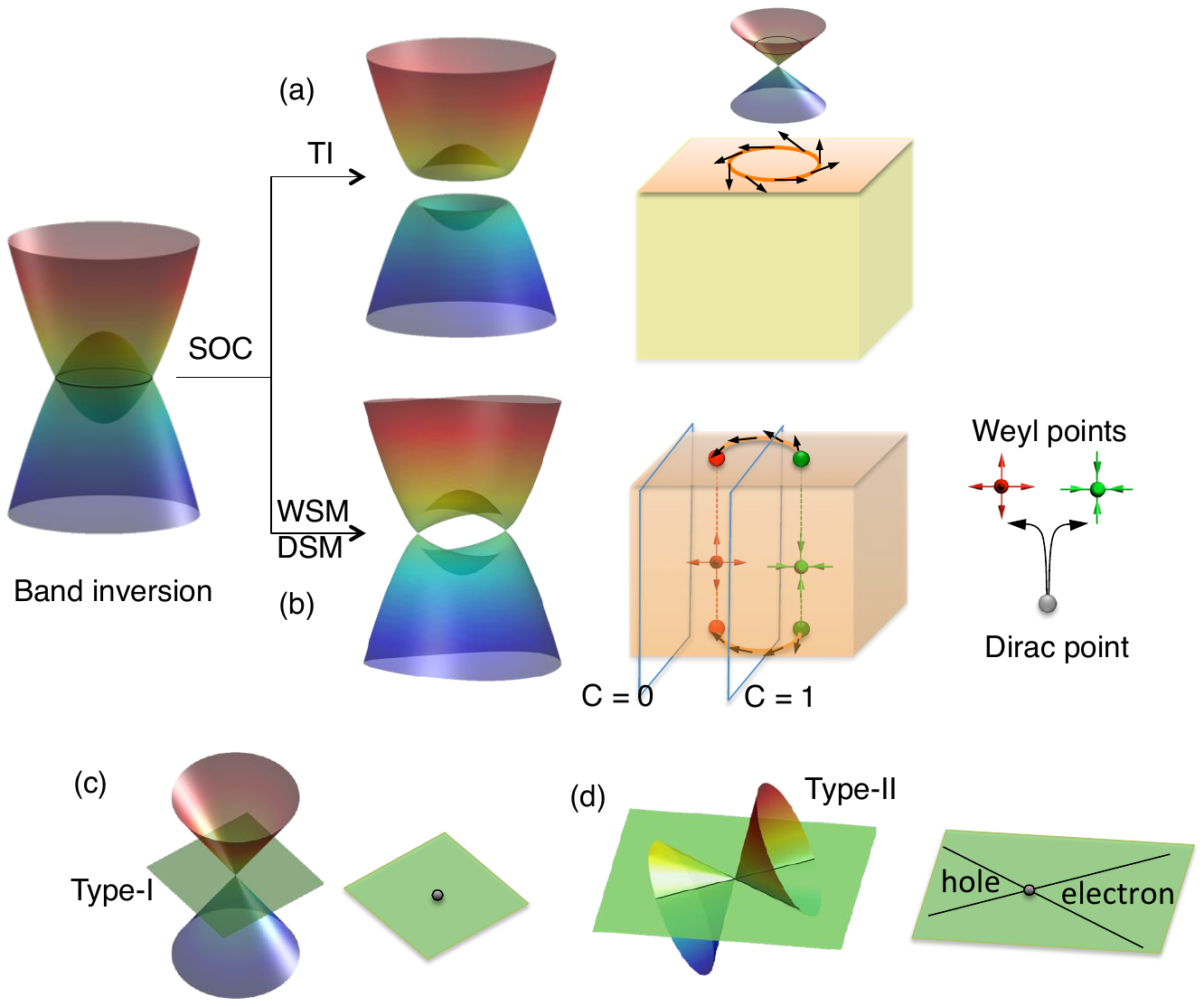}
    \caption{The topologies of Weyl semimetals (WSMs) and Dirac semimetals (DSMs) arise from similar inverted band structures. (a) In the presence of spin-orbit coupling (SOC), a full energy gap opens following the band inversion, leading to a topological insulator. 
    (b) In three-dimensional momentum space, the bulk bands are gapped by spin-orbit coupling except at isolated, linearly crossing points (Weyl or Dirac points). This results in the emergence of a Weyl or Dirac semimetal, which can be viewed as a three-dimensional analog of graphene. 
    Owing to the topology of the bulk bands, topological surface states appear and form exotic Fermi arcs on the surface of materials. In a DSM, all bands are doubly degenerate, while in a WSM, the degeneracy is lifted due to the breaking of inversion symmetry, time-reversal symmetry, or both. 
    (c) Type-I WSM: When the Fermi energy is sufficiently close to the Weyl points, the Fermi surface shrinks to zero at the Weyl points. (d) Type-II WSM: The strong tilting of the Weyl cone causes the Weyl point to act as the touching point between electron and hole pockets in the Fermi surface. The figures are adapted from Ref.~\cite{yan2017topological}.}
    \label{fig9}
\end{figure}

To this end, we first briefly introduce the concept of Weyl semimetals as a typical example of topological semimetals, emphasizing their nontrivial topological properties characterized by gapless points. Further details on this topic can be found in specialized review articles~\cite{Armitage2018rmp,vafek2014dirac}. The key distinction between ordinary metals and insulators lies in whether the Fermi energy lies within an bulk energy gap~\cite{sachdev2023quantum}. 
In contrast, semimetals are defined by the presence of band-touching points or nodes at or near the Fermi energy, 
where two or more bands become exactly degenerate at specific crystal momentum values in the first Brillouin zone (BZ). 
At first glance, such band-touching points may seem highly unstable, as degeneracies are generally lifted unless protected by symmetry, as dictated by fundamental quantum mechanics. However, this naive perspective overlooks the possibility of an accidental degeneracy between a single pair of bands in a three-dimensional material. 
To understand this, consider two bands that touch at a point $\vec{k}_0$ in the first BZ at an energy $\epsilon_0$. Near this point, the momentum-space Hamiltonian can be expanded as a Taylor series around $\vec{k}_0$. 
Neglecting possible anisotropies for simplicity, this expansion takes the form: $ H(\vec{k}) = \epsilon_0 \sigma_0 \pm \hbar v_F (\vec{k} - \vec{k}_0) \cdot \vec{\sigma}$, where $\sigma_0$ is the $2 \times 2$ identity matrix, and $\vec{\sigma} = (\sigma^x, \sigma^y, \sigma^z)$ are the three Pauli matrices. 
This expression represents a general $2 \times 2$ Hermitian matrix written in terms of the identity matrix and the Pauli matrices.  
The crucial observation is that the band-touching point in $H(\vec{k})$ cannot be removed. Adjusting $\epsilon_0$ or $\vec{k}_0$ merely shifts the location of the point in momentum space or modifying the parameter $v_F$, which 
only changes the slope of the band dispersion. 
The point itself remains intact, as illustrated in Fig.~\ref{fig9}. This robustness arises from the fact that the number of crystal momentum components matches the number of Pauli matrices, both being three. Consequently, unremovable band-touching points naturally require three spatial dimensions and nondegenerate bands.  
This second condition cannot be satisfied in materials that preserve both inversion symmetry $P$ and time-reversal symmetry $\mathcal{T}$, as these symmetries ensure that all bands are at least doubly degenerate at every $\vec{k}$. 
This degeneracy follows from the fundamental property of fermionic systems, where $(P \mathcal{T})^2 = -1$. Therefore, unremovable band-touching points can only exist in materials that are either noncentrosymmetric or magnetic. 
Furthermore, if we simply redefine the zero of energy as $\epsilon_0 = 0$, the Hamiltonian transforms into the Weyl Hamiltonian, describing a massless relativistic particle with right-handed or left-handed chirality, as indicated by the $\pm$ sign. Weyl fermions are fundamental building blocks in the Standard Model of particle physics, although all known elementary fermions acquire mass through interactions among fundamental Weyl fermions~\cite{weinberg1995quantum}.  

Generally, the coefficient of the identity matrix $\sigma_0$ is not merely a constant $\epsilon_0$ but rather a function of the crystal momentum, with $\epsilon_0$ representing the zeroth-order term in its Taylor expansion. 
A crucial question is whether this expansion contains a term linear in $\vec{k} - \vec{k}_0$. 
Such a linear term is absent only if $\vec{k}_0$ is a time-reversal-invariant momentum (TRIM), which occurs only at the center or certain high-symmetry points of the BZ. For generic Weyl points that appear away from TRIM, the linear term exists, leading to the following expression:
\begin{equation}
\label{eq:3}
H = \hbar \tilde v_F (\vec{k} - \vec{k}_0) \sigma_0 \pm \hbar v_F (\vec{k} - \vec{k}_0) \cdot \vec{\sigma}, 
\end{equation}
where we set $\epsilon_0 = 0$. When $\tilde{v}_F > v_F$, the two touching bands form electron and hole pockets, and the Weyl point becomes the point where these pockets meet, which is now known as type-II Weyl semimetals~\cite{soluyanov2015type}. An example is MoTe$_2$~\cite{Huang2016NM} (see Fig.~\ref{fig9}). 
In contrast, TaAs, the first discovered Weyl semimetal~\cite{xu2015discovery,Lv2015prx}, belongs to the type-I Weyl semimetals. An alternative approach to realizing a Weyl semimetal is by breaking inversion or time-reversal symmetry, which can be captured by the following minimal Hamiltonian
\begin{equation}
\label{eq:4}
H = \left(
\begin{array}{cc}
\hbar v_F \vec{\sigma} \cdot \vec{k} & m \\
m & - \hbar v_F \vec{\sigma} \cdot \vec{k}
\end{array}
\right), 
\end{equation}
where $m$ is an off-diagonal term that couples two Weyl fermions of opposite chirality located at the same TRIM in the BZ. 
This coupling opens a gap of magnitude $2m$, and the Hamiltonian describes a transition between a topological insulator and an ordinary insulator in three dimensions. 
At the transition point, where $m = 0$, the system exhibits a Dirac semimetal phase with fourfold degeneracy at the Dirac point, rather than the twofold degeneracy at a Weyl point. 
Such a degeneracy requires fine-tuning or additional symmetries to enforce $m = 0$. If $m = 0$ is protected by symmetry, the result is a stable Dirac semimetal phase, distinct from a Weyl semimetal. 
Another scenario involves creating two Dirac points along a rotational axis, protected by rotational symmetry. The corresponding Hamiltonian is given by:  
\begin{equation}
\label{eq:5}
H = \hbar v_F(\tau^x \sigma^z k_x - \tau^y k_y) + m(k_z) \tau^z, 
\end{equation}
where $\vec{\tau}$ and $\vec{\sigma}$ are Pauli matrices representing orbital and spin degrees of freedom, respectively, and $m(k_z) = -m_0 + m_1 k_z^2$. For a given eigenvalue of the spin operator $\sigma = \pm 1$, this Hamiltonian describes a Weyl semimetal with two Weyl points at $\vec{k}_{\pm} = (0, 0, \pm \sqrt{m_0 / m_1})$. 
These Weyl semimetals, corresponding to $\sigma = \pm 1$, are related by time-reversal symmetry. Consequently, each of the two band-touching points $\vec{k}_{\pm}$ contains two Weyl points of opposite chirality, forming Dirac points. This type of Dirac semimetal, characterized by two Dirac points, differs from the previously discussed case of a single Dirac point at a TRIM. Experimental realizations of this phase include materials such as Na$_3$Bi and Cd$_2$As$_3$~\cite{Wang2012prb,Wang2013prb,Liu2014Science,Neupane2014NC}.  

\begin{figure}[h]
    \centering
\includegraphics[width=0.65\linewidth]{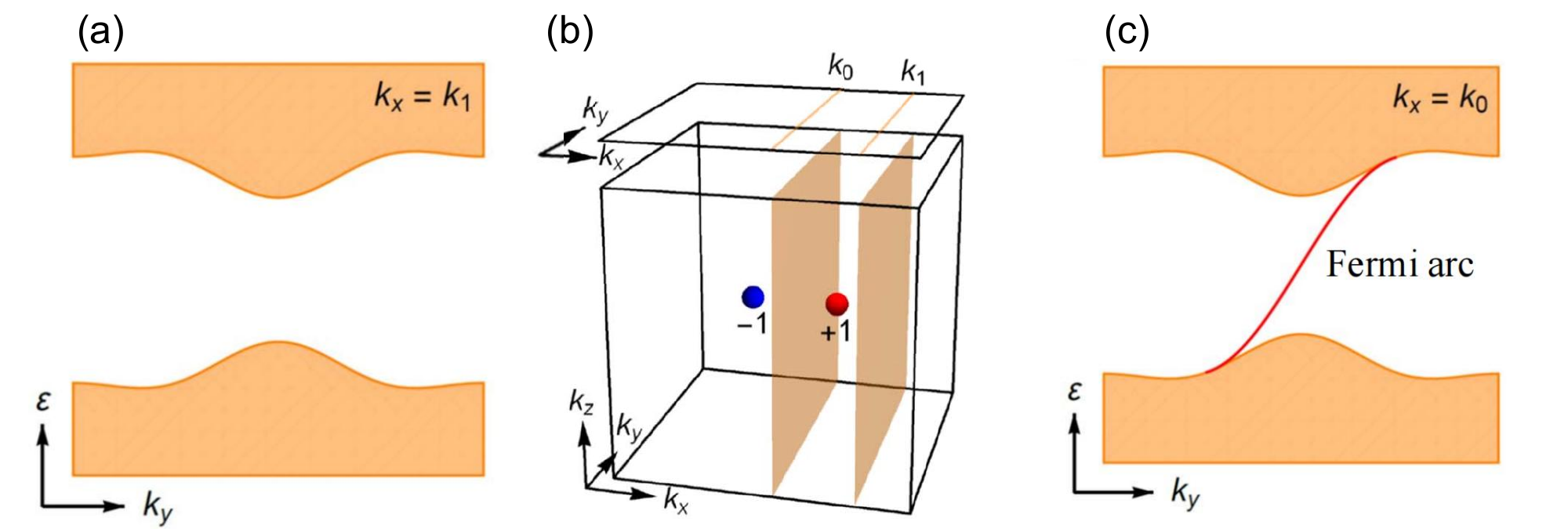}
    \caption{(a) For slices with a Chern number, the one-dimensional edge of the two-dimensional slice is gapped. (b) A system with two Weyl points of chirality $\pm1$ and when a slice is swept through a Weyl point, the two-dimensional system undergoes a topological phase transition, and the Chern number changes by $\pm 1$. (c) While slices with a Chern number $\nu = +1$ host a protected gapless chiral edge mode. The Fermi arc can be understood as arising from all of the chiral edge states assembled into a surface state. The figures are adapted from Ref.~\cite{hasan2017discovery}.}
    \label{fig10}
\end{figure}

The integer quantum Hall effect is characterized by gapless chiral edge modes protected by a topological invariant---the Chern number. 
What boundary states are ensured by the nontrivial topological response in topological semimetals? To address this, consider two-dimensional slices of the bulk Brillouin zone, as illustrated in Fig.~\ref{fig10} (a)-(c). 
Any slice that does not intersect the singular points (Weyl or Dirac points) remains gapped, allowing us to compute a well-defined Chern number for that slice. This perspective treats the three-dimensional band structure as a collection of two-dimensional slices parameterized by $k_x$. As $k_x$ varies, the bulk band gap undergoes a series of closures and reopenings, driving the system through topological phase transitions that modify the Chern number. 
The critical slices, located at $ k_x = \pm k_0$, correspond to those containing a singular point.  
Similarly, the surface states of a topological semimetal can be decomposed into one-dimensional edge states within the surface Brillouin zone, as depicted in Fig.~\ref{fig10} (b). 
The slices associated with a nonzero bulk Chern number host gapless chiral edge modes, as shown in Fig.~\ref{fig10} (c), whereas slices corresponding to a zero bulk Chern number are gapped, as in Fig.~\ref{fig10} (a). These one-dimensional edge states collectively form a continuous sheet of surface states, terminating at the surface projections of the singular points and giving rise to a \emph{topological Fermi arc}. 
In a constant-energy cut of the surface band structure, this Fermi arc appears as an open, disconnected curve.  
Much like the Dirac cone surface state of a three-dimensional $ \mathbb{Z}_2$ topological insulator~\cite{shen2012topological,bernevig2013topological}, the Fermi arc of a topological semimetal is anomalous---it cannot exist in an isolated two-dimensional system and is unique to the boundary of a three-dimensional bulk, serving as a manifestation of the chiral anomaly. 
However, the Fermi arc provides an even more striking example of an anomalous band structure. 
Unlike the Dirac cone or any conventional two-dimensional band structure, its constant-energy contours do not form closed loops, making it a distinctive hallmark of topological semimetals
~\cite{WangChong2020PRL,Gioia2021PRR,WangChong2021PRB}. 
Furthermore, \emph{it is crucial to emphasize that a topological semimetal can be regarded as a fermionic gSPT phase protected by translational symmetry, where the Fermi arcs as boundary states exist only at gapped momentum slices}. The key distinction between topological semimetals and conventional gSPT phases lies in their topological robustness: the former relies on space-translational symmetry and is thus vulnerable to destabilization by disorder~\cite{syzranov2018high,Sbierski2014PRL,Chen2015PRL,Altland2015PRL,Shapourian2016PRB,Yu2022PRB} 
, whereas the topological edge modes in gSPT states remain robust even in the presence of symmetry-preserving disorder.

The chiral U(1) symmetry (often called the chiral charge symmetry or axial symmetry) in a Weyl semimetal is not a fundamental, microscopic symmetry of the underlying lattice Hamiltonian. Instead, it is an emergent, low-energy, approximate symmetry that becomes valid only in the IR limit, meaning at energies very close to the Weyl nodes. This is a typical manifestation of emergent symmetry, which commonly arises in gapless quantum phase transitions. However, canonical gSPT states (as discussed in Sec.~\ref{section3}) usually do not require emergent symmetry. Even for the more interesting intrinsically gSPT states (see Sec.~\ref{section3.2}) with global $\Gamma$ symmetry, the nontrivial symmetry extension $1 \rightarrow H \rightarrow \Gamma \rightarrow G \rightarrow 1$ implies that the $H$ symmetry associated with stacking SPT states acts trivially in the low-energy subspace. The truly nontrivial symmetry at low energies is the $G$ symmetry, a subgroup of $\Gamma$, which carries a quantum anomaly in the IR limit. In this context, intrinsically gSPT states exhibit emergent anomalies encoded in a “smaller” subgroup of the global symmetry $\Gamma$, which is conceptually distinct from emergent symmetries.

The true meaning of treating "Weyl semimetals as a special class of fermionic gSPT” is that the existence of Weyl semimetals is protected by lattice spatial symmetry. Specifically, a three-dimensional Weyl semimetal is gapless only at isolated points in momentum space, while any two-dimensional slice that avoids these points forms a gapped lower-dimensional system. The Chern number of such slices protects the well-known Fermi arcs (the “topological edge modes”). Topological semi-metals thus rely on a momentum-dependent energy gap, meaning that disorder—which connects momenta—destabilizes them. Therefore, in analogy to gSPT states protected by global onsite symmetry, topological semimetals can be regarded as gSPT states protected by lattice translation symmetry. Beyond topological semimetals, gSPT phases protected by more general crystalline symmetries are expected to exhibit richer physics and merit further study in future work.

\subsection{Other applications}
\label{section2.4}
In the remainder of Sec.~\ref{section2}, we explore further generalizations of nontrivial topology in free-fermion quantum critical systems, including the systematic construction of higher-dimensional free-fermion gSPT states using the idea of multiplicative construction~\cite{Cook2022cp,florescalderon2023topologicalquantumcriticalitymultiplicative}, the stability of topological edge modes in critical fermionic models with long-range hopping and pairing terms~\cite{Zhong2024pra}, and the characterization of topological edge states at nonequilibrium phase transition points~\cite{zhou2024topologicaledgestatesfloquet}.

\subsubsection{Multiplicative construction of higher dimensional free fermion gSPT phases}
\begin{figure}
    \centering
   \includegraphics[width=0.65\linewidth]{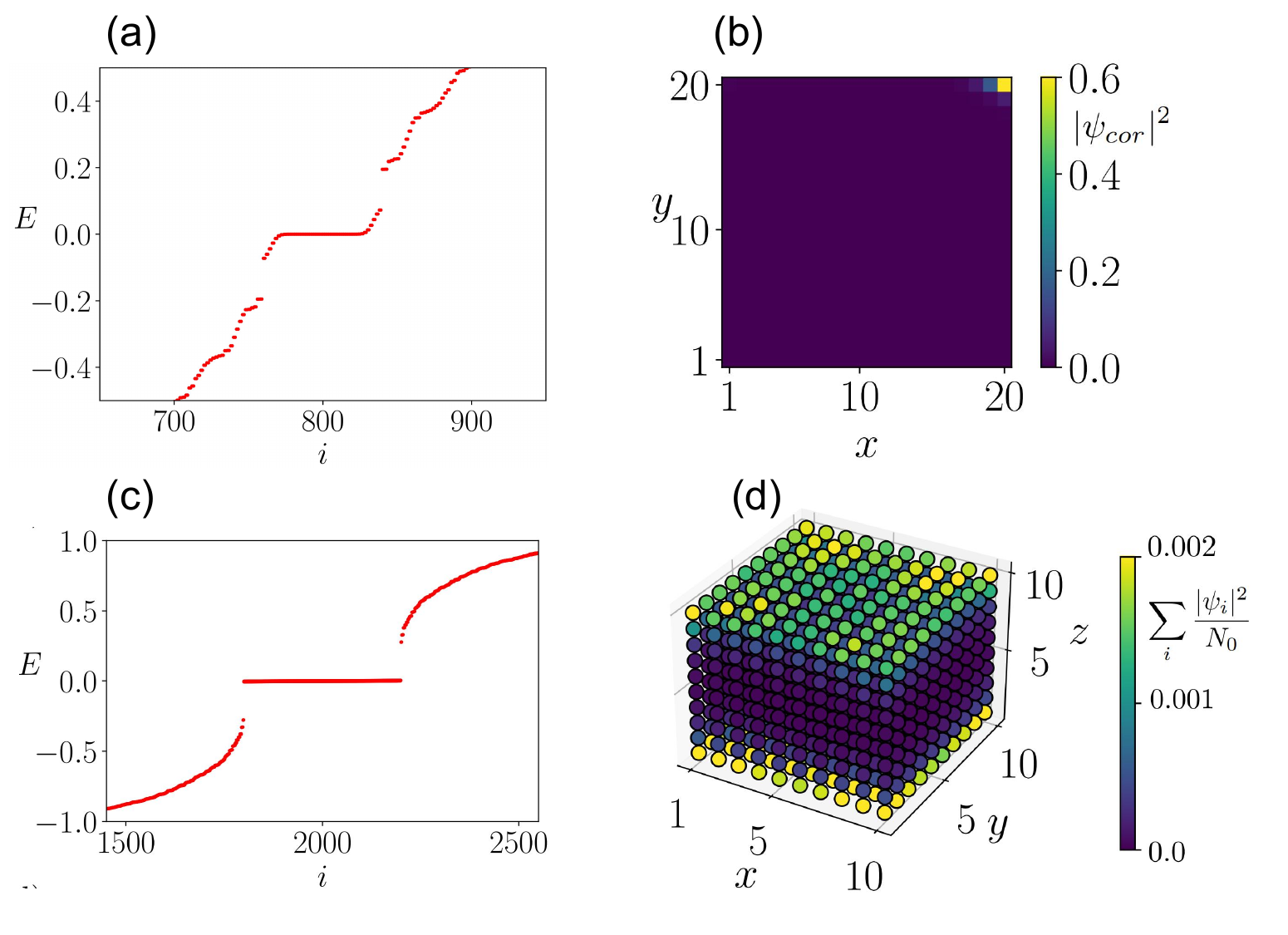}
    \caption{(a), (c) Energy spectrum as a function of the eigenvalue index around zero energy in the presence of disorder for two- and three-dimensional systems, respectively. 
    (b), (d) Zero-energy modes distributed along two edges of the two- and three-dimensional systems, respectively. The figures are adapted from Ref.~\cite{florescalderon2023topologicalquantumcriticalitymultiplicative}.}
    \label{fig11}
\end{figure}

The multiplicative construction provides a systematic approach to generalizing free fermion gapless topological phases to higher dimensions by combining two mutually perpendicular parent topological chains, as first proposed and detailed in Ref.~\cite{Cook2022cp,florescalderon2023topologicalquantumcriticalitymultiplicative}. For example, in constructing two-dimensional gSPT states, this approach takes the tensor product of two perpendicular Kitaev chains, $h_{\mu}(k_\nu) = (\text{cos}(k_{\nu}-\mu)\tau_z+\text{sin}(k_{\nu})\tau_y$ with chemical potential $\mu$ along each spatial direction. The corresponding Hamiltonian is given by: 
\begin{align}  
     h(k_x,k_y) &= h_\mu(k_x) \otimes h_{1}(k_y).  
     \label{perpMKC}  
\end{align}  
To diagonalize the Hamiltonian, we switch to a Majorana basis, redefining the annihilation operators as $\tilde{c}_{\textbf{k}A}=(\tilde{\pi}_\textbf{k}+i\pi_\textbf{k})/2$ and $\tilde{c}_{\textbf{k}B}=(\tilde{\gamma}_\textbf{k}+i\gamma_\textbf{k})/2$, where $A$ and $B$ label the pseudospin degrees of freedom. 
In this basis, the Hamiltonian for one block takes the form:  
\begin{align}  
      \hat{H}_\pi &= \frac{i}{2} \sum_{x,y} \big( \tilde{\pi}_{x,y} \pi_{x+1,y+1} - \tilde{\pi}_{x,y} \pi_{x+1,y} + \mu \tilde{\pi}_{x,y} \pi_{x,y+1} - \mu \tilde{\pi}_{x,y} \pi_{x,y} \big).  
     \label{perppi}  
\end{align}  
The Hamiltonian for the other sector, $\hat{H}_\gamma$, is obtained as its mirror conjugate.
To explicitly reveal the nontrivial topology, we reformulate Eq.~(\ref{perppi}) within the $\alpha$-chain formalism~\cite{Verresen2018prl}, as discussed in Sec.~\ref{section2.1} and generalized to two dimensions in this case. The Hamiltonian $ h(k_x,k_y) $ is expressed as:  
\begin{align}
\label{alpha2d}
    \hat{H} = \sum_\alpha \sum_\beta t_{\alpha,\beta} \hat{H}_{\alpha,\beta}, \quad \hat{H}_{\alpha,\beta} = \frac{i}{2} \sum_{x,y} \tilde{\pi}_{x,y} \pi_{x+\alpha,y+\beta}.  
\end{align}  
Following Ref.~\cite{Verresen2018prl}, the topological invariant for such two-dimensional gapless systems is defined in terms of a complex function obtained via the two-dimensional Fourier transform of the Majorana tight-binding coefficients: $f(z_1,z_2) = \sum_{\alpha=-\infty}^\infty \sum_{\beta=-\infty}^\infty t_{\alpha,\beta} \, z_1^\alpha z_2^\beta.\label{fz2D}$ For simplicity, we reduce the Hamiltonian \eqref{perpMKC} to a one-dimensional case by setting $ k_x = \pm k_y$, thereby restricting the analysis to this submanifold of the BZ. Diagonalizing the Hamiltonian for representative parameters reveals a robustly gapless low-energy spectrum with huge degeneracy, even in the presence of disorder (Fig.~\ref{fig11}(a)), which breaks translational invariance and is fundamentally different from topological semimetals mentioned before (Sec.~\ref{section2.3}). 
The subset of this degenerate manifold consists of zero energy states, localized at corners of the system for open boundary conditions along both the $x$ and $y$ directions, referred to as topological corners states, as clearly visible in Fig.~\ref{fig11} (b). 
These corner states originate from the gSPT phase in each $\pi$ and $\gamma$ sector. 
Importantly, only the corner modes are topologically protected, while the presence of additional zero-energy states is topologically trivial and merely reflects the inherent gapless nature of the system~\cite{florescalderon2023topologicalquantumcriticalitymultiplicative}. 

This framework can be further generalized to three dimensions by combining a Kitaev chain parent Hamiltonian $h_{\mu} (k_z)$ with a WSM parent Hamiltonian $h_{\mathrm{WSM}}(k_x,k_y,k_z) = ((\text{cos}(k_x)+\text{cos}(k_y)+\text{cos}(k_z)-2))\sigma_z+\text{sin}(k_x)\sigma_x+\text{sin}(k_y)\sigma_y$, resulting in a child Hamiltonian given by~\cite{florescalderon2023topologicalquantumcriticalitymultiplicative}:  
\begin{align}  
    h(k_x,k_y,k_z) = h_{\mathrm{WSM}}(k_x,k_y,k_z) \otimes h_{\mu}(k_z).  
\end{align}  
Here, the Kitaev chain parent is gapped, while the WSM parent remains gapless due to the presence of Weyl nodes at $ \left(k_x,k_y,k_z\right) = \left(0,0,\pm \pi/2\right) $. 
By preserving internal symmetries, this multiplicative construction stabilizes gSPT phases even in the absence of translational symmetry in the Kitaev chain parent. Extending the $ \alpha $-chain formalism to three dimensions, we define a complex function $ f(z_1,z_2,z_3) $ analogous to Eq.~\eqref{alpha2d}. Similar to the one-dimensional case discussed in Sec.\ref{section2.1}, the edge states can be determined by the number of zeros and poles in the complex function $f(z_1, z_2, z_3)$; further details can be found in Refs.~\cite{florescalderon2023topologicalquantumcriticalitymultiplicative,Verresen2018prl}. Speicifically, diagonalizing the Hamiltonian under open boundary conditions along all three directions reveals a zero-energy manifold that is separated from higher-energy states by a finite gap (Fig.~\ref{fig11}(c)). 
A subset of these states forms topological corner states, while the rest consists of surface-localized states (Fig.~\ref{fig11}(d)). 
The two-dimensional surface states originate from the zero-dimensional states of the Kitaev chain parent, which are extended across the $ xy $-plane through coupling with the WSM parent Hamiltonian. 

\subsubsection{The stability of topological edge modes in long-range critical free fermion models}
Recently, there has been a surge of theoretical interest in the long-range (LR) extension of free fermion topological phases, incorporating LR superconducting pairing and hopping terms~\cite{Dutta2017prb,Francica2022prb,Jager2020prb,Gong2016prb,Patrick2017prl,Kartik2021prb,Gong2023prl}. These models exhibit exotic topological properties, including massive edge modes~\cite{Vodola2014prl,Viyuela2016prb,Sarkar2025SR}, anomalous scaling of correlation functions~\cite{Gong2016prb,Vodola_2016}, and a novel bulk-boundary correspondence~\cite{Jones2023prl,Gong2023prl}. 
From an experimental perspective, significant attention has been devoted to studying LR models in various physical systems, such as trapped ions~\cite{Britton2012Nature}, magnetic impurities~\cite{Zhang2019SR}, and atoms coupled to multimode cavities~\cite{Douglas2015NP}. 
Moreover, in the Bogoliubov–de Gennes representation, extended Kitaev chains with LR pairing and hopping can be simulated using Shiba chains~\cite{Pientka2013prb}. Additionally, digital simulations of the extended LR Kitaev chain can be implemented on current noisy intermediate-scale superconducting quantum processors~\cite{low2019hamiltonian,Tao2020prb,Barends2015NC}. These developments highlight the fundamental importance of investigating LR interacting models both theoretically and experimentally. 
Given this context, a natural and intriguing question arises: What is the stability of topological edge modes in LR gapless systems? More importantly, can LR interactions induce a crossover in topological edge modes from massless to massive in the free fermion gSPT states, analogous to the effects observed in LR topological insulators?

\begin{figure}
    \centering
   \includegraphics[width=0.65\linewidth]{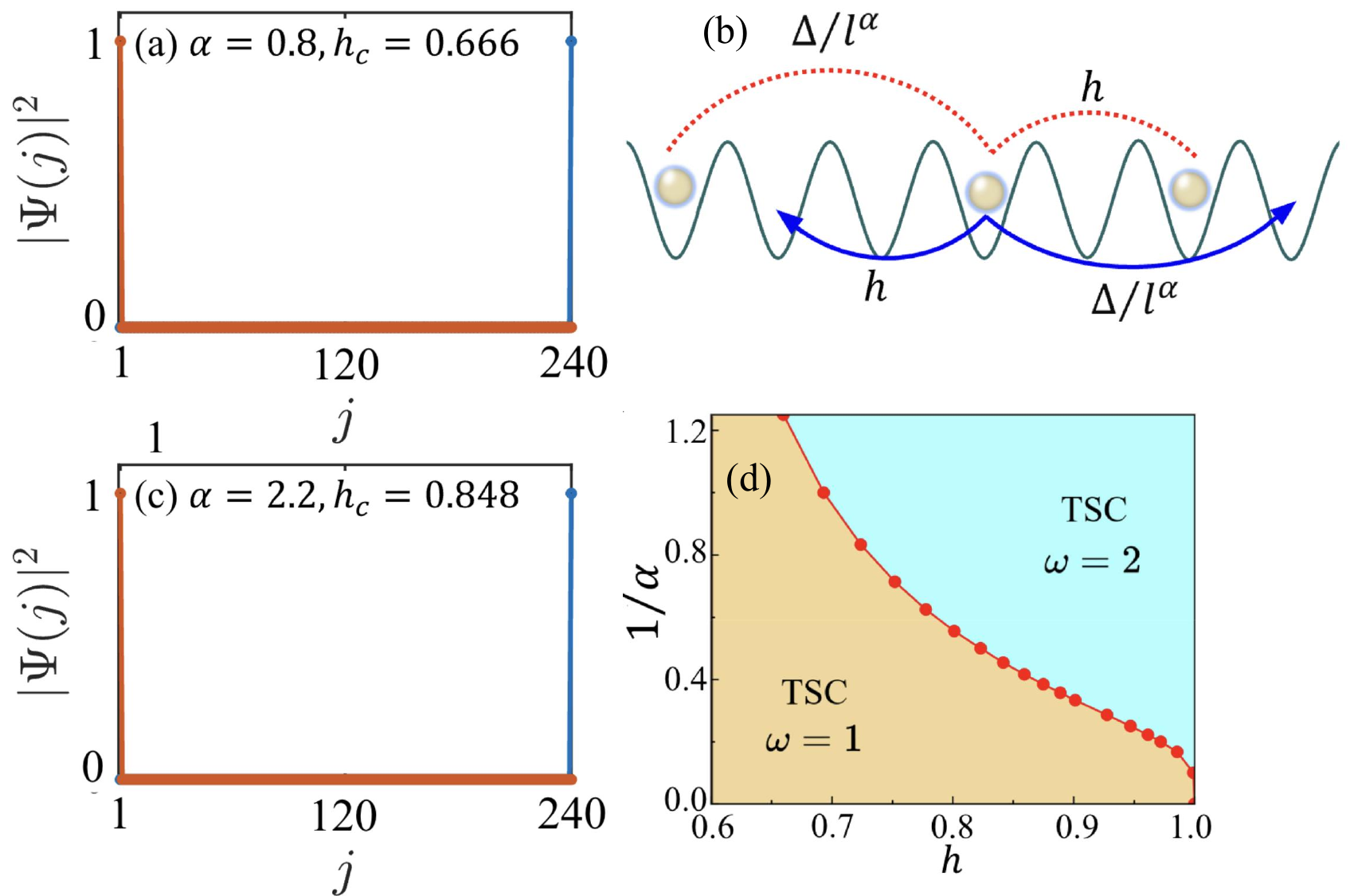}
    \caption{ (a), (c) Probability distributions of the zero-energy state at the critical point for $\alpha = 0.8$ and $2.2$, respectively. (b) A schematic representation of the LR interacting fermionic chain. The brown circles represent fermions, while the blue arrows (solid lines) and red dashed lines represent LR and next-nearest-neighbor hopping (pairing) terms, respectively. (d) Global phase diagram of the extended Kitaev chain with LR interactions as a function of the power-law exponent $\alpha$ and the driving parameter $h$. The diagram delineates topological superconducting (TSC) phases with winding numbers $\omega$, including the TSC phase with $\omega=1$ (brown region) and $\omega=2$ (light blue region). The red points denote critical points $h_c^*$ for different $\alpha$, forming a critical line (red curve) separating the two phases. The figures are adapted from Ref.~\cite{Zhong2024pra}.}
    \label{fig12}
\end{figure}
To address the aforementioned questions, Ref.~\cite{Zhong2024pra} investigates the topological behavior at the critical point in extended Kitaev chains with LR hopping and pairing, which can be reduced to the cluster Ising model via the Jordan-Wigner transformation in the short-range limit (as depicted in Figs.\ref{fig12} (b) and (d)). The Hamiltonian of the LR free fermion lattice model is given by: 
\begin{equation}
\label{Efl}
\begin{split}
&H_{\text{LR}} = -\Delta\sum_{j=1}^{L} \sum_{l=1}^{L-1} d_{l}^{-\alpha} (c_j^\dagger c_{j+l} + c_j^\dagger c_{j+l}^\dagger+ h.c.) \\
& - h\sum_{j=1}^{L-2} (c_j^\dagger c_{j+2} + c_j^\dagger c_{j+2}^\dagger + h.c.)-\sum_{j=1}^{L} \mu_j(c_j^\dagger c_j -\frac{1}{2}).
\end{split}
\end{equation}
Here, $c^{\dagger}_{j}$ ($c_{j}$) represents the fermionic creation (annihilation) operator at site $j$. The parameter $h$ denotes the strength of the next-nearest-neighbor fermion $p$-wave pairing and hopping amplitude. For a periodic chain, we define $d_l = l$ ($d_l = L-l$) if $l < L/2$ ($l > L/2$), using antiperiodic boundary conditions. For an open chain, we set $d_l = l$ and omit terms involving $c_{j>L}$. 
The $\Delta=1$ sets the energy unit and 
the chemical potential $\mu_j = 0$. 
From Fig.~\ref{fig12}, it was found that topological edge modes at the critical point between TSC phases with different winding numbers remain stable under LR hopping and pairing. 
Moreover, these edge modes remain massless even when LR hopping and pairing become considerable strong, in stark contrast to the case of LR gapped topological phases, as illustrated in Fig.~\ref{fig12} (a) and (c). 
Additionally, finite-size scaling analysis~\cite{Zhong2024pra} reveals that the critical exponent of the LR fermionic lattice model remains consistent with that of the short-range Majorana universality class, which is fundamentally different from the novel LR universality class typically found in usual LR interacting lattice models~\cite{Defenu2023rmp}.

\subsubsection{Topological edge states at Floquet quantum criticality}
Our discussion so far pertains to the topological edge states of equilibrium quantum critical points. 
However, recent advances in quantum simulation experiments~\cite{Blatt2012NP,Christian2017Science,Browaeys2020NP,kjaergaard2020superconducting} have increasingly drawn attention to the study of nonequilibrium phases of matter and their transitions. A prominent class of nonequilibrium systems, known as periodically driven (Floquet) systems, exhibits fascinating topological phenomena that have no counterparts in equilibrium systems and host a variety of exotic states of matter~\cite{oka2019floquet,else2020discrete,RODRIGUEZVEGA2021168434}, including Floquet topological phases~\cite{Jiang2011prl,Rudner2013prx,Chen2014prl,Lee2018prl}, Floquet quantum criticality and conformal field theory~\cite{William2018PNAS,wen2018floquetconformalfieldtheory,Fan2021SciPost,Fan2020PRX}, Floquet many-body localization~\cite{Ponte2015prl,Zhang2016prb,Decker2020prl}, and discrete-time crystals~\cite{Else2016prl}, among others. A particularly intriguing question is whether topologically protected edge states—some of which may have no static analogs—can coexist with the Floquet gap closing, thereby persisting through Floquet quantum criticality.

\begin{figure}
    \centering
   \includegraphics[width=0.65\linewidth]{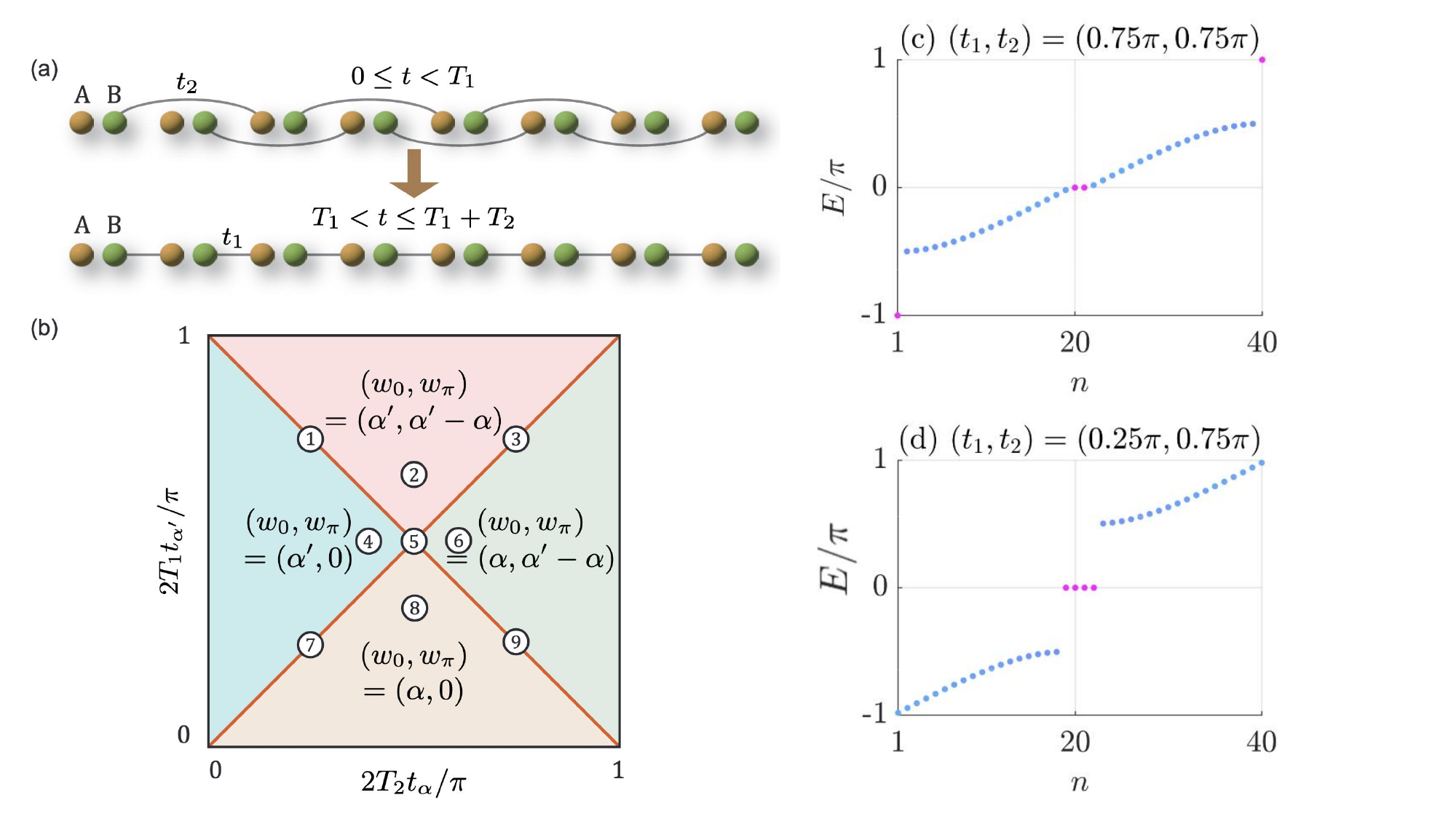}
    \caption{(a) A schematic illustration of the Floquet two-step driving protocol in a Majorana fermion chain with $(\alpha, \alpha') = (1,2)$. (b) The universal phase diagram of Eq.~(\ref{Eqfloquet}), showing four gapped Floquet phases, each characterized by a pair of topological winding numbers $(w_{0}, w_{\pi}) \in \mathbb{Z} \times \mathbb{Z}$. Transitions between these gapped phases correspond to either topologically trivial or nontrivial Floquet quantum critical lines, as detailed in the text. Numbered circles indicate representative points within each phase and at phase transitions. (c) Floquet quasienergy spectra under open boundary conditions along quantum critical lines at $(t_1,t_2) = (0.75\pi, 0.75\pi)$. The bulk spectrum is gapless at $E=0$ and gapped at $E = \pm \pi$, featuring two degenerate Majorana edge modes at $E=0$ and another two at $E=\pi$. (d) For $(t_1,t_2) = (0.25\pi, 0.75\pi)$, the bulk spectrum is gapless at $E=\pi$ and gapped at $E=0$, with four degenerate Majorana zero modes. The figures are adapted from Ref.~\cite{zhou2024topologicaledgestatesfloquet}.}
    \label{fig13}
\end{figure}

To develop a general theory of Floquet gapless topology, recent works~\cite{zhou2024topologicaledgestatesfloquet,cardoso2024gaplessfloquettopology,zhou2025gaplesshigherordertopologycorner} have analytically investigated the existence of topological edge modes—both within gapped phases and precisely at topological phase boundaries—for a broad class of free Majorana $\alpha$ chains~\cite{Verresen2018prl,Verresen2017prb} under periodic driving. The time-periodic Hamiltonian is given by:  
\begin{equation}
\label{Eqfloquet}
    H(t)=\begin{cases}
H_{\alpha'}=-i\sum_{n}t_{\alpha'}\widetilde{\gamma}_{n}\gamma_{n+\alpha'}, & t\in[0,T_{1}),\\
H_{\alpha}=-i\sum_{n}t_{\alpha}\widetilde{\gamma}_{n}\gamma_{n+\alpha}, & t\in[T_{1},T_{1}+T_{2}),
\end{cases}
\end{equation}  
where $\gamma_{n}$ and $\widetilde{\gamma}_{n}$ represent Majorana fermions on odd and even sublattice sites (depicted as orange and green balls in Fig.~\ref{fig13} (a)). The hopping strengths $t_{\alpha^{\prime}(\alpha)}$ are nonzero only for finite values of $\alpha^{\prime}(\alpha)$. The total driving period is $T = T_1 + T_2 = 1$, and we assume $\alpha^{\prime} > \alpha$ (a similar analysis applies for $\alpha^{\prime} < \alpha$). The Hamiltonian satisfies the periodic condition $H(t) = H(t+T)$, and its time evolution operator, given by $F = \hat{\mathsf{T}}e^{-i\int_0^T H(t)dt}$, defines the Floquet states as $|\psi(t)\rangle = e^{-iEt} |\Phi(t)\rangle$. Here, $\hat{\mathsf{T}}$ is the time-ordering operator, and the Floquet states satisfy $|\Phi(t+T)\rangle = |\Phi(t)\rangle$, analogous to Bloch’s theorem in systems with discrete translational symmetry. The quasienergy $E$ is conserved stroboscopically and is defined modulo $2\pi$.  Furthermore, the Floquet operator $F$ exhibits chiral symmetry, allowing the nontrivial topology can be characterized by winding numbers. Remarkably, the topological phase diagrams of models Eq.~(\ref{Eqfloquet}) exhibit a universal structure, enabling the identification of systems that host arbitrarily many Majorana edge modes, as shown in Fig.~\ref{fig13} (b). These Majorana zero modes appear not only within gapped phases but also persist along topological phase boundaries (see Fig.~\ref{fig13} (d)), coexisting with a gapless Floquet bulk spectrum. These modes are referred to as \emph{critical Majorana modes}. At these phase boundaries, a new type of critical Majorana $\pi$ mode—absent in static systems—emerges (see Fig.~\ref{fig13} (c)). Using a set of generalized winding numbers, a simple rule for bulk-edge correspondence at Floquet quantum criticality can be established.  To validate these analytical predictions, Ref.~\cite{zhou2024topologicaledgestatesfloquet} provides several lattice simulation examples and extends the findings to $(2+1)$-dimensional nonequilibrium systems. These results deepen our understanding of Floquet quantum criticality and open new avenues for exploring non-equilibrium topological phenomena.

\section{Topological physics in many body quantum critical systems}
\label{section3}
\subsection{Overview}
\label{section3.1}
Section~\ref{section2} mainly focuses on the general topological theory and recent developments in analytically more tractable free-fermion critical systems across various dimensions. In this section, we turn to the topological properties of more challenging many-body quantum critical systems in one dimension.
A notable subset of such systems exhibits conformal invariance in the low-energy limit and is described by CFT. These systems, include both critical points and stable critical phases, which are relatively more tractable from a field theory perspective. Specifically, following our discussion of free fermion systems in Section~\ref{section2}, we focus specifically on the emergence of nontrivial topology in gapless many-body systems protected by global symmetry, known as gSPT phases. These systems reveal a rich interplay between topology and quantum criticality, offering new insights into the fundamental nature of topological phases beyond gapped counterparts. 
In this section, we mainly focus on SPT physics in strongly interacting bosonic systems in one dimension, such as quantum spin chains. The discussion of interacting fermionic systems and higher-dimensional cases is more involved and will be briefly addressed in the last section. 

To clarify the notation, we use $\Gamma$ to denote the global symmetry group of the system, with $H$ as a normal subgroup of $\Gamma$, and the corresponding quotient group written as $G_{\rm quot} $. The emergent anomaly associated with the symmetry group $G$ is denoted by $\omega_G$.

We begin with an overview of the current classification of gSPT phases based on whether they contain a gapped sector and whether they have gapped counterparts (i.e., whether they are intrinsic). 
This classification yields four distinct categories: gSPT, purely gSPT, intrinsically gSPT, and intrinsically purely gSPT phases, as summarized in Table~\ref{tab:gSPT}. 
The following detailed explanations are primarily based on the literature~\cite{li2023intrinsicallypurelygaplesssptnoninvertibleduality}.

\begin{table}[h]
\centering
\begin{tabular}{|c|c|c|}
    \hline
    & Contains gapped sector & No gapped sector  \\
    \hline
    Non-intrinsic  & gSPT  & purely gSPT \\
    \hline
    Intrinsic & intrinsically gSPT  & intrinsically purely gSPT\\
    \hline
\end{tabular}
\caption{Classification of the gSPTs by whether they contain gapped sector (horizontal direction) or whether they have gapped counterparts (vertical direction). 
The table is taken from Ref.~\cite{li2023intrinsicallypurelygaplesssptnoninvertibleduality}.}
\label{tab:gSPT}
\end{table}

i) \emph{The gSPT phases}: The first example of a gSPT phase, which is non-intrinsic and includes a gapped sector, was systematically investigated in Ref.~\cite{Scaffidi2017prx}. 
Specifically, a general construction of gSPT phases was introduced based on the decorated domain wall or defect picture~\cite{chen2014symmetry}. 
The central idea is to decorate the $G$-symmetry defects of a $G$-symmetric conformal invariant gapless system with an $H$-symmetric gapped SPT. 
This decorated defect construction creates a gapped sector that acts both on $G$ and $H$ symmetry, resulting in a gSPT phase whose topological properties can also manifest in gapped counterparts. 
As a result,these features are not unique to the gapless system~\cite{Scaffidi2017prx,Li2024scipost}, and the topological properties of these gSPT phases arise from the gapped sector and can be interpreted as a gapped SPT stack with a CFT. 
Typically, recent progress~\cite{Scaffidi2016prb,Verresen2021prx} have shown that such gSPT states can emerge at conformally invariant critical points separating spontaneously symmetry-breaking (SSB) phases and gapped SPT phases, also known as symmetry-enriched CFT or topologically nontrivial quantum critical points. 

ii) \emph{The intrinsically gSPT phases}: More importantly, there exists a remarkable class of gSPT, known as intrinsically gSPT phases~\cite{Thorngren2022prb}, whose topological features are fundamentally forbidden in gapped counterparts.
Specifically, recent works~\cite{li2023intrinsicallypurelygaplesssptnoninvertibleduality,Li2024scipost} propose a systematic construction of intrinsically gSPT phases, which we briefly outlined below: The total symmetry group is denoted by $ \Gamma $, fitting into the group extension $ 1 \to H \to \Gamma \to G \to 1 $. 
The construction begins with a $ G $-symmetric gapless system or CFT characterized by a self-anomaly $ \omega_G $. 
An $H$-symmetric gapped SPT phase is then stacked on top of the $G$ symmetry domain walls or defects. 
Due to the non-trivial group extension, the resulting gapped sector exhibits an (emergent) anomaly $ -\omega_G $ that cancels the anomaly in the gapless sector, rendering the combined system an intrinsically gSPT phase that is $ \Gamma $-anomaly-free. 
By construction, intrinsically gSPT phases also include a gapped sector, leading to exponentially localized edge modes near the boundaries. 
Importantly, the topological features of intrinsically gSPT phases cannot be realized in any $ \Gamma $-symmetric gapped systems, thereby justifying the term ``intrinsic''. 
Furthermore, Li et al.~\cite{li2023intrinsicallypurelygaplesssptnoninvertibleduality,Li2024scipost} utilized the decorated defect construction and the Kennedy-Tasaki (KT) transformation to construct analytically tractable (1+1)D spin models of both $ \mathbb{Z}_2 \times \mathbb{Z}_2 $ symmetric gSPT and $ \mathbb{Z}_4 $ symmetric intrinsically gSPT phases. 
Additionally, the recently developed topological holography principle (known as symmetry topological order) was employed to provide a unified classification of gapped and gapless SPT phases from a new perspective~\cite{huang2023topologicalholographyquantumcriticality,wen2023classification11dgaplesssymmetry}, as briefly discussed in a later section. Experimentally, these intrinsically gSPT phases can emerge at the transition point between a quantum spin Hall insulator and an $s$-wave superconducting phase, which has recently been proposed to be realizable in materials such as WTe$_2$~\cite{song2024unconventional}.

iii) \emph{The purely and intrinsically purely gSPT phases}: From the previous discussion, we have established that both gSPT and intrinsically gSPT phases typically include a gapped sector, which gives rise to exponentially localized topological edge modes. 
In contrast, Verresen et al.~\cite{Verresen2017prb,Verresen2021prx} studied the critical cluster Ising model with time-reversal symmetry and demonstrated that this model lacks a gapped sector. 
This is evidenced by the algebraic decay of the energy splitting of the edge mode, a behavior that is forbidden in gapped systems. 
This result suggests the existence of a gSPT phase without any gapped sector, referred to as a ``purely gSPT phase''~\cite{wen2023classification11dgaplesssymmetry,Wen2023prb,li2023intrinsicallypurelygaplesssptnoninvertibleduality,Li2024scipost}. 
However, to the best of our knowledge, the ground state of the critical cluster Ising chain remains the only known lattice realization of a purely gSPT phase. 
Thus, constructing additional lattice Hamiltonians that realize such novel phases is highly desirable. Furthermore, it is natural to explore the possibility of an intrinsically gSPT phase that also lacks any gapped sector, a phase termed an ``intrinsically purely gSPT phase''~\cite{li2023intrinsicallypurelygaplesssptnoninvertibleduality}. 
Unfortunately, despite ongoing efforts~\cite{li2023intrinsicallypurelygaplesssptnoninvertibleduality,Li2024scipost}, explicitly constructing a lattice model that exhibits such a phase remains an open challenge, making it an important direction for future research.

\color{red}
For convenience, Table~\ref{tab:1} provides a concise summary and comparison of the methods used in the literature to characterize gSPT, highlighting their scope, underlying assumptions, and advantages. A more detailed discussion of these methods will be presented in the following sections.

\begin{table}[]
\caption{Comparison of different frameworks for gSPT classification.\\}
\label{tab:1}
\begin{tabular}{@{} p{4cm}p{4cm}p{4cm}p{4cm} @{}}
\toprule
Method                   & Assumptions & Scope &Advantages \\ 
\midrule
Decorated defect~\cite{Li2024scipost}    & \parbox{4cm}{The processes of defect decoration and fluctuation commute with each other}    &\parbox{4cm}{The gSPT and intrinsically gSPT} & \parbox{4cm}{Unified construction of gapped and gapless SPT} \\ 
\\ &&
\\
Kennedy-Tasaki transformation~\cite{li2023intrinsicallypurelygaplesssptnoninvertibleduality} &     \parbox{4cm}{Always possible to find two decoupled systems}           &    \parbox{4cm}{The gSPT protected by on-site symmetry} & \parbox{4cm}{Easy to understand and construct gSPT without a gapped sector}       \\
&& \\
Anomaly-based methods~\cite{Thorngren2022prb,yang2025deconfinedcriticalityintrinsicallygapless} &           \parbox{4cm}{Nontrivial symmetry group extension}             &    \parbox{4cm}{The intrinsically gSPT} & \parbox{4cm}{A general mechanism for constructing and understanding gSPT physics}       \\
&& \\
Entanglement-based diagnostics~\cite{Yu2022prl,Yu2024prl,zhong2025quantumentanglementfermionicgapless} &           \parbox{4cm}{Conformal invariant quantum critical points}             &    \parbox{4cm}{The gSPT, intrinsically gSPT, purely gSPT, and intrinsically purely gSPT} & \parbox{4cm}{Both bosonic and fermionic gSPT can be identified}       \\
&& \\
Topological holography~\cite{huang2023topologicalholographyquantumcriticality,wen2023classification11dgaplesssymmetry} &           \parbox{4cm}{Bulk-boundary equivalence}             &    \parbox{4cm}{Bosonic and fermionic gSPT in all dimensions} & \parbox{4cm}{Unified classification of gapped and gapless SPT in all dimensions, beyond group cohomology}       \\
&& \\ \bottomrule
\end{tabular}
\end{table}

\color{black}

\subsection{Symmetry-enriched quantum criticality}
\label{section3.2}
To illustrate the nontrivial interplay between topology and quantum criticality, we begin with conformally invariant critical points in quantum spin chains that host robust topological edge modes protected by global symmetry. 
These are known as symmetry-enriched or topologically nontrivial quantum critical points~\cite{Verresen2021prx}, which often arise at phase transitions where all adjacent gapped phases are nontrivial, either SPT or SSB phases. 
As discussed above, these novel critical points in spin systems can be regarded, in a broader sense, as bosonic gSPT states, where nontrivial topological properties emerge at the critical point.  
As a concrete example, we briefly introduce the basic concepts and physical properties of symmetry-enriched Ising criticality in (1+1) dimensions. For a more detailed discussion, we refer to the pioneering work by Verresen et al.~\cite{Verresen2021prx}.

We first review the well-known order parameters for gapped phases in one dimension, which can be categorized as either local (for SSB phases) or nonlocal (for SPT phases) operators. 
In the following, we explain that critical points can similarly host both local and nonlocal order parameters, and the presence of nontrivial nonlocal order parameters is closely tied to the existence of topological edge modes.

For concreteness, we consider a specific example where the global symmetry group is $\Gamma = \mathbb{Z}_2 \times \mathbb{Z}_2^T$, where $\mathbb{Z}_2^T$ represents an anti-unitary symmetry that squares to the identity. These symmetries are commonly realized in spin-$\frac{1}{2}$ chains, with Pauli matrices by $\sigma^x$, $\sigma^y$, and $\sigma^z$. The symmetry group $\mathbb{Z}_2 \times \mathbb{Z}_2^T$ is represented by the spin-flip operator $P \equiv \prod_n \sigma^x_n$ and an anti-unitary symmetry operator given by complex conjugation in this context, $T \equiv K$.

\paragraph{\textbf{Local and nonlocal order parameters for gapped phases}}: In Landau symmetry breaking theory, phases that spontaneously break a symmetry can naturally be characterized by the long-range order of a local operator carrying a nontrivial charge, i.e., the operator that does not commute with the symmetry. 
A standard example is the Ising chain, $ H_{\rm Ising} = -\sum_n \sigma^z_n \sigma^z_{n+1} $, where the ground state satisfies $\lim_{|n-m| \to \infty} \langle \sigma^z_m \sigma^z_n \rangle \neq 0$. This implies that $\langle \sigma^z_n \rangle \neq 0$, indicating that the ground state is no longer invariant under the $\mathbb{Z}_2$ symmetry, which in turn leads to ground-state degeneracy. Indeed, since $\sigma^z_n$ has charge $-1$ under the $\mathbb{Z}_2$ Ising symmetry $P$, this phase is distinct from the trivial phase, where long-range correlations exist only for operators with charge $+1$. 

In the presence of only $ \mathbb{Z}_2 $ symmetry, there is a single symmetry-breaking ``Ising'' phase. 
However, if complex conjugation symmetry $ \mathbb{Z}_2^T $ is also imposed, this phase splits into two distinct phases. For example, consider the Hamiltonian $ H_{\rm Ising}' = -\sum_n \sigma^y_n \sigma^y_{n+1} $. With only $ \mathbb{Z}_2 $ symmetry, $ H_{\rm Ising} $ and $ H_{\rm Ising}' $ can be smoothly connected by rotating $ \sigma^y_n $ into $ \sigma^z_n $. 
However, when both $ \mathbb{Z}_2 \times \mathbb{Z}_2^T $ symmetries are enforced, this connection is forbidden because it violates $\mathbb{Z}_{2}^{T}$ symmetry---since the Ising order parameter $ \sigma^y_n $ is imaginary, whereas $ \sigma^z_n $ is real. 
Consequently, the two Hamiltonians must be separated by a quantum phase transition.

For SPT phases that preserve all global symmetries, there exists no local observable that can distinguish them from a trivial phase. 
Instead, they can be detected through a nonlocal string order parameter, derived from the concept of symmetry fractionalization~\cite{Pollmann2010prb,Turner2011prb}. For instance, the trivial paramagnet $H_{\rm triv} = -\sum_n \sigma^x_n$ has long-range order for nonlocal operator $\lim_{|n-m| \to \infty} \langle \sigma^x_m \sigma^x_{m+1} \cdots \sigma^x_{n-1} \sigma^x_n \rangle \neq 0$. 
In comparison, the paradigmatic cluster SPT model~\cite{Son_2011,Smacchia2011pra,Verresen2017prb,Guo2022pra,Li2025pra}, $H_{\rm cluster} = - \sum_n \sigma^z_{n-1} \sigma^x_n \sigma^z_{n+1}$ has long-range order for other non-local operator $\lim_{|n-m| \to \infty} \langle \sigma^z_{m-1} \sigma^y_m \sigma^x_{m+1} \cdots \sigma^x_{n-1} \sigma^y_n \sigma^z_{n+1} \rangle \neq 0$. In both cases, the string is constructed using the unbroken $\mathbb{Z}_2$ symmetry generator $P$, but the endpoint operators exhibit different $T$-symmetry charges: in the trivial case, the endpoints are real with charge $+1$, while in the nontrivial case, they are imaginary with charge $-1$. 
This discrete distinction in endpoint charges implies that the two Hamiltonians must be separated by a quantum phase transition.

More generally, the string order parameter~\cite{Pollmann2012prb,Else2013prb,Jones2021PRR} for an on-site symmetry $\prod_n U_n$ takes the form: $\mathcal S_n = \cdots U_{n-2} U_{n-1} \mathcal O_n$, 
where $\mathcal{O}_n$ is a local operator chosen such that $\langle \mathcal{S}_m^\dagger \mathcal{S}_n \rangle$ exhibits long-range order. In reference~\cite{Verresen2021prx}, these string order operators are termed ``symmetry flux'' operators, as they act like sources of flux for operators charged under the symmetry. 
Such operators provide a unified characterization of both gapped and gapless SPT phases. 
For example, the symmetry properties of the endpoint operator $\mathcal{O}_n$ of symmetry fluxes encode the projective representation (or equivalently, the second group cohomology class) that labels the gapped phase~\cite{Pollmann2012prb,Else2013prb}. Furthermore, as shown in the next paragraph, the concept of the string order parameter (or symmetry fluxes) can be generalized to gapless cases, where long-range order is replaced by ``slowest algebraic decay''~\cite{Verresen2021prx}.

\paragraph{\textbf{Local and nonlocal order parameters for gapless phases}}: For gapped phases, we previously introduced the key idea that the symmetry properties of both local and nonlocal operators allow the definition of discrete invariants. This concept can be generalized to gapless systems, as illustrated by a typical example of the critical Ising spin chain, which we discuss below.

The first example, which aims to explain local order parameters at criticality, includes the following two critical transverse field Ising chains:
\begin{equation}
\begin{array}{ccc}
H &= &- \sum_n (\sigma^z_n \sigma^z_{n+1} + \sigma^x_n), \\
H'&= &- \sum_n (\sigma^y_n \sigma^y_{n+1} + \sigma^x_n).
\end{array}
\end{equation}
Both critical systems belong to the Ising universality class, or equivalently, a CFT with central charge $ c = 1/2 $~\cite{francesco2012conformal}. 
This universality class contains a unique local operator $\sigma(x)$ with a scaling dimension $\Delta_\sigma = 1/8$, such that $\langle \sigma(x) \sigma(0) \rangle \sim 1/x^{2\Delta_\sigma}$~\cite{francesco2012conformal}. 
The lattice correspondence of this continuum primary field $\sigma(x)$ is the Ising order parameters: $\sigma(x) \sim \sigma^z_n$ for $H$ and $\sigma(x) \sim \sigma^y_n$ for $H'$. 
However, these operators transform differently under antiunitary symmetry $T$: for $H$, $T \sigma T = +\sigma$, while for $H'$, $T \sigma T = -\sigma$. 
Thus, the two Ising CFTs are enriched by the $\mathbb Z_2^T$ symmetry, with the sign difference ($T \sigma T = \pm \sigma$) serving as a discrete invariant. 
Furthermore, this invariant remains unchanged as long as the system stays within the Ising universality class.

What does this discrete invariant mean? It ensures that any $\Gamma$-symmetric path (where $\Gamma = \mathbb Z_2 \times \mathbb Z_2^T$) of gapless Hamiltonians connecting $H$ and $H'$ must pass through a transition with a different universality class. 
For example, the interpolation $\lambda H + (1-\lambda) H'$ (with $0 \leq \lambda \leq 1$) lies within the Ising universality class \emph{except} at $\lambda = 1/2$, where the system passes through a multicritical point with dynamical critical exponent $z_{\rm dyn} = 2$. Alternatively, the path $\lambda H - (1-\lambda) H'$ transitions through a Gaussian fixed point (central charge $c = 1$) at $\lambda = 1/2$. 
At these non-Ising points, the property $T \sigma T = \pm \sigma$ changes. 

\begin{figure}
    \centering
   \includegraphics[width=0.55\linewidth]{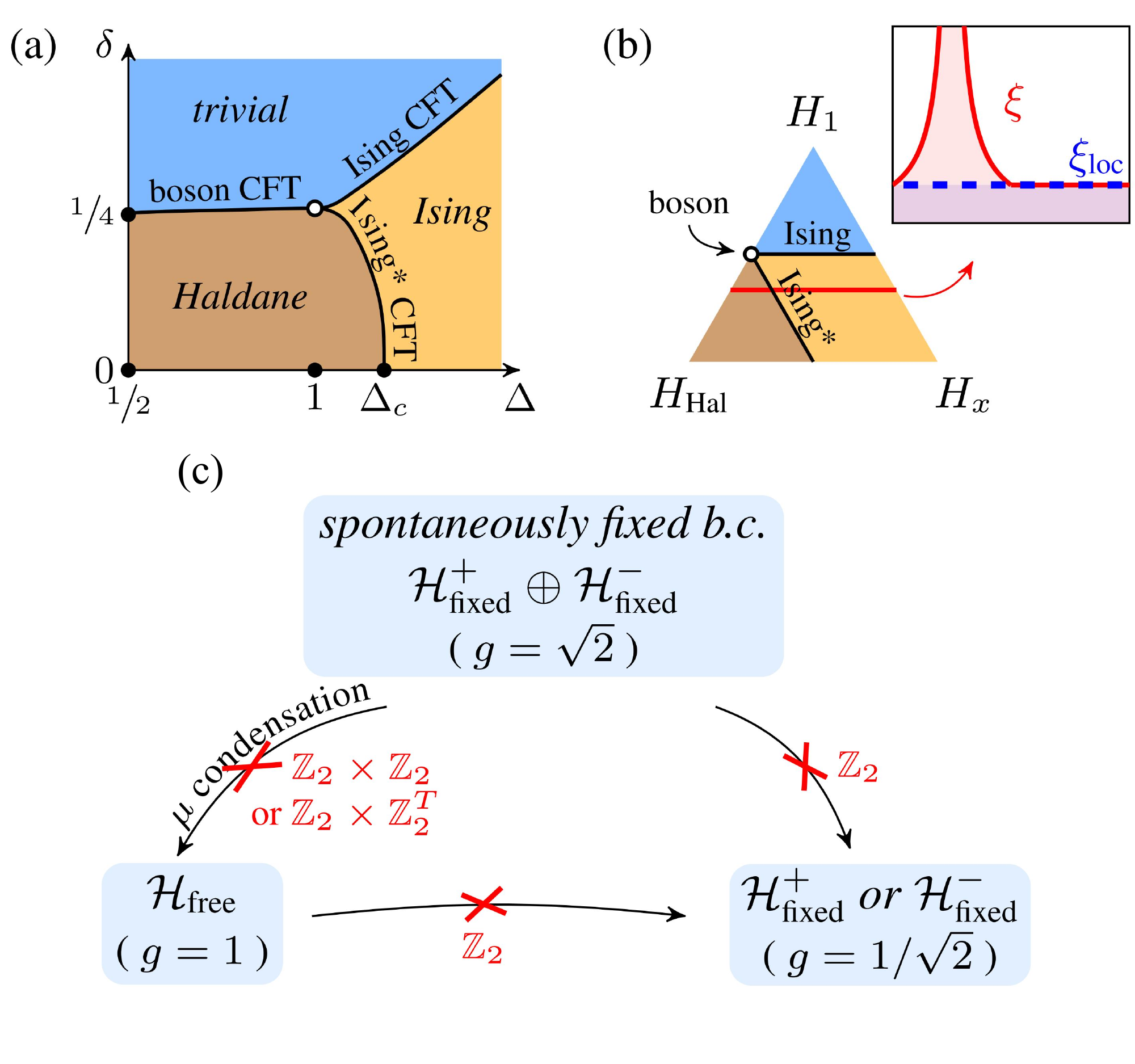}
    \caption{(a) Phase diagram of the bond-alternating $S=1$ XXZ chain. 
    A $c=1$ transition separates two topologically distinct $c=1/2$ Ising transitions. 
    The tricritical point (hollow marker) corresponds to a WZW $SU(2)_1$ CFT, where both the trivial and Haldane string order parameters exhibit a scaling dimension of $1/8$. 
    (b) A similar phase diagram is shown for an exactly solvable $S=1/2$ model. In this case, the $c=1$ boson CFT belongs to the free Dirac universality class. (c) Topologically protected edge modes in the Ising CFT. The boundary renormalization group flow for the Ising CFT is depicted. Typically, the free boundary condition (B.C.) is stable when preserving the global $\mathbb{Z}_2$ symmetry. However, this stability can be disrupted if the operator $\mu$ is charged under additional symmetries. In such cases, the spontaneously fixed boundary condition (characterized by a global twofold degeneracy) becomes stable. The figures are adapted from Ref.~\cite{Verresen2021prx}.}
    \label{fig14}
\end{figure}

Now turn to the more intriguing cases, which arise when two enriched critical points can only be distinguished by the symmetry properties of \emph{nonlocal} operators. To this end, we notice that critical transverse field Ising chain $H$ describes a phase transition between a trivial paramagnet (PM) and a ferromagnet (FM) phase, while critical cluster Ising chain $H''= - \sum_n (\sigma^z_n \sigma^z_{n+1} + \sigma^z_{n-1}\sigma^x_n\sigma^z_{n+1})$ describes a transition between a ferromagnet (FM) and the cluster SPT phase. 
Both systems belong to the Ising universality class with local operator $\sigma(x) \sim \sigma^z_n$, and in both cases, $T \sigma T = +\sigma$, which cannot be distinct from these critical points. However, the Ising CFT also features a \emph{nonlocal} operator $\mu(x)$ with scaling dimension $\Delta_\mu = 1/8$, related to $\sigma(x)$ by Kramers-Wannier duality. For $H$, $\mu(x)$ corresponds to the string operator $\mu(x) \sim \cdots \sigma^x_{n-2} \sigma^x_{n-1} \sigma^x_n$. In contrast, for $H''$, it takes the form: $\mu(x) \sim \cdots \sigma^x_{n-2} \sigma^x_{n-1} \sigma^y_n \sigma^z_{n+1}$~\cite{Verresen2017prb}. To reveal the distinction between these two critical chains, we construct symmetry fluxes of the $\mathbb Z_2$ symmetry generator $P$ : $\mathcal S_n = \big(\prod_{m<n} \sigma^x_m\big) \mathcal O_n$, where the endpoint operator $\mathcal O_n$ is chosen to ensure that $\langle \mathcal S^\dagger_m \mathcal S_n \rangle$ exhibits the slowest possible algebraic decay. 
For the Ising CFT, the smallest scaling dimension is $\Delta = 1/8$~\cite{francesco2012conformal}. Thus, the symmetry flux of $P$ is given by $\mathcal O_n = \sigma^x_n$ for $H$ and $\mathcal O_n = \sigma^y_n \sigma^z_{n+1}$ for $H''$, with distinct charges under $T$. This difference defines a discrete topological invariant of the Ising universality class, giving rise to symmetry-enriched Ising CFTs, as illustrated in Fig.~\ref{fig14} (a) and (b). 
In this context, while the bulk correlation length diverges, the edge mode localization length remains finite at the critical point, indicating the coexistence of a stable edge mode with bulk critical fluctuations.

We therefore conclude that different symmetry-enriched CFTs can be distinguished by how the symmetry fluxes of $ g \in \Gamma $ transform under other symmetries in $ \Gamma $.
In the case of the cluster Ising model, the discrete invariant is the sign change that occurs when the symmetry flux of $ P $ is conjugated by $ T $. 
This definition of symmetry flux is convenient and naturally reduced to the usual string order parameter in the gapped SPT phase, where the slowest algebraic decay is replaced by long-range order.

\paragraph{\textbf{The consequence: topological edge modes from charged symmetry fluxes}}: With charged symmetry fluxes in hand, a natural question arises: what are the physical consequences of these string operators at criticality? The authors of reference~\cite{Verresen2021prx} argue that in the Ising CFT if a symmetry flux carries a nontrivial charge under another symmetry, this can be linked to ground state degeneracies in open boundary conditions. While this phenomenon is well known for gapped systems~\cite{Pollmann2012prb}, it also extends to gapless systems. To illustrate this, consider the lattice model $H''$, defined on a half-infinite system with sites $n = 1, 2, \dots$. The boundary of this critical chain spontaneously magnetizes because, at the lattice level, $\sigma^z_1$ commutes with $H''$, leading to spontaneous edge magnetization $\sigma^z_1 = \pm 1$ and resulting in a twofold degeneracy. This magnetization is not merely an artifact of a fine-tuned model but is stable within the Ising CFT, as we illustrate below. For critical transverse field Ising model $H$, the $P$-symmetry flux is invariant under $T$ symmetry, meaning it can be added to the Hamiltonian. 
As a result, the symmetry flux can condense near the boundary ($\langle \mu(0) \rangle \neq 0 $), enforcing a symmetry-preserving boundary condition $\langle \sigma(0) \rangle = 0$ and no spontaneous magnetization (see Fig.~\ref{fig14} (c)). However, if $T \mu T = -\mu$, the $\mathbb{Z}_2^T$ symmetry prohibits such a perturbation, making the edge magnetization stable. 
Additionally, all symmetry-allowed perturbations correspond to operators with scaling dimensions greater than one, rendering them irrelevant in the boundary renormalization group flow~\cite{Verresen2021prx}. 
In summary, the edge spontaneously magnetization breaks $P$ symmetry, stabilized by the $\mathbb{Z}_2 \times \mathbb{Z}_2^T$-enriched bulk CFT.  

So far, we have focused on a single boundary, applicable to a half-infinite system. For a finite system of length $ L $, one must account for the finite-size splitting of symmetry-preserving states, which entangle both edges:  
\begin{equation}
|\uparrow_l \uparrow_r \rangle \pm |\downarrow_l \downarrow_r \rangle \quad \textrm{and} \quad |\uparrow_l \downarrow_r\rangle \pm |\downarrow_l \uparrow_r \rangle. \label{eq:overviewstates}
\end{equation}  
To analytically determine this splitting, it is useful to start from the scale-invariant renormalization group fixed point and then consider perturbations that differentiate the states. For gapped systems, all four states in Eq.~\eqref{eq:overviewstates} remain degenerate at the fixed point, and local perturbations couple the edges only at the $ L^\text{th} $ order in perturbation theory. As a result, the finite-size splitting in gapped SPT phases is exponentially small in system size~\cite{Verresen2017prb}. In contrast, for critical systems, the renormalization group fixed point is governed by a CFT, leading to richer behavior: the two antiferromagnetic states in Eq.~\eqref{eq:overviewstates} are split from the ferromagnetic ones at an energy scale $ \sim 1/L $, which matches the finite-size bulk gap. This occurs because the spontaneous boundary magnetizations interact through the critical bulk. The remaining two states, $ |\uparrow_l \uparrow_r \rangle \pm |\downarrow_l \downarrow_r \rangle $, remain exactly degenerate within the CFT. 
Perturbations away from the fixed point, introduced by renormalization group-irrelevant operators, can further split these states, typically in second order in perturbation theory, since each edge couples to the critical bulk. 
Interestingly, the dominant contribution to this splitting scales as $ \sim 1/L^{14} $, arising from the seventh descendant of $ \mu $~\cite{Verresen2021prx}.  

The twofold degeneracy observed with open boundaries is a generic feature of symmetry-enriched Ising CFTs. The nature of the finite-size splitting depends on the protecting symmetry. 
In other cases, $ \mu $ may be odd under symmetries associated with additional gapped degrees of freedom, leading to exponentially small finite-size splitting. While algebraically localized edge modes are less common in prior literature, they remain physically meaningful. 
To the best of our knowledge,the first identification of an algebraically localized edge mode appeared in~\cite{Verresen2021prx}, where it was shown to be unique to gapless systems, absent in short-range gapped phases, and destroyed if the symmetry and universality class is not preserved.

\begin{figure*}
    \centering
   \includegraphics[width=0.9\linewidth]{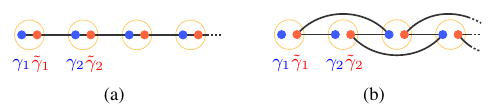}
    \caption{ (a) In the Majorana representation, for the standard translation-invariant Kitaev Majorana chain ($H$), the symmetry flux of fermion parity $ P $ is even under spinless time-reversal symmetry $T$.  (b) In the $ H''$ chain, the symmetry flux of $P$ is odd under $T$, which protects a Majorana edge mode with a finite-size splitting of $\sim 1/L^{14} $. If the system is non-interacting, this finite-size splitting becomes exponentially small. The figures are adapted from Ref.~\cite{Verresen2021prx}.}
    \label{fig15}
\end{figure*}

\paragraph{\textbf{The Jordan-Wigner mapping between bosonic and fermionic gSPT states in one dimension}}: Alternatively, as shown in Section~\ref{section2}, in one dimension, a Jordan-Wigner transformation maps a symmetry-enriched Ising CFT in bosonic spin systems to a fermionic gSPT state in free Majorana fermion systems, where the edge mode becomes more apparent. Specifically, the bulk transforms into a free Majorana $ c=\frac{1}{2} $ CFT, and each boundary hosts a localized zero-energy Majorana edge mode. This system exhibits the same ground state degeneracy as the well-known gapped Kitaev chain~\cite{Kitaev_2001}. Moreover, similar to the Kitaev chain, an edge Majorana operator that toggles between the two stable ground states exists. 
The system is characterized by the symmetry flux of fermionic parity being odd under spinless time-reversal symmetry, as illustrated in Fig.~\ref{fig15} (a) and (b). 
This critical Majorana chain emerges as a phase transition between two gapped phases: one with two Majorana modes per edge (protected by spinless time-reversal symmetry) and the Kitaev chain phase with one Majorana mode per edge. At the transition, one edge mode delocalizes and becomes the bulk critical mode, while the other edge mode remains localized~\cite{Verresen2018prl}.  
Mapping this behavior back to the spin chain, we start in the gapped cluster SPT phase, where two of the four degenerate ground states under open boundary conditions experience a splitting determined by the bulk correlation length. As the system approaches the critical point toward the Ising SSB phase, these states delocalize, leaving only the twofold degeneracy described earlier. 
Upon entering the gapped Ising phase, an edge magnetization emerges, which gradually merges with the bulk magnetization as the system moves deeper into this phase.

\subsection{Gapless symmetry protected topological phases}
\label{section3.3}
The concept of nontrivial topology at quantum critical points can be further extended to stable critical phases, now known as gSPT phases, first proposed by Keselman and Berg~\cite{Keselman2015prb} and later systematically investigated by Scaffidi et al~\cite{Scaffidi2017prx}. The symmetry-enriched quantum critical points discussed earlier represent a special class of gSPT order (phase), where all nontrivial topological phenomena occur at a single critical point. 
In this section, we review the simplest gSPT phases, which can be regarded as stacks of gapped SPT with CFTs. Their topological edge modes originates from the gapped SPT phases, making them ``non-intrinsic'' and presence of a gapped sector (as categorized in Table~\ref{tab:gSPT}). In the following two subsections, following Ref.~\cite{Scaffidi2017prx}, we first introduce the decorated domain wall construction, a technique originally used to construct gapped SPT phases~\cite{Chen2014NC}, which can also be applied to construct non-intrinsic gSPT phases. The gSPT phases built through this approach can be viewed as ``twisted'' versions of trivial quantum critical points or critical phases, similar to how gapped SPT phases can be interpreted as ``twisted'' quantum paramagnets. 
Finally, we present a concrete lattice realization of these gSPT phases, namely, the topological Luttinger liquid.
 
\subsubsection{Decorated domain construction for non-intrinsic gSPT phases}
\begin{figure}
    \centering
   \includegraphics[width=0.45\linewidth]{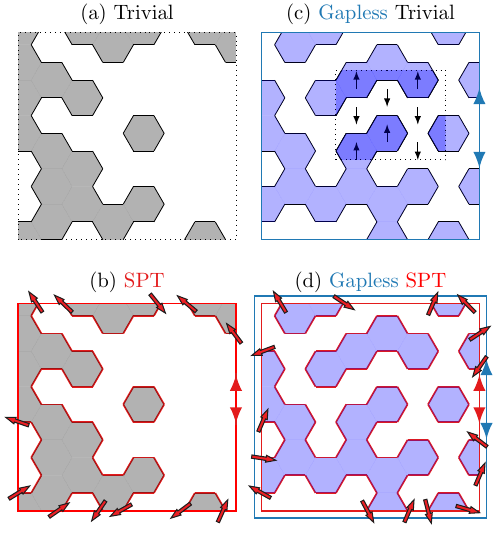}
    \caption{ (a) Trivial PM: Paramagnetic spins on a triangular lattice with fluctuating domain walls. (b) Gapped SPT: Decorating the domain walls transforms the system into an SPT phase with a $c=1$ edge mode. (c) Gapless Trivial: Tuning the domain walls to criticality by constraining them to fully-packed loop configurations closes the bulk gap and produces a $c=1$ edge mode. (d) Gapless SPT: A CFT docorated with a gapped SPT, resulting in a gapless SPT phase with $c=1+1=2$ edge modes. The figures are adapted from Ref.~\cite{Scaffidi2017prx}.}
    \label{fig16}
\end{figure}

To set the stage, we consider a bosonic system (e.g. spin model) in $ d $ dimensions, consisting of $\sigma$ and $\tau$ degrees of freedom and governed by the global symmetry group $ \Gamma = G_\sigma \times G_\tau $, where we take $ G_\sigma = \mathbb{Z}_2 $ for simplicity.
The construction follows the decorated domain wall approach to gapped SPT phases~\cite{Chen2014NC}. 
In this context, a trivial paramagnet (see Fig.~\ref{fig16} (a)) is described as a condensate of domain walls in a gapped system. 
However, a topological paramagnet (or SPT phase, see Fig.~\ref{fig16} (b)) arises when the domain walls of $ G_\sigma = \mathbb{Z}_2 $ are ``decorated'' with $(d-1)$-dimensional SPT phases protected by the symmetry $ G_\tau $. This construction naturally leads to topologically protected edge modes, as the terminating domain walls at the system boundary host the edge modes of the lower-dimensional SPT phase.

To generalize for gapless systems, we need to tune the underlying $\sigma$ degrees of freedom to criticality, thereby driving the domain wall condensate to the criticality. Similar to the gapped case, when the domain walls remain undecorated, this tuning typically leads to a topological trivial quantum critical point (gapless trivial (gTrivial) in Fig.~\ref{fig16} (c)). For example, in the one-dimensional critical Ising model, the phase transition can be understood as a domain wall condensation transition. In such cases, the edges of gTrivial systems generally lack symmetry protection and may or may not exhibit additional gapless modes.

The key idea behind constructing topologically nontrivial gapless states using the decorated domain wall method is that a trivial critical system can be ``decorated'' with lower-dimensional gapped SPT phases (see Fig.~\ref{fig16} (d)). Analogous to the gapped case, the resulting topological nontrivial state shares the same bulk properties as the trivial gapless system but exhibits fundamentally different edge physics. These exotic edge phenomena manifest as topological edge modes near the boundary, which may take the form of anomalous edge magnetization or the emergence of ballistic dynamics at the edge of a diffusive system. It is important to note that the domain wall construction reviewed in this section can be applies to a broad class of gapless systems. In the next subsection, we use the decorated domain wall method to construct a concrete lattice model of a topological Luttinger liquid in one dimension~\cite{Scaffidi2017prx, Parker2018prb} and provide numerical evidence to verify the nontrivial topological properties of these phases.

\subsubsection{Example: topological Luttinger liquid in one dimension}
In the previous section, we have repeatedly discussed the gSPT phase with $\mathbb{Z}_2 \times \mathbb{Z}_{2}$ symmetry, exemplified by the critical cluster Ising chain or purely gSPT phase. To highlight the generality of the decorated domain wall construction, we now impose an additional $U(1)$ symmetry on the gapless $\sigma$ spins to stabilize them in a Luttinger-liquid (LL) phase. We then investigate the possible realization of a gSPT state with $U(1) \rtimes \mathbb{Z}_{2} \times \mathbb{Z}_{2}$ symmetry in one dimension, postponing discussions of higher-dimensional cases to a later section.

\begin{figure}
    \centering
   \includegraphics[width=0.85\linewidth]{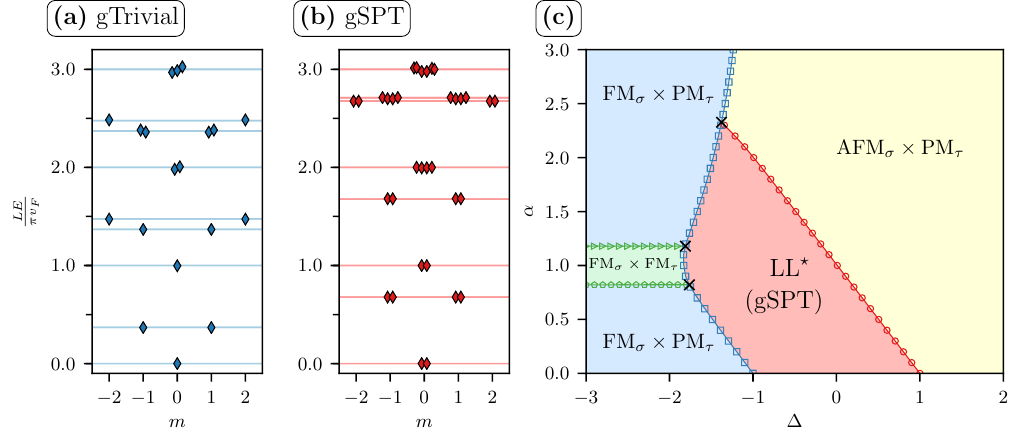}
    \caption{The rescaled energy Spectrum of $H_{\text{gTrivial}}^{\text{LL}}$(a) and gSPT $ H_{\text{gSPT}}^{\text{LL*}}$ (b). In both cases, the spectra are normalized to allow the operator dimensions of the CFT to be directly read off. The conformal blocks are labeled by the magnetic charge sector $m$ and arranged horizontally, with small horizontal spacings indicating degenerate eigenvalues (up to exponential splitting). In the gSPT case, all states are doubly degenerate due to the presence of topological edge modes, and the operator dimensions differ from those in the gTrivial case. The numerical spectra were obtained using DMRG~\cite{White1992prl}, incorporating finite-size scaling. Solid lines represent the exponents predicted by boundary CFT, calculated using $ \Delta_{\rm eff} = - \cos \pi g $~\cite{francesco2012conformal}. (c) Quantum phase diagram of $ H_{\text{gSPT}}^{\text{LL*}}$, computed via DMRG. Each line corresponds to a different eigenvalue crossing that marks a phase transition, while black crosses indicate multicritical points. The figures are adapted from Ref.~\cite{Scaffidi2017prx}.}
    \label{fig17}
\end{figure}

We begin with a trivial gapless system in one dimension, specifically a Luttinger liquid~\cite{giamarchi2003quantum}. The starting Hamiltonian is defined as follows~\cite{Scaffidi2017prx}:  
\begin{equation}
	\begin{aligned}
	& H^{\rm LL}_\text{gTrivial} \ =\ \sum_i \sigma_i^x \sigma_{i+1}^x + \sigma_i^z \sigma_{i+1}^z + \Delta \sigma_i^y \sigma_{i+1}^y \\
	 &- \sum_i \tau_{i-\frac{1}{2}}^x + g_\tau \tau_{i-\frac{1}{2}}^z \tau_{i+\frac{1}{2}}^z + u_\tau \tau_{i-\frac{1}{2}}^x \tau_{i+\frac{1}{2}}^x + \alpha  \sigma_i^y \tau_{i+ \frac{1}{2}}^x \sigma_{i+1}^y,\\ 
\end{aligned}
	\label{eq:H_gTrivialXXZ}
\end{equation}
where the $\sigma$ spins form a gapless XXZ model, coupled via the $\alpha$-term to gapped Ising $\tau$ spins, which are deep in their paramagnetic phase (with $u_\tau$ and $g_\tau$ small). The parameter $\Delta$ controls the $\sigma^y \sigma^y$ interaction, and the Hamiltonian exhibits a $ U(1) \rtimes \mathbb{Z}_2^{(\sigma)} \times \mathbb{Z}_2^{(\tau)} $ symmetry, generated by $ U_\theta = \prod_i e^{i \theta \sigma^y_i} $, $\mathcal{C}_\sigma = \prod_i \sigma_i^x$, and $\mathcal{C}_\tau = \prod_i \tau_{i+1/2}^x$, respectively. For small $\alpha$, the gapped $\tau$ spins can be integrated out, effectively renormalizing the anisotropy parameter as $ \Delta_{\rm eff} = \Delta - \alpha \langle \tau^x \rangle $. The $\sigma$ spins remain gapless for $ -1 < \Delta_{\rm eff} \leq 1 $, forming a Luttinger liquid phase. The low-energy effective Lagrangian is given by ${\cal L} = \frac{g}{4 \pi} (\partial_\mu \phi )^2 $, where $ \Delta_{\rm eff} = -\cos (\pi g) $, and \(\phi\) is a compact boson field. To obtain a topologically nontrivial Luttinger liquid, we apply the unitary twist $ U $~\cite{Scaffidi2017prx,Parker2018prb} to the Hamiltonian and wavefunction of the ordinary Luttinger liquid: $U = \prod_{{\rm DW}(\sigma)} (-1)^{\frac{1-\tau_{i-1/2}^z}{2}},\label{eq:twist_operator}$ where the product runs over all domain walls of the $\sigma$ spins in the $z$ basis. The operator $ U $ effectively attaches a $\mathbb{Z}_2^{(\tau)}$ charge to the domain walls of $\mathbb{Z}_2^{(\sigma)}$~\cite{Chen2014NC}. Concretely, $ U $ introduces a factor of $ (-1) $ for each pair of consecutive down spins (in the $z$ basis) in a classical spin configuration. The resulting Hamiltonian is given by $H^{\rm LL^\star}_\text{gSPT} = U H^{\rm LL}_\text{gTrivial} U^\dagger$. Importantly, the $ U(1) $ symmetry is also modified by the twist. 
In the special limit $ \alpha = g_\tau = u_\tau = 0 $, the twisted Hamiltonian exhibits topological edge modes, which remain robust as long as the gap of the $\tau$ spins remains open. These edge modes induce spontaneous edge magnetization along the $z$-direction, modifying the conformally invariant boundary conditions~\cite{francesco2012conformal}, and more importantly, the presence of nontrivial edge modes can modify the boundary critical exponents, which is a sharp contrast with the $\mathbb{Z}_2 \times \mathbb{Z}_{2}$ gSPT case. The stability of this topological Luttinger liquid (LL$^\star$) has been verified through exact diagonalization and DMRG calculations~\cite{Scaffidi2017prx,Parker2018prb}, as illustrated in Fig.~\ref{fig17}.  The phase diagram includes all perturbations from $H_\text{int}$ to avoid fine-tuning, confirming that LL$^\star$ is a stable critical phase. 
Various limiting cases can provide further insights into its proximate phases and phase transitions.  

i). \textbf{Along the line $ \alpha = 0 $:} The model reduces to the well-known XXZ model, which exhibits ferromagnetic order for $ \Delta < -1 $ and antiferromagnetic order for $ \Delta > 1 $. The intermediate region $ -1 < \Delta < 1 $ corresponds to an ordinary Luttinger liquid, whose low-energy properties are described by the compact free boson CFT.  

ii). \textbf{For $ \alpha \to \infty $:} The Hamiltonian is dominated by the interaction term $ \alpha \sigma^z \tau^x \sigma^z $, leading to $ 2^{L/2} $ degenerate ground states. Any classical configuration of $ \sigma^z $ is allowed, while the $ \sigma $ spins constrain the $ \tau $ spins via $ \langle\sigma^y_i \tau^x_{i+1/2} \sigma^y_{i+1}\rangle = -1 $. 
The perturbative analysis yields an effective low-energy Hamiltonian $H_\text{eff} = \sum_i \left( \Delta +1 \right) \sigma^y_i \sigma_{i+1}^y$, implying a direct transition from a ferromagnetic to an antiferromagnetic phase at $ \Delta = -1 $.  

iii). \textbf{For $ \Delta \to \infty $:} The system forms a trivial antiferromagnet for $\sigma$ and a paramagnet for $\tau$. Conversely, for $ \Delta \to -\infty $, the $ \sigma $ spins are perfectly ferromagnetic, enforcing $ \langle\sigma^y_i \sigma^y_{i+1}\rangle = 1 $. The effective Hamiltonian for the $ \tau $ spins is then $H_\text{eff} = -\sum_i (1-\alpha) \tau^x_{i-1/2} + g_\tau \tau^z_{i-1/2} \tau^z_{i+1/2} + u_\tau \tau^x_{i-1/2} \tau^x_{i+1/2}$. When $ u_\tau = 0 $, the $\tau$ spins undergo Ising transitions at $ \alpha = 1 \pm g_\tau $. The presence of $ u_\tau $ breaks integrability, shifting the transition points to $ \alpha = 1 \pm ( g_\tau - O(u_\tau) ) $.  

To analyze phase transitions out of the LL$^\star$ phase, we integrate out the gapped $\tau$ spins, leading to the effective anisotropy parameter$\Delta_\text{eff}(\alpha) = \Delta + \alpha \langle \tau^x \rangle \approx \Delta + \alpha$.
This predicts transitions at $ \Delta_\text{eff}(\alpha) = \pm 1 $, or equivalently, $ \Delta(\alpha) = \pm 1 - \alpha $, which aligns with the phase boundaries in Fig.~\ref{fig17} (c) for $ \alpha \lesssim 1 $.  Therefore, the LL$^\star$ is a well-defined phase within the considered parameter regime and does not require fine-tuning. Notably, both LL and LL$^\star$ phases can be gapped out by dimerizing the $ \sigma $ spins, implying that translation symmetry by a single unit cell must be included in the protecting symmetry group. Unlike previously studied topological Luttinger liquids~\cite{Keselman2015prb,JIANG2018753}, the decorated domain wall construction does not rely on spin-charge separation. Instead, it provides a systematic framework for generating many-body gSPT phases while making their topological properties transparent, analogous to gapped SPT systems.  

\subsection{Intrinsically gapless SPT phase}
\label{section3.4}
In the previous section, we introduced the remarkable stability of SPT physics upon closing the bulk energy gap. The nontrivial topological classifications present in critical systems can also be found in their gapped counterparts with the same symmetry and dimensionality. However, a more intriguing and fundamental question arises: can entirely new topological phases emerge exclusively in gapless systems, exhibiting nontrivial topological properties beyond gapped counterparts? To address this, a pioneering work by Thorngren et al.~\cite{Thorngren2022prb} proposed a novel mechanism leading to topological phases that are not only gapless but also fundamentally require the absence of the bulk energy gap. These phases, known as intrinsically gSPT phases, have no gapped counterparts and are thus distinct from previously discussed gSPT phases. Intrinsically gSPT phases exhibit several unique properties, including: (i) Protected edge modes that cannot be realized in any gapped system with the same symmetries. (ii) String order parameters that are likewise forbidden in gapped phases. (iii) Constraints on the phase diagram that arise when perturbing the system. In the following subsections, we briefly introduce two lattice models and field theories, illustrating the key concepts underlying intrinsically gSPT states.

\subsubsection{Example 1: dopped Ising-Hubbard chain}
In the doped Ising chain example we will discuss, electrons are gapped, causing the fermion parity symmetry to act trivially on all gapless degree of freedom. The effective low-energy symmetry is therefore given by the quotient of the full symmetry group by the subgroup that acts only on the gapped degrees of freedom. A defining feature of intrinsically gSPT phases is the emergent anomaly in the low-energy symmetry, which can be diagnosed by examining the charges of string order parameters. These string order parameters become nontrivial when they are inconsistent with any gapped SPT phase. Importantly, we emphasize that emergent anomalies associated with on-site symmetries lead to nontrivial edge modes that are protected by both symmetry and gaplessness.

\begin{figure}
    \centering
   \includegraphics[width=0.55\linewidth]{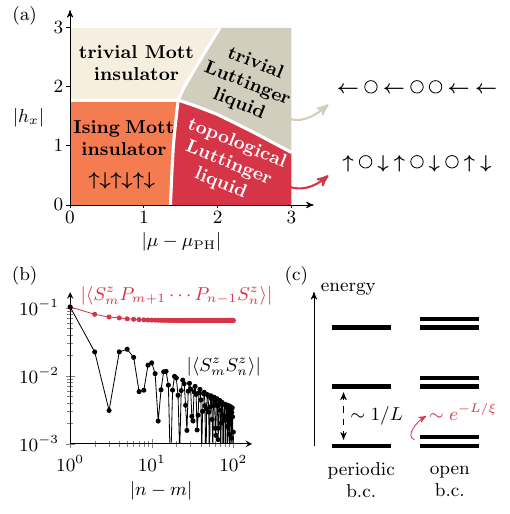}
    \caption{(a) Phase diagram for the doped Ising-Hubbard model with parameters $ t = J_z = 1 $ and $ U = 5 $. The chemical potential is measured relative to the particle-hole symmetric value $ \mu_\textrm{PH} = U/2 $. (b) The topological properties (or hidden symmetry breaking) of the topological Luttinger liquid can be detected using the string order parameter, shown here for $ |\mu - \mu_\textrm{PH}| = 2 $ and $ h_x = 0 $.  (c) While the topological Luttinger liquid has a unique ground state under periodic boundary conditions, it exhibits a twofold degeneracy under open boundary conditions, with an exponentially small energy splitting determined by the spin correlation length. The figures are adapted from Ref.~\cite{Thorngren2022prb}.}
    \label{fig18}
\end{figure}

To illustrate this phenomenon, we consider a simple lattice model---the doped Ising-Hubbard model~\cite{Thorngren2022prb}. This model describes a chain of spinful fermions $ c^{\dagger}_s $ with the Hamiltonian $ H = H_\textrm{Ising}+ H_\textrm{Hub} $, where
\begin{align}
H_\textrm{Ising} &= \sum_n \left( J_z S^z_{n} S^z_{n+1} + h_x S^x \right), \label{eq:Ising}\\
H_\textrm{Hub} &= -t \sum_{j,s} \left(c^\dagger_{j+1,s} c_{j,s} + h.c.\right) + U \sum_j n_{j,\uparrow} n_{j,\downarrow} - \mu N, \label{eq:Hub}
\end{align}
with $ n_{j,s} = c^\dagger_{j,s} c_{j,s} $ and $ S^\alpha_j = \frac{1}{2} c^\dagger_{j,s} \sigma^\alpha_{s,s'} c_{j,s'} $. The spin rotation symmetry is explicitly broken by the $ J_z $ and $ h_x $ terms down to a $\pi$-rotation $ R_x $ around the $ x $-axis. Importantly, this defines a $\mathbb{Z}_4$ symmetry, since $ R_x^2 = (-1)^F = P $. At half-filling, the Hubbard interaction $ U $ drives the system into a Mott insulating phase ($\langle n_j \rangle = 1$), effectively reducing it to a spin-$1/2$ chain. Depending on the value of $ h_x $, the system can be in either an Ising phase, where $ R_x $ is spontaneously broken down to its $\mathbb{Z}_2$ fermion parity subgroup, or in a symmetry-preserving paramagnetic phase. However, upon doping ($\langle n_j\rangle \neq 1$), a gapless Luttinger liquid emerges, while the spin degrees of freedom remain gapped. The numerical phase diagram for $ t=J_z=1 $ and $ U=5 $, obtained using the DMRG \cite{White1992prl}, is shown in Fig.~\ref{fig18} (a).

A key consequence of doping the Ising antiferromagnet is the destruction of long-range magnetic order in $ S^z $, as an arbitrary number of holes can separate antiferromagnetically aligned spins. Nevertheless, as illustrated in Fig.~\ref{fig18} (a), the resulting Luttinger liquid exhibits a hidden symmetry-breaking pattern: in the ``squeezed state'', where holes are removed, antiferromagnetic long-range order restore~\cite{Kruis2004prb,Zhang1997prb}. 
This hidden order can be probed by measuring the $ S^z $-correlation function while taking into account the positions of all holes, which is achieved by inserting a string of fermion parity operators:$ \langle S^z_m \big( \prod_{m<k<n}P_k \big) S^z_n \rangle$. Figure~\ref{fig18} (b) shows that this string operator exhibits long-range order in the topological Luttinger liquid phase adjacent to the Ising phase, whereas the local Ising order parameter decays algebraically.  Moreover, the presence of a nontrivial string order implies that: (i) the system hosts protected edge modes, and (ii) no gapped SPT phase can support this particular type of string order~\cite{Thorngren2022prb}. Thus, this phase corresponds to an intrinsically gSPT state.

Furthermore, the fractionalization of the $ R_x $ symmetry can be observed in the string order parameter, where its endpoints, carrying an $ S^z $ insertion, are charged. This is a hallmark feature of SPT phases~\cite{Pollmann2012prb}. However, in this case, we observe long-range order in a string order parameter even the system being gapless. Although the concept of gSPT phases has been discussed in previous sections, the example presented in this section is novel since the charge of the $\mathbb{Z}_{4}$ symmetry string operator remains well-defined as long as $ R_x $ symmetry is preserved. This makes it the first example that extends beyond the gapped classification, since $ H^2(\mathbb{Z}_4,U(1)) = 0 $, implying that such string order cannot exhibit long-range order in any gapped phase. 

What are the consequences of the charged string operator in this case? Unlike the critical cluster Ising model discussed above, where long-range string operator implies bulk symmetry-breaking degeneracy, in the context of intrinsically gSPT phases, charged string operator leads to degeneracies on an interval with open boundaries. To see this, we note that the string operator maintains long-range order even as its endpoints approach the boundaries. Applying the global symmetry transformation $ P $, the nonlocal correlator transforms into a correlation function of a pair of locally charged operators near the boundaries. This implies spontaneous symmetry breaking at each boundary, resulting in an exponentially split ground-state degeneracy, with a correlation length determined by the spin gap. In contrast, the energy splitting in the bulk is much larger ($\sim 1/L$), meaning this degeneracy can be detected in the finite-size spectrum, as sketched in Fig.~\ref{fig18} (c).  These distinctive properties of intrinsically gSPT phases can be understood in terms of an emergent anomaly. Since fermions are gapped, the parity subgroup $ \mathbb{Z}_2 \subset \mathbb{Z}_4 $ acts only on the gapped degrees of freedom, effectively reducing $ R_x $ to a $ \mathbb{Z}_2 $ symmetry in the low-energy theory. However, this symmetry action is incompatible with an on-site microscopic $ \mathbb{Z}_2 $ symmetry, which is the essence of the anomaly. This anomaly is also illustrated using $ R_x $ string correlators of the form  $\langle O_m (\prod_{m < k < n} R_x ) O_n \rangle$, where $ O $ is a local endpoint operator. This correlation function either decays algebraically or vanishes exponentially. As argued in Ref.~\cite{Thorngren2022prb}, the algebraic case occurs if and only if $ O $ has odd fermion parity, as it carries a nontrivial charge under $ R_x^2 $. Therefore, the $ R_x $ strings, which are associated with an effective $ \mathbb{Z}_2 $ symmetry, acquire a fractional charge from the perspective of the gapless degrees of freedom, which is a hallmark of the anomaly.

Additionally, to provide a unified description of the intrinsically gSPT phase and the emergent anomaly, Thorngren et al.\cite{Thorngren2022prb} develop a field theory starting from free spinful fermions ($ U=J_z=h_x=0 $) and introduce a single perturbation that drives the system into one of two topologically distinct Luttinger liquids, as shown in Fig.~\ref{fig18}. One of these phases exhibits protected edge modes and an emergent anomaly. Within this framework, the free spinful fermion serves as a critical transition point where the fermions become gapless, and the emergent anomaly undergoes a discontinuous change.

We represent the spinful fermion in Abelian bosonization as a pair of $ 2\pi $-periodic compact boson fields $ (\varphi_s,\theta_s) $, satisfying the commutation relation $[\partial_x \varphi_s(y),\theta_{s'}(x)] = 2\pi i \delta_{ss'} \delta(x-y),s=\uparrow,\downarrow$. 
The fermion operators are expressed as $\psi^\dagger_{s,\pm} = U_s e^{\pm i\varphi_s/2 + i\theta_s}$, where $ \pm $ denote the left- and right-moving components, and $ U_{1,2} $ are Klein factors ensuring that the operators anticommute \cite{von1998bosonization}. In this formulation, the $ \mathbb{Z}_4 $ symmetry acts as $\psi^\dagger_{s,\pm} \mapsto i \psi^\dagger_{-s,\pm}$, leading to the following transformation of the compact boson fields:  
\[\label{eqnsymmactions} 
R_x: \begin{cases}
\varphi_s \mapsto \varphi_{-s} \\
\theta_\uparrow \mapsto \theta_\downarrow +\pi/2 \\
\theta_\downarrow \mapsto \theta_\uparrow-\pi/2  \end{cases} \qquad \begin{cases} U_\uparrow \mapsto U_\downarrow \\ U_\downarrow \mapsto -U_\uparrow \end{cases}\]
Thus, rotation about the $ x $-axis exchanges opposite-spin fermions while introducing a phase, ensuring that $ R_x^2 = (-1)^F $.  

The perturbation that drives the gapless spinful fermion into two topologically distinct Luttinger liquids is: $ \mathcal{O}_{zz} = \cos (\varphi_\uparrow - \varphi_\downarrow) = \cos(\Phi_1) $, which gaps out all states with odd fermion parity~\cite{Thorngren2022prb}. The remaining low-energy degrees of freedom correspond to a Luttinger liquid of spinless Cooper pairs, with $ \psi_{\uparrow,+}^\dagger \psi_{\downarrow,-}^\dagger \sim \exp\left(i(\theta_\uparrow + \theta_\downarrow - \varphi_\uparrow/2 + \varphi_\downarrow/2)\right)$. This can be rewritten in terms of the conjugate compact boson fields $\Phi_2 = \varphi_\downarrow $ and $ \Theta_2 = \theta_\uparrow + \theta_\downarrow - \varphi_\uparrow/2 + \varphi_\downarrow/2$.  

To determine the symmetry action on these fields, we replace $ \Phi_1$ with its expectation value~\cite{Thorngren2022prb}. In the effective low-energy theory, $ R_x $ acts as a $ \mathbb{Z}_2 $ symmetry:  
\[\label{eqneffsymm}
R_x:\begin{cases} \Phi_2 \mapsto \Phi_2 + \langle \Phi_1 \rangle \\
\Theta_2 \mapsto \Theta_2 + \langle \Phi_1 \rangle,
\end{cases}
\]
where $ \langle \Phi_1 \rangle = 0 $ or $ \pi $ depending on the sign of the $ \cos \Phi_1 $ perturbation. If we take the sign to be positive, then $ \langle \Phi_1 \rangle = \pi $, and the resulting $ R_x $ action matches the anomalous boundary action of the Levin-Gu $ \mathbb{Z}_2 $-SPT phase \cite{Levin2012prb}. This confirms that the field theory describes an intrinsically gSPT phase with $ R_x $ acting as the unique anomalous $ \mathbb{Z}_2 $ symmetry. Conversely, if the sign of $ \mathcal{O}_{zz} $ is negative, we obtain a trivial Luttinger liquid phase with $ \langle \Phi_1 \rangle = 0 $ and a trivial $ R_x $ action.  

To derive the string operator for fermion parity in the effective field theory, we note that the generator of fermion parity given by $\exp(i \int dx( \partial_x \varphi_\uparrow/2 + \partial_x \varphi_\downarrow/2))$. Although fermion parity is a gapped symmetry, the factor $ e^{i\Phi_2(x)} $ in its correlation function leads to algebraic decay. To construct a string operator with long-range order, the endpoint operator must be $ O(x) = e^{-i\Phi_2(x)} $, which is charged under $ R_x $ in the topological phase. We can further reveal the edge modes by studying a spatial interface between the topological and trivial Luttinger liquids. Following Ref.~\cite{Keselman2015prb}, we tune the coefficient of the $ \mathcal{O}_{zz} = \cos \Phi_1 $ perturbation from positive to negative. The transition across the interface creates an edge mode associated with the path $ \Phi_1 $ takes from 0 to $ \pi $. Since any continuous energy-minimizing path has a degenerate partner under the exchange $ R_x:\Phi_1 \to - \Phi_1 $, the edge mode forms a spin-$\frac{1}{2}$ qubit. Together with the bulk gapless charges, this ensures that boundary fermions remain gapless. Thus, our $ \mathbb{Z}_4 $  case provides a clearer example of an intrinsically gapless SPT phase, as no subgroup supports a gapped SPT phase. This represents the first instance of an emergent anomaly in gSPT phases, which has not been previously explored in the literature.

\color{red}
Here, we present a complete field-theoretic analysis of the $\mathbb{Z}_4$-symmetric intrinsic gSPT state in the Ising–Hubbard chain, demonstrating how topology persists in critical theories.

Let $\Gamma = \mathbb{Z}_4$ denote the microscopic on-site symmetry, and let $G_{\rm gap} = \mathbb{Z}_2$ be the (normal) subgroup of $\Gamma$ that acts trivially on the gapless degrees of freedom. The effective symmetry of the low-energy theory is then $G_{\rm low} = \Gamma/G_{\rm gap} = \mathbb{Z}_2$, with $\pi: \Gamma \to G_{\rm low}$ being the quotient map.

For any system with such a gapped sector but a gapless low-energy theory, the partition function $Z(X, A)$ on a spacetime $X$ coupled to a background $\Gamma$ gauge field $A$ takes the form
\begin{equation}
\label{eqnpartfun}
Z(X,A) = Z_{\rm low}(X,A_{\rm low})e^{2\pi i \int_X \alpha(A)},
\end{equation}
where $Z_{\rm low}(X, A_{\rm low})$ is the partition function of the gapless degrees of freedom coupled to background $G_{\rm low}$ gauge field derived from $A$ by $A_{\rm low} = \pi(A)$. More explicitly, the $\mathbb{Z}4$-valued $A$ can be decomposed into $\mathbb{Z}_2$ gauge fields $A_{\rm low}$ and $A_{\rm gap}$ as $A = 2A_{\rm gap} + A_{\rm low}$, with the $\mathbb{Z}_4$ constraint encoded in $dA = 0$ mod 4 $\Leftrightarrow$ $2dA_{\rm gap} = dA_{\rm low}$ mod 4. This relation implies that a $2\pi$-flux of $A_{\rm low}$ corresponds to a $\pi$-flux of $A_{\rm gap}$, reflecting the relation $R_x^2 = (-1)^F$. The term $\alpha(A)$ is a topological term obtained after integrating out the gapped degrees of freedom. In the case that there is no emergent anomaly, $Z_{\rm low}(X, A_{\rm low})$ and $\alpha(A)$ are both gauge invariant, and $\alpha(A)$ describes a $\mathbb{Z}_4$-SPT phase in the gapped sector.

When there is an emergent anomaly, on the other hand, $Z_{\rm low}(X, A_{\rm low})$ and $\alpha(A)$ are not separately gauge invariant, and instead transform in such a way that only their combination $Z(X, A)$ is gauge invariant. We cannot interpret $\alpha$ as an SPT class in this case. Instead, invoking the bulk-boundary correspondence, we can express the emergent anomaly in terms of a topological term $\omega(A_{\rm low}) = \frac{1}{4}A_{\rm low} d A_{\rm low}$ for a $G_{\rm low}$ SPT in one higher dimension \cite{Chen2013prb,kapustin2014anomaliesdiscretesymmetriesvarious}. This means that for $\partial X = 0$, $Z_{\rm low}(X,A_{\rm low})\exp\left(2\pi i \int_{X \times \mathbb{R}_{\ge 0}} \omega(A_{\rm low})\right)$ is gauge invariant. By standard arguments, gauge invariance of \eqref{eqnpartfun} on closed spacetime manifolds is then equivalent to the anomaly vanishing equation $d\alpha = \omega$. Solving this condition, we find that the unique topological term is $\alpha(A) = \frac{1}{2} A_{\rm gap} A_{\rm low}$. Under a gauge transformation $A_{\rm gap} \mapsto A_{\rm gap} + dg$, there is a boundary term $\frac{1}{2} g \; A_{\rm low}$, indicating that there must be an extra boundary contribution which makes the combination \eqref{eqnpartfun} gauge invariant again. This extra boundary contribution must come from some kind of edge mode. 

\color{black}

\subsubsection{Example 2: Intrinsically gSPT phases in critical spin chain}
As a second example, we introduce a simpler one-dimensional spin model~\cite{Li2024scipost,li2023intrinsicallypurelygaplesssptnoninvertibleduality} that also exhibits a $ \mathbb{Z}_4 $-symmetric intrinsically gSPT phase and can be easily constructed from familiar spin models (e.g., the Ising and XY models). Consequently, this model is more accessible for both theoretical analysis and numerical computations.

To this end, we first notice that intrinsically gSPT states form a class of gapless systems that exhibit an emergent anomaly at a low-energy limit. While the full global symmetry $ \Gamma $ remains anomaly-free, it does not act faithfully on the low-energy gapless degrees of freedom. Instead, a normal subgroup $ H $ of $ \Gamma $ exists that acts only on the gapped sector, such that the quotient group $ \Gamma/H $ faithfully represents the low-energy sector. If $ \Gamma $ is a nontrivial extension of $ \Gamma/H $ by $ H $, the quotient group $ \Gamma/H $ acquires a nontrivial ’t Hooft anomaly~\cite{Wang2018prx,wang2021domainwalldecorationsanomalies}, referred to as the emergent anomaly. The emergent anomaly of $ \Gamma/H $ is precisely canceled by the quantum anomalous in the gapped SPT phase with $ H $ symmetry, ensuring that the total global symmetry $ \Gamma $ remains anomaly-free. Therefore, the gapped sector plays a crucial role in protecting the nontrivial topological properties of intrinsically gSPT phases.

We begin with a Hamiltonian describing an Ising $\sigma$ spin SSB phase stacked with an XY $\tau$ spin chain:  
\begin{eqnarray}\label{eq:SSBXX}
 H_{\text{SSB}+\text{XY}} = - \sum_{i=1}^{L} \left(\tau^z_{i-\frac{1}{2}} \tau^z_{i+\frac{1}{2}}  + \tau^y_{i-\frac{1}{2}} \tau^y_{i+\frac{1}{2}}  + \sigma^z_{i-1} \sigma^z_{i}\right).
\end{eqnarray}  
Here, each pair ($\tau_{i-1/2},\sigma_i$) represents the ith unit cell, where the two species of spins per unit cell are described by Pauli operators $\sigma$ and $\tau$, acting on the sites and links of a one-dimensional chain, respectively. This Hamiltonian possesses a larger global symmetry group $ U(1)^\tau \times \mathbb{Z}_2^\sigma $, where $ \mathbb{Z}_2^\sigma $ is generated by $ U_\sigma = \prod_{i}\sigma^{x}_{i}$, and $ U(1)^\tau $ is generated by $ \prod_{i=1}^{L} e^{\frac{i \alpha}{2} (1-\tau^x_{i-\frac{1}{2}})}, \hspace{1cm} \alpha \simeq \alpha+2\pi$. The normal subgroup $ \mathbb{Z}_2^\tau $ of $ U(1)^\tau $ is generated by $ U_\tau  = \prod_i \tau^{x}_i$. Instead of considering the full $ U(1)^\tau $ symmetry, we focus on its $ \mathbb{Z}_4^\tau $ subgroup ($\alpha=\pi/2$), whose generator is  $V_\tau= \prod_{i=1}^{L} e^{\frac{i \pi}{4} (1-\tau^x_{i-\frac{1}{2}})}$, which satisfies $ V_\tau^2 = U_\tau $. We are thus interested in the symmetry group $ \mathbb{Z}_4^\tau \times \mathbb{Z}_2^\sigma $, generated by $ V_\tau $ and $ U_\sigma $.  

Now, we apply the KT transformation~\cite{Kennedy1992CMP,Kennedy1992prb}, using the $ \mathbb{Z}_2^\tau \times \mathbb{Z}_2^\sigma $ symmetry generated by $ U_\tau $ and $ U_\sigma $. Under this transformation, the operators transform as follows:  
\begin{eqnarray}
\begin{split}
    \mathcal{N}_{\text{KT}} \tau^z_{i-\frac{1}{2}} \tau^z_{i+\frac{1}{2}} &=  \tau^z_{i-\frac{1}{2}}\sigma^x_i \tau^z_{i+\frac{1}{2}} \mathcal{N}_{\text{KT}}, \\
    \mathcal{N}_{\text{KT}} \tau^y_{i-\frac{1}{2}} \tau^y_{i+\frac{1}{2}} &=  \tau^y_{i-\frac{1}{2}} \sigma^x_i \tau^y_{i+\frac{1}{2}}\mathcal{N}_{\text{KT}},  \\
    \mathcal{N}_{\text{KT}}  \sigma^z_{i-1} \sigma^z_{i}  &= \sigma^z_{i-1}\tau^x_{i-\frac{1}{2}} \sigma^z_{i}\mathcal{N}_{\text{KT}}.
\end{split}
\end{eqnarray}  
As a result, the Hamiltonian precisely describes an intrinsically gSPT phase with $ \mathbb{Z}_4^\tau \times \mathbb{Z}_2^\sigma $ symmetry~\cite{Li2024scipost,li2023intrinsicallypurelygaplesssptnoninvertibleduality}:  
\begin{eqnarray}\label{eq:igSPTHam}
 H_{\text{igSPT}} = - \sum_{i=1}^{L} \left(\tau^z_{i-\frac{1}{2}} \sigma^x_{i} \tau^z_{i+\frac{1}{2}}  + \tau^y_{i-\frac{1}{2}} \sigma^x_{i} \tau^y_{i+\frac{1}{2}}  + \sigma^z_{i-1} \tau^x_{i-\frac{1}{2}} \sigma^z_{i}\right).
\end{eqnarray}  

The low-energy effective theory of Eq.~(\ref{eq:igSPTHam}) is described by a $c=1$ free boson CFT. 
The model possesses a $\mathbb{Z}_{4}$ symmetry generated by $V_\tau U_\sigma= U = \prod_{i}\sigma^{x}_{i}e^{i\frac{\pi}{4}(1-\tau^{x}_{i-1/2})}$, which exhibits an emergent anomaly at low energies. The reasoning is as follows: in the low-energy sector, where $\sigma^z_{i-1} \sigma^z_{i} = \tau_{i-1/2}^x$, the $\mathbb{Z}_{4}$ is approximately $U \sim \prod_{i}\sigma^{x}_{i}e^{i\frac{\pi}{4}(1-\sigma^z_{i-1} \sigma^z_{i})}$, which corresponds to the same anomaly present on the boundary of a (2+1)D Levin-Gu SPT phase~\cite{Levin2012prb}.
This anomaly prohibits the realization of a gapped phase with the same symmetry and dimensionality.
Moreover, in an open chain with a length $L$, the square of the low-energy symmetry operator fractionalizes onto each end of the boundary~\cite{Wen2023prb,wen2023classification11dgaplesssymmetry}, $U^2 \sim \tau^x_{1/2} \sigma_1^z \sigma^z_{L}$. 
This fractional charge locally anticommutes with the $U$ symmetry, thereby protecting a two-fold ground-state degeneracy.

\subsection{Universal properties of gSPT phases in one dimension}
\label{section3.5}

\subsubsection{General construction}
\label{section3.5.1}
After introducing several typical lattice models that exemplify the nature of gSPT phases, we now summarize the general construction methods for (1+1)D gSPT phases in strongly interacting quantum spin systems. These methods include the decorated defect construction, the KT transformation, and the pivot Hamiltonian approach.

\paragraph{Decorated defect construction}:
A useful method for constructing both gSPT and intrinsically gSPT phases is the decorated defect construction (DDC), which generalizes the decorated domain wall method previously discussed in Sec.~\ref{section3.3} and Refs.~\cite{Chen2014NC, Scaffidi2017prx}. The DDC was initially developed to construct gapped SPT states~\cite{Chen2014NC} and was later extended to discover the first examples of gSPT ~\cite{Scaffidi2017prx}. The goal of this section is to review this construction and apply it to bosonic spin chains with on-site symmetries that exhibit both gSPT and intrinsically gSPT phases. These models are relatively simple, allowing for the derivation of certain analytical results regarding their symmetry properties.

We first briefly review the DDC approach for constructing gapped SPT phases, which starts from known lower-dimensional gapped SPT phases \cite{Chen2014NC}. Suppose one aims to construct a gapped SPT system with a global symmetry $\Gamma$. Assume that $\Gamma$ fits into the symmetry extension  $1\to H\to \Gamma\to G\to 1 $, where $ H $ is a normal subgroup of $ \Gamma $, and $ G := \Gamma/H $. For simplicity, we assume that the extension is central, meaning that $ G $ does not act on $ H $. The construction begins with a phase where $ G $ symmetry is spontaneously broken. On codimension-$ p $ $ G $-defects, one decorates a $(d+1-p)$-dimensional gapped SPT phase protected by the symmetry $ H $ (i.e., an $ H $ symmetric gapped SPT). To restore the full $ \Gamma $ symmetry, the decorated $ G $-defects must proliferate in a way that is consistent with the absence of $ H $-anomalies. If nontrivial gapless degrees of freedom remain localized on the $ G $-defects, their proliferation will not lead to a gapped phase with a unique ground state. The system forms a gapped SPT phase protected by $ \Gamma $ when defect proliferation.  Notably, a given symmetry $ \Gamma $ can admit different symmetry extensions characterized by different choices of the pair $ (H, G) $. For a fixed extension $ (H, G) $, once all possible ways of decorating $H $ symmetric gapped SPT phases on $ G $-defects are enumerated, proliferating the $ G $-defects will yield all possible $ \Gamma $-symmetric gapped SPTs. Consequently, different choices of $ (H, G) $ produce the same set of $ \Gamma $ symmetric gapped SPTs, allowing one to select the most convenient pair for the construction.

Let us now construct $\Gamma$-symmetric gSPT states by extending the decorated defect construction described above. The following discussion mainly follows Ref.~\cite{Li2024scipost}. We continue to assume that the global symmetry $\Gamma$ fits into the symmetry extension and begins with a gapped phase where $ G $ is spontaneously broken. On a codimension-$ p $ $ G $-defect, we decorate a $(d+1-p)$-dimensional gapped SPT phase protected by symmetry $ H $. More importantly, instead of keeping the $ G $-defect gapped, we drive it to a critical point. As a result, the system realizes a nontrivial critical point, which we identify as a gSPT phase. Compared to the decorated defect construction of gapped SPTs, the construction of gSPTs introduces several key distinctions. In particular, since the full proliferation of the $G$-defect no longer necessarily results in a gapped SPT phase, the consistency condition for the decoration can be relaxed. Depending on whether this consistency condition is maintained or relaxed, the resulting gSPTs are classified as non-intrinsic or intrinsic, respectively, as summarized in Table~\ref{tab:gSPT}.

\begin{figure*}
    \centering
   \includegraphics[width=0.8\linewidth]{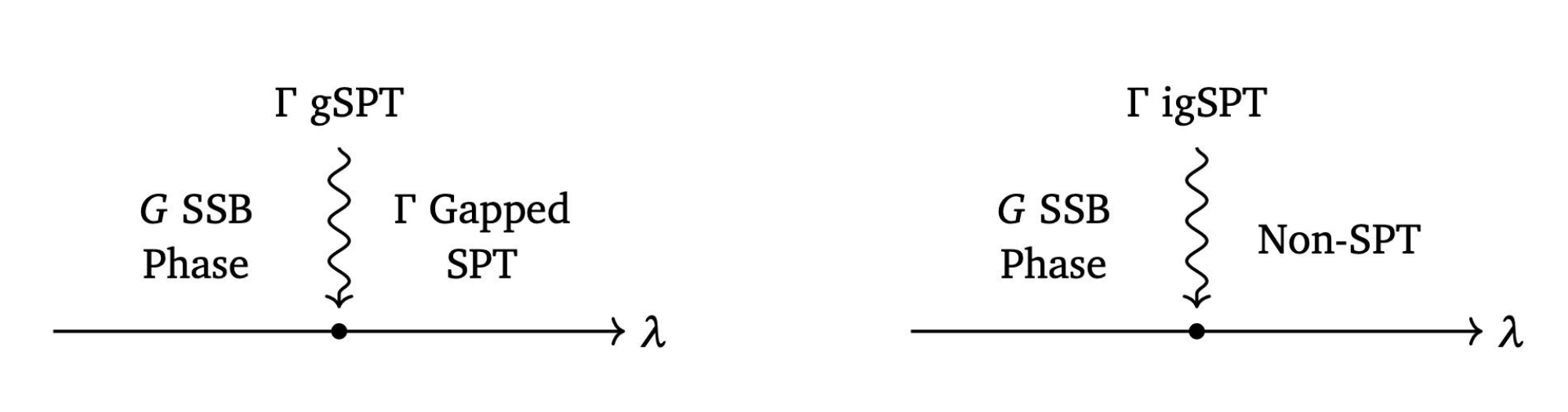}
    \caption{Phase diagram of non-intrinsic and intrinsically gSPTs. The horizontal axis represents the strength of $G$-defect fluctuation. For the non-intrinsic case (left panel), the $G$-defects can be fully proliferated, resulting in a $\Gamma$ gapped SPT. For the intrinsic case (right panel), the $G$-defects can only fluctuate to the critical point. Further increasing the fluctuation strength does not drive the system into a $\Gamma$-symmetric gapped SPT phase. The figures are adapted from Ref.~\cite{Li2024scipost}.}
    \label{fig19}
\end{figure*}

\begin{enumerate}
    \item \emph{The non-intrinsic gSPT:} if an \textcolor{red}{$ H $-symmetric gapped SPTs} decorated on the $ G $-defects satisfy the same consistency condition as those used in the construction of gapped SPTs~\cite{Chen2014NC}. Specifically, the $ G $-defect must be free of any $ A $-anomaly. This ensures that further increasing the fluctuation strength of the $ G $-defect ultimately leads to a $ \Gamma $-symmetric gapped SPT. This is the key reason why the gSPT phase is non-intrinsic. In this scenario, the gSPT corresponds to the critical point separating the $ G $-symmetry SSB phase and $ \Gamma $-symmetric gapped SPT phase, as illustrated in the left panel of Fig.~\ref{fig19}.  In particular, when the symmetry extension is trivial, i.e., \textcolor{red}{$ \Gamma = H \times G $}, this construction reduces to the approach discussed in Refs.~\cite{Scaffidi2017prx,Verresen2021prx}.
    
    \item \emph{The intrinsically gSPT:} On the other hand, if an \textcolor{red}{$ H $-symmetric gapped SPTs} decorated on the $ G $-defects satisfy a weaker, modified consistency condition~\cite{Li2024scipost}. Specifically, the $G$ symmetry SSB phase we start with has a particular anomaly associated with a \textcolor{red}{quotient group $ G_{\rm quot} $ of $ \Gamma $, where $ G \subset G_{\rm quot} $. The choice of $ G_{\rm quot} $ and its anomaly should be considered as part of the input data for the construction. The defect decoration is constrained such that the anomaly of $ G_{\rm quot} $ in the $ G $ SSB phase is precisely canceled by the anomaly induced by the defect decoration.} After the decoration, the total symmetry group $ \Gamma $ is anomaly-free, and fluctuating the $ G $-defect to the critical point yields a $ \Gamma $-anomaly-free intrinsically gSPT~\cite{Thorngren2022prb}, as shown in the right panel of Fig.~\ref{fig19}.
\end{enumerate}

The gapless critical system is obtained by first fluctuating the $ G $-defects of the $ G $ symmetry SSB phase before decorating them with \textcolor{red}{$ H $-symmetric gapped SPTs.} For gSPTs, we must start with a critical point that is anomaly-free, whereas for intrinsically gSPTs, \textcolor{red}{we must begin with a critical point carrying a specific $ G_{\rm quot} $ anomaly. Moreover, for a given $ \Gamma $ symmetry, there may be multiple choices for the symmetry extension. While different choices of the symmetry extension $ (H, G) $ yield the same set of gapped SPTs, this is no longer true for intrinsically gSPTs. In the latter case, we must specify an anomaly of $ G_{\rm quot} $ (which includes symmetry $ G $) as part of the input data for the decorated defect construction. By definition, the resulting intrinsically gSPT depends on the choice of symmetry extension, the selection of $ G_{\rm quot} $, and the anomaly associated with $ G_{\rm quot} $.}

\paragraph{KT transformation}:
Another alternative approach to constructing gSPT states is the KT transformation, originally proposed by Kennedy and Tasaki~\cite{Kennedy1992prb,Kennedy1992CMP}. It was first discovered as a mapping that transforms Haldane’s spin-1 chain in the $\mathbb{Z}_2 \times \mathbb{Z}_2$ gapped SPT phase into a hidden $\mathbb{Z}_2 \times \mathbb{Z}_2$ SSB phase. This transformation was first found to have a simple and compact form, given by~\cite{Oshikawa_1992}:
\begin{eqnarray}\label{eq:KToriginal} 
U_{\text{KT}}=\prod_{i>j} \exp\left(i\pi S^{z}_i S^x_j\right), 
\end{eqnarray}
where $U_{\text{KT}}$ is a unitary and highly nonlocal operator. The KT transformation in Eq.~\eqref{eq:KToriginal} was originally formulated for spin-1 systems with open boundary conditions. However, its definition of a ring remained an open question until recently~\cite{Li2024scipost}. Interestingly, on a ring, the KT transformation is implemented by the non-unitary operator $\mathcal{N}_{\text{KT}}$, which obeys a non-invertible fusion rule.

This section provides a brief review of the application of the KT transformation in constructing various examples of gSPT states, both with and without a gapped sector (purely gSPT). Furthermore, we show that when gapped sectors are present, this construction naturally leads to the decorated defect construction discussed earlier.

The main results are summarized as follows~\cite{li2023intrinsicallypurelygaplesssptnoninvertibleduality}. 
\begin{eqnarray}
\label{eq:gappedSPTmap}
    \mathbb{Z}_2^\sigma \text{ SSB} + \mathbb{Z}_2^\tau \text{ SSB} &\stackrel{\mathcal{N}_{\text{KT}}}\Longleftrightarrow&  \mathbb{Z}^\sigma_2\times\mathbb{Z}^\tau_2 \text{ gapped SPT}
\\
\label{eq:SSBtrivialmap}
 \mathbb{Z}_2^\sigma \text{ SSB} + \mathbb{Z}_2^\tau \text{ trivial} &\stackrel{\mathcal{N}_{\text{KT}}}\Longleftrightarrow&   \mathbb{Z}_2^\sigma \text{ SSB} + \mathbb{Z}_2^\tau \text{ trivial}
\\
\label{eq:trivialmap}
  \mathbb{Z}_2^\sigma \text{ trivial} + \mathbb{Z}_2^\tau \text{ trivial} &\stackrel{\mathcal{N}_{\text{KT}}}\Longleftrightarrow&   \mathbb{Z}_2^\sigma \text{trivial} + \mathbb{Z}_2^\tau \text{trivial}
\\
\label{eq:gSPTmap}
    \mathbb{Z}_2^\sigma \text{ Ising CFT} + \mathbb{Z}_2^\tau \text{ SSB} &\stackrel{\mathcal{N}_{\text{KT}}}\Longleftrightarrow& \mathbb{Z}^\sigma_2\times\mathbb{Z}^\tau_2 \text{  gSPT}
\\
\label{eq:IsingCFT2map}
  \mathbb{Z}_2^\sigma \text{ Ising CFT} + \mathbb{Z}_2^\tau \text{ Ising CFT} & \stackrel{\mathcal{N}_{\text{KT}}}\Longleftrightarrow& \text{SPT-trivial critical point}
\\
\label{eq:igSPTmap}
     \mathbb{Z}_2^\sigma \text{ SSB} + \mathbb{Z}_4^\tau \text{ free boson CFT} & \stackrel{\mathcal{N}_{\text{KT}}}\Longleftrightarrow& \mathbb{Z}^\Gamma_4 \text{  intrinsically gSPT}
\\
\label{eq:pgSPTmap}
    \mathbb{Z}_2^\sigma \text{ free boson CFT} + \mathbb{Z}_2^\tau \text{ free boson CFT} & \stackrel{\mathcal{N}_{\text{KT}}}\Longleftrightarrow & \mathbb{Z}^\sigma_2\times\mathbb{Z}^\tau_2 \text{  purely gSPT}
\\
\label{eq:ipgSPTmap}
   \mathbb{Z}_2^\sigma \text{ free boson CFT} + \mathbb{Z}_4^\tau \text{ free boson CFT}  & \stackrel{\mathcal{N}_{\text{KT}}}\Longleftrightarrow& \mathbb{Z}^\Gamma_4 \text{  intrinsically purely gSPT}
\end{eqnarray}
Equations~\eqref{eq:gappedSPTmap}, \eqref{eq:SSBtrivialmap}, and~\eqref{eq:trivialmap} describe the construction of a gapped SPT phase with $\mathbb{Z}_2 \times \mathbb{Z}_2$ symmetry. In particular, the $\mathbb{Z}_2^\sigma \times \mathbb{Z}_2^\tau$ gapped SPT phase can be obtained by starting from decoupled $\mathbb{Z}_2^\sigma$ and $\mathbb{Z}_2^\tau$ SSB phases and applying the KT transformation. By replacing the $\mathbb{Z}_2^\sigma$ SSB phase in \eqref{eq:gappedSPTmap} with a $\mathbb{Z}_2^\sigma$ Ising CFT, we obtain the $\mathbb{Z}_2^\sigma \times \mathbb{Z}_2^\tau$ gSPT after applying the mapping in \eqref{eq:gSPTmap}. Furthermore, replacing the $\mathbb{Z}_2^\tau$ SSB phase in \eqref{eq:gappedSPTmap} with a $\mathbb{Z}_4^\tau$ symmetric free boson CFT (realized by the XY chain on the lattice) leads to the $\mathbb{Z}_4^\Gamma$ intrinsically gSPT, where $\mathbb{Z}_4^\Gamma$ is generated by the product of the $\mathbb{Z}_2^\sigma$ and $\mathbb{Z}_4^\tau$ generators, as mentioned earlier. If both $\mathbb{Z}_2^\sigma$ and $\mathbb{Z}_2^\tau$ SSB phases are replaced by $\mathbb{Z}_2^\sigma$ and $\mathbb{Z}_2^\tau$ free boson CFTs, respectively, we obtain the $\mathbb{Z}_2^\sigma \times \mathbb{Z}_2^\tau$ purely gSPT. Finally, replacing the $\mathbb{Z}_2^\sigma$ SSB phase in \eqref{eq:gappedSPTmap} with a $\mathbb{Z}_2^\sigma$ free boson CFT and the $\mathbb{Z}_2^\tau$ SSB phase with a $\mathbb{Z}_4^\tau$ free boson CFT leads to the $\mathbb{Z}_4^\Gamma$ intrinsically purely gSPT. A common feature of these constructions is that they start from a decoupled system, either gapped or gapless, which is then mapped by the KT transformation to a coupled system with nontrivial topological properties. This approach not only reproduces known models of gSPT and intrinsically gSPT phases but also identifies new instances of purely gSPT and intrinsically purely gSPT phases. Moreover, it provides a framework to study the stability of various gSPT phases, ranging from \eqref{eq:gSPTmap} to \eqref{eq:ipgSPTmap}, under symmetric perturbations. In particular, if a perturbation to a gSPT is such that, when the KT transformation is undone, the resulting theory remains decoupled, we can analytically investigate its topological properties on both a ring and an interval~\cite{li2023intrinsicallypurelygaplesssptnoninvertibleduality}. Additionally, gSPT phases are often characterized by edge states, which appear as low-energy states in the spectrum of an open chain and are distinguishable from bulk gapless excitations. While this distinction is more subtle than in gapped SPT phases, the KT transformation provides valuable insight by relating these edge states to the quasi-degenerate ground states associated with spontaneous symmetry breaking in finite-size systems. This helps clarify the identification and stability of edge states in gSPTs. Furthermore, since the theories on the left-hand sides of \eqref{eq:gSPTmap} to \eqref{eq:ipgSPTmap} admit field-theoretic descriptions, the KT transformation enables the derivation of corresponding field theories for the gSPT under consideration.

\paragraph{Pivot Hamiltonian construction}: 
Based on previous experience studying gSPT phases, particularly symmetry-enriched quantum criticality, these phases typically emerge at transition points between gapped SPT and SSB-ordered phases~\cite{Scaffidi2017prx,Verresen2021prx}. Therefore, to systematically obtain such novel critical phases, it is essential to construct these SPT transitions~\cite{Verresen2017prb,Lu2014prb,CHEN2013248,TSUI2015330,TSUI2017470}. Fortunately, Nathanan et al ~\cite{Nathanan2023scipost_a,Nathanan2023scipost_b} introduced a new approach, the Pivot Hamiltonian method,  which opens a new avenue for exploring SPT transitions. Below, we briefly outline the key idea of this method, with further details available in reference~\cite{Nathanan2023scipost_a,Nick2025SciPostCore}.

\begin{figure}
    \centering
   \includegraphics[width=0.55\linewidth]{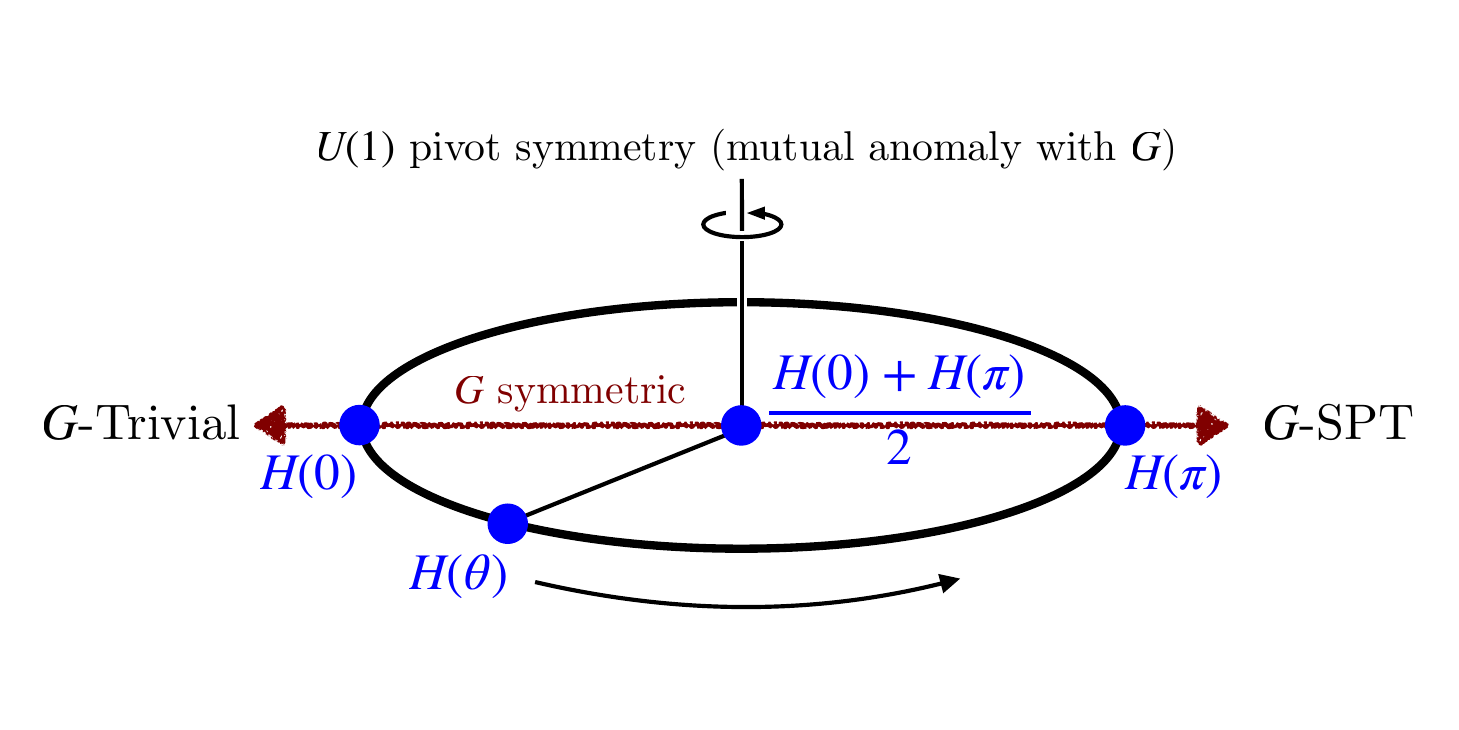}
    \caption{By evolving the system with a pivot Hamiltonian, we generate a one-parameter family of Hamiltonians, $ H(\theta) = e^{-i \theta H_{\text{piv}}} H_0 e^{i \theta H_{\text{piv}}} $, which is $ 2\pi $-periodic (see Eq.~\eqref{eq:pivoted}). Among these, only $ H(0) $ and $ H(\pi) $ remain symmetric under a \textcolor{red}{given symmetry group $ \Gamma = G $, allowing these special points to correspond to distinct SPT phases. In contrast, the other points along the circular trajectory—illustrated schematically by the compass—lie outside the space of $ \Gamma $-symmetric models. The $ U(1) $ symmetry generated by the pivot exhibits a mutual anomaly with $ \Gamma $}. The figures are adapted from Ref.~\cite{Nathanan2023scipost_a}.}
    \label{fig20}
\end{figure}

To begin with, we consider two Hamiltonians: $H_0$, which possesses a symmetry group $\Gamma$, and $H_{\text{piv}}$, which may have a lower symmetry. The latter is used to evolve $H_0$ into a new Hamiltonian, as illustrated in Fig.~\ref{fig20}:  
\begin{equation}
H(\theta) = e^{-i \theta H_{\text{piv}}} H_0 e^{i \theta H_{\text{piv}}}.  
\label{eq:pivoted}
\end{equation}  
What makes pivoting particularly interesting is that $H_{\text{piv}}$ is chosen to function as an SPT entangler. More precisely, we impose the following two conditions:  i) $H_{\text{spt}} := H(\pi)$ \textcolor{red}{realizes a nontrivial SPT phase protected by $\Gamma$.}  
ii) $H(2\pi) = H(0)$, meaning there exists a normalization of $H_{\text{piv}}$ such that conjugation by a $2\pi$-rotation leaves $H_0$ invariant.  

One might be tempted to strengthen the second condition by requiring that $e^{-2\pi i H_{\text{piv}}}$ be the identity operator, which would trivially imply $H(2\pi) = H(0)$. To systematically construct SPT transitions, we implement the pivoting process in two aspects:

$\bullet$ \emph{Constructing Gapped SPT Models:} Pivoting provides a new framework for generating gapped SPT models with interesting interrelations. Specifically, as outlined above, starting with two Hamiltonians, $H_0$ and $H_{\text{piv}}$, one can generate a new model, $H_{\text{spt}}$. This process can be iterated—using $H_{\text{spt}}$ as a new pivot to rotate either $H_0$ or $H_{\text{piv}}$, thereby producing another new model. As a result, pivoting generates an entire family of models residing in distinct SPT phases, interconnected through a web of dualities~\cite{Nathanan2023scipost_a}.  

$\bullet$ \emph{SPT Transitions and $U(1)$ Pivot Symmetry:} The pivoting process also provides insights into SPT transitions, particularly when the pivot acquires symmetry. To see this, consider the one-parameter family of Hamiltonians: $H(\alpha) = (1-\alpha) H_0 + \alpha H_{\text{spt}}$. By construction, the entangler $U = e^{-i\pi H_{\text{piv}}}$ induces a $\mathbb{Z}_2$ duality transformation $\alpha \to 1-\alpha$. Moreover, at the midpoint ($\alpha = 1/2$), this duality becomes an exact $\mathbb{Z}_2$ symmetry, satisfying $[H_0 + H_{\text{spt}}, U] = 0$. In our case, the $\mathbb{Z}_2$ unitary $U$ is generated by $H_{\text{piv}}$, raising the question of whether the stronger condition $[H_0 + H_{\text{spt}}, H_{\text{piv}}]=0$ also holds. Indeed, at the SPT transition, the pivot Hamiltonian generates a $U(1)$ symmetry, leading us to define a $U(1)$ pivot symmetry. Even if the above condition does not strictly hold, the renormalization group fixed point governing the critical behavior still exhibits an emergent $U(1)$ pivot symmetry. This insight sheds new light on the nature of SPT transitions and provides guidance for designing lattice models that support stable, direct SPT transitions. Furthermore, gauging a finite subgroup of a global symmetry can transform conventional phases and phase transitions into unconventional ones, including gSPT phases and DQCPs~\cite{Umberto2021scipost,Su2024prb}.

\subsubsection{General topological invariant at criticality}
\label{section3.5.2}
The most fundamental aspects of characterizing the topological phase of matter are the topological invariant and the bulk-boundary correspondence. However, at critical point, the topological invariants used for gapped topological phases (e.g., the winding number) are generally no longer applicable. Despite previous efforts~\cite{Verresen2018prl,Verresen2021prx,verresen2020topologyedgestatessurvive}, there remains ongoing debate and a lack of a general topological invariant at criticality. Fortunately, for gapless systems with conformal symmetry, such as quantum critical points described by CFT, recent progress~\cite{Yu2022prl,Parker2018prb,Chulliparambil2023PRB} suggests that the Affleck-Ludwig boundary $g$-function~\cite{Affleck1991PRL} serves as a general topological invariant at critical points, at least in (1+1) dimensions. Specifically, topologically distinct quantum critical points or gSPT states correspond uniquely to different values of the boundary $g$-function. Moreover, the topological edge modes in gSPT states manifest in the entanglement spectrum of the bulk wavefunction, establishing a novel bulk-boundary correspondence in critical systems~\cite{Yu2024prl,Zhang2024pra,zhong2025quantumentanglementfermionicgapless}. In the present and following section, we briefly review topological invariants and the novel bulk-boundary correspondence at quantum criticality, focusing on (1+1) dimensions. 

The first question we address is how to classify topologically distinct critical systems and assess the generality of such discrete invariant. Reference~\cite{Yu2022prl} proposed a "strange" discrete invariant related to surface criticality and boundary CFT, which enables the classification of nontrivial topology at criticality, at least in (1+1) dimensions. A key concept in boundary CFT is the conformal boundary condition (b.c.), which determines the operator content of the critical system~\cite{CARDY1986186,CARDY1989581}. Specifically, the Hamiltonian eigenstates, which are organized into conformal families consisting of a primary state and its conformal descendants. Additionally, the conformal b.c. dictates the universality class of surface criticality~\cite{Cardy1984,CARDY1986186,song2025boundaryphasetransitionstwodimensional,Zhang2017PRL,Ding2018PRL,Zhu2021PRB,Ding2023SciPost,Xu2022PRB,Zhang2022PRB,Zhang2023PRB,shen2024newboundarycriticalitytopological,sun2025boundaryoperatorexpansionextraordinary,jiang2025boundarycriticalitygrossneveuyukawamodels,ge2025boundarycriticalitytwodimensionalinteracting,Max2022SciPost,Parisen2022PRL,Abijith2023SciPost,Jaychandran2022SciPost,Weber2018PRB,Weber2019PRB,Weber2021PRB,Parisen2021PRL,Parisen2023PRB,przetakiewicz2025boundaryoperatorproductexpansion,Parisen2025PRR,zhu2022exoticsurfacebehaviorsinduced,Wang2022PRB,Wang2023PRB,Wang2024PRB,Hu2021PRL,Sun2022PRB,Sun2022PRB_b,Sun2023PRL}.  

To explore these aspects, we study two families of quantum spin chains that generalize the 1D Ising and three-state Potts models~\cite{Verresen2018prl,Jones2019JSP,Brien2020prb}. Specifically, the first family of models we examine consists of the transverse-field Ising (TFI) chain and the cluster Ising (CI) chain, as previously discussed in Sec.~\ref{section3.2}. Both models exhibit a $\mathbb{Z}_{2} \times \mathbb{Z}_{2}^{T}$ symmetry, generated by spin-flip symmetry $P=\prod_{l=1}^{L}\sigma_{l}^{x}$ and time-reversal symmetry $T=K$ (complex conjugation). A quantum critical point arises at $h_{c}=1$ in both models, separating the FM order phase at $|h|<h_{c}$ from the disordered phase at $h>h_{c}$, both of which are described by the 2D Ising CFT. 

\begin{figure}
    \centering
   \includegraphics[width=0.55\linewidth]{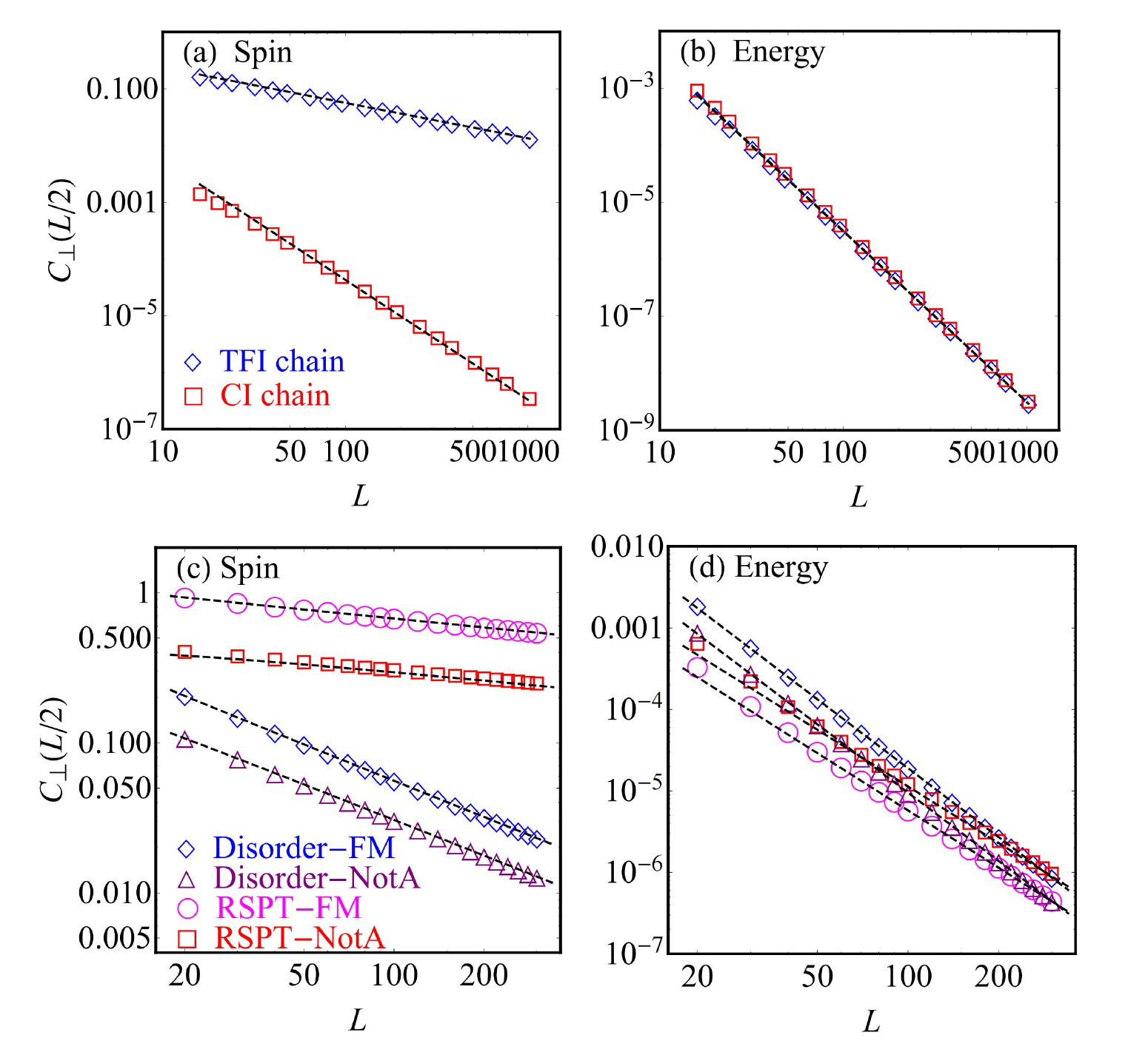}
    \caption{Connected correlation functions $ C_{\perp}(L/2) $ of (a) the spin operator $ \sigma_{l}^{z} $ and (b) the energy operator $ \epsilon_{l}=\sigma_{l}^{z}\sigma_{l+1}^{z} $ in critical Ising chains. (c, d) Connected correlation functions $ C_{\perp}(L/2) $ of the spin operator $ \sigma_{l} $ and the energy operator $ \epsilon_{l}=\tau_{l}+\tau_{l}^{\dag} $ in critical Potts chains. The dashed lines represent power-law fits based on Eq. (\ref{eq:Cperp}), including a correction-to-scaling term, $ bL^{-\Delta_{\phi}-\Delta_{\phi}^{b}-1} $. The figures are adapted from Ref.~\cite{Yu2022prl}.}
    \label{fig21}
\end{figure}

To analytically construct the topological invariant, we note that both models can be exactly solved using the Jordan-Wigner transformation (see Sec.~\ref{section2} for details), given by $\sigma_{l}^{z}=\prod_{k=1}^{l-1}(i\gamma_{k}\tilde{\gamma}_{k})\gamma_{l}, \quad \sigma_{l}^{x}=i\gamma_{l}\tilde{\gamma}_{l}$, where the Majorana fermion operators satisfy the anticommutation relations $\{\gamma_{k},\gamma_{l}\}=\{\tilde{\gamma}_{k},\tilde{\gamma}_{l}\}=2\delta_{kl}$ and $\{\gamma_{k},\tilde{\gamma}_{l}\}=0$. In the Majorana representation, the Hamiltonian takes the form  
\begin{equation}
H=-\sum_{l=1}^{L-1}i\tilde{\gamma}_{l}\gamma_{l+1}-h\sum_{l=1}^{L-\alpha}i\tilde{\gamma}_{l}\gamma_{l+\alpha},
\end{equation}
where $\alpha=0$ for the TFI chain and $\alpha=2$ for the CI chain. At the critical point, both models map to 1D massless Majorana fermions. However, in the CI chain, two decoupled Majorana modes, denoted by $\gamma_{1}$ and $\tilde{\gamma}_{L}$, give rise to a two-fold degeneracy in the energy spectrum~\cite{Verresen2018prl,Jones2019JSP,Yu2022prl}, which are protected by the \(\mathbb{Z}_{2}^{T}\) symmetry~\cite{Verresen2021prx}. In the spin representation, this degeneracy originates from the conservation of the edge spin operators, given by $\sigma_{1}^{z}=\gamma_{1}$ and $\sigma_{L}^{z}=-i\tilde{\gamma}_{L}P$, which characterize the spontaneous edge magnetization. As a result, the energy spectrum splits into four sectors labeled by $\sigma_{1}^{z}$ and $\sigma_{L}^{z}$, indicating that the $\mathbb{Z}_{2}$ symmetry is spontaneously broken at the edges. Therefore, the boundary properties of critical spin chains play a key role in determining the topological invariant at criticality.

To characterize the boundary critical behavior, we compute the connected correlation functions $C_{\perp}(L/2)$ for the spin operator $\sigma_{l}^{z}$ and the energy operator $\epsilon_{l}=\sigma_{l}^{z}\sigma_{l+1}^{z}$. These correlations are defined as follows: For a local operator at the boundary [denoted as $\phi(r)$] and in bulk [$\phi_{b}(R)$], the connected correlation functions obey the following scaling relations:  
\begin{align}
&C_{\perp}(r-R) = \langle\phi(r)\phi_{b}(R)\rangle_{c} \propto |r-R|^{-\Delta_{\phi}-\Delta_{\phi}^{b}}, \label{eq:Cperp}
\end{align}  
where $\langle AB\rangle_{c}=\langle AB\rangle-\langle A\rangle\langle B\rangle$. Here, $r_{1}-r_{2}$ is parallel to the surface, while $r-R$ is perpendicular to it. The exponents $\Delta_{\phi}$ and $\Delta_{\phi}^{b}$ represent the scaling dimensions of the boundary and bulk operators, respectively. In practice, the boundary correlations are evaluated in the Majorana representation and fitted to Eq.~(\ref{eq:Cperp}) using the bulk scaling dimensions $\Delta_{\sigma}^{b}=1/8$ and $\Delta_{\epsilon}^{b}=1$ (see Fig. \ref{fig21} (a) and (b)). The extracted scaling dimensions of the boundary operators are summarized in Table \ref{tab:exponents}. While the TFI chain corresponds to the Ising CFT with free boundary conditions, the exponents of the CI chain align with those of a fixed boundary condition. This behavior can be attributed to spontaneous edge magnetization, which is expected to be a general feature of other 1D quantum critical states exhibiting edge degeneracy. Thus, it appears that the conformal b.c. can be regarded as a topological invariant, allowing for the classification of topologically distinct quantum critical points (gSPT states) that feature degenerate edge modes.

\begin{table}[tb]
\caption{Scaling dimensions of the boundary spin and energy operators. Scaling dimensions in the Ising and Potts boundary CFTs are also listed for comparison~\cite{francesco2012conformal,Yu2022prl}.}
\label{tab:exponents}
\begin{tabular}{cc|cc}
Class	& Model/b.c.			& $\Delta_{\sigma}$	& $\Delta_{\epsilon}$	\\
\hline
Ising	& TFI					& 0.4992(3)			& 1.99957(8)			\\
		& CI					& 1.9984(2)			& 2.00008(2)			\\
CFT		& Free					& 1/2				& 2						\\
		& Fixed					& 2					& 2						\\
\hline
Potts	& Disorder-FM			& 0.66598(3)		& 0.7993(2)				\\
		& Disorder-NotA	    	& 0.6629(1)			& 0.7915(5)				\\
		& RSPT-FM				& 0.0661(3)			& 0.204(9)				\\
		& RSPT-NotA				& 0.0601(1)			& 0.21(1)				\\
CFT		& Free					& 2/3				& 4/5					\\
		& Dual-mixed			& 1/15				& 1/5
\end{tabular}
\end{table}

To verify the generality of conformal boundary conditions as a topological invariant at criticality, we further consider the one-dimensional generalized three-state Potts model, which exhibits a topologically nontrivial criticality but lacks degenerate edge modes. The Hamiltonian is given by~\cite{Brien2020prb}
\begin{equation}
H=H_{\text{P}}+\lambda H_{0}, \label{eq:Potts}
\end{equation}  
where \( H_{\text{P}} \) represents the standard quantum Potts model:  
\begin{equation}
H_{\text{P}}=-J\sum_{l=1}^{L-1}(\sigma_{l}^{\dagger}\sigma_{l+1}+\sigma_{l}\sigma_{l+1}^{\dagger})-f\sum_{l=1}^{L}(\tau_{l}+\tau_{l}^{\dagger}),
\end{equation}  
and \( H_{0} \) is given by  
\begin{equation}
H_{0}=\sum_{l=1}^{L-1}3\big((S_{l}^{+}S_{l+1}^{-})^{2}-S_{l}^{+}S_{l+1}^{-}+\mathrm{H.c.}\big)-\sum_{l=1}^{L}(\tau_{l}+\tau_{l}^{\dagger}).
\end{equation}  
The operators are defined as $ \tau = \text{diag}(1,\omega,\omega^{2}) $ with $ \omega=e^{2\pi i/3} $, and  
\begin{equation}
\sigma=
\begin{pmatrix}
0 & 1 & 0 \\
0 & 0 & 1 \\
1 & 0 & 0
\end{pmatrix},\quad
S^{+}=
\begin{pmatrix}
0 & 0 & 1 \\
1 & 0 & 0 \\
0 & 0 & 0
\end{pmatrix}=(S^{-})^{\dagger}.
\end{equation}  
The operators $ S^{\pm} $ act as ladder operators for $ S^{z} = \text{diag}(0,1,-1) $. The model in Eq. (\ref{eq:Potts}) is invariant under the $ S_{3} $ symmetry, generated by the $ \mathbb{Z}_{3} $ rotation $ R=\prod_{l=1}^{L}\tau_{l} $ and the charge conjugation operator $ C=\prod_{l=1}^{L}c_{l} $, where  
\begin{equation}
c=
\begin{pmatrix}
1 & 0 & 0 \\
0 & 0 & 1 \\
0 & 1 & 0
\end{pmatrix}.
\end{equation}

To begin, we briefly summarize the phase diagram of the model (\ref{eq:Potts}), referring to Ref. \cite{Brien2020prb} for a detailed derivation. The quantum phase diagram is parameterized by $ \lambda=1-\alpha $, $ J=\alpha+\beta $, and $ f=\alpha-\beta $, revealing four distinct gapped phases. Each phase is adiabatically connected to a special point within the phase, namely $ (\alpha,\beta)=(\pm 1,\pm 1) $, where the exact ground states can be constructed.  The Potts FM ordered phase exhibits threefold degenerate ground states, which continuously connect to the fully polarized FM states $ \otimes_{l}\ket{A}_{l} $, $ \otimes_{l}\ket{B}_{l} $, and $ \otimes_{l}\ket{C}_{l} $, where $\ket{A} $, $ \ket{B} $, and $ \ket{C} $ are eigenstates of $ \sigma $ with eigenvalues $ 1 $, $ \omega $, and $ \omega^{2} $, respectively. The FM order is characterized by the order parameter $ \langle{\sigma_{l}}^{3}\rangle > 0 $. The disordered phase, in contrast, is adiabatically connected to the $ S_{3} $-symmetric state $ \otimes_{l}\ket{0}_{l} $, where $\ket{0}=\frac{1}{\sqrt{3}}(\ket{A}+\ket{B}+\ket{C}) $ is an eigenstate of $ S^{z} $. The other two phases exhibit unconventional properties. The “not-$ A $” ordered phase~\cite{Brien2020prb} includes the point $ (\alpha,\beta)=(-1,-1) $, where the threefold degenerate ground states are given by $ \otimes_{l}\ket{\bar{A}}_{l} $, $ \otimes_{l}\ket{\bar{B}}_{l} $, and $ \otimes_{l}\ket{\bar{C}}_{l} $, with  
\begin{equation}
\ket{\bar{A}}=\frac{1}{\sqrt{2}}(\ket{B}+\ket{C}),\quad
\ket{\bar{B}}=\frac{1}{\sqrt{2}}(\ket{C}+\ket{A}),\quad
\ket{\bar{C}}=\frac{1}{\sqrt{2}}(\ket{A}+\ket{B}).
\end{equation}  
Here, the $ S_{3} $ symmetry is spontaneously broken, with the order parameter satisfying $ \langle{\sigma_{l}}^{3}\rangle < 0 $. The final phase is known as the representation symmetry-protected topological (RSPT) state. At $ (\alpha,\beta)=(-1,1) $, its ground state can be represented as a matrix-product state $ R=\otimes_{l}R_{l} $, where  
\begin{equation}
R_{l}=
\begin{pmatrix}
\ket{0}_{l} & \ket{+}_{l}	\\
\ket{-}_{l}	& \ket{0}_{l}
\end{pmatrix}.
\end{equation}  
Here, $ \ket{0} $, $ \ket{+} $, and $ \ket{-} $ are the three eigenstates of $ S^{z} $. The ground state is given by $ \operatorname{tr}(R) $ for periodic boundary conditions, forming a unique, $ S_{3} $-symmetric state. However, under open boundary conditions, the matrix elements of $ R $ yield fourfold degenerate ground states, which together form a linear representation of the $ S_{3} $ symmetry. The duality transformation $\tau_{l} \mapsto \sigma_{l}^{\dagger} \sigma_{l+1}$, $\sigma_{l} \mapsto \prod_{j=1}^{l} \tau_{j}$ exchanges the two terms in $H_{\text{P}}$ while leaving $H_{0}$ invariant. Consequently, this transformation maps the FM phase to the disordered phase, the not-$A$ ordered phase to the RSPT phase, and vice versa. Each ordered phase undergoes a continuous quantum phase transition to its corresponding disordered phase, all of which involve spontaneous $S_{3}$ symmetry breaking and belong to the 2D three-state Potts universality class. These transition lines intersect at the multicritical point $(\alpha, \beta) = (0,0)$, which is self-dual and possesses a U(1) symmetry generated by $Q = \sum_{l} S^{z}_{l}$.


Similarly, we first examine the surface critical behavior. The connected correlation functions of the spin operator $\sigma_{l}$ and the energy operator $\epsilon_{l} = \tau_{l} + \tau_{l}^{\dag}$ are computed using the DMRG algorithm~\cite{White1992prl,schollwock2011density} and displayed in Fig. \ref{fig21} (c) and (d). The scaling dimensions of the boundary operators, extracted by fitting Eq. (\ref{eq:Cperp}) with $\Delta_{\sigma}^{b} = 2/15$ and $\Delta_{\epsilon}^{b} = 2$, are summarized in Table \ref{tab:exponents}. The four quantum critical points fall into two distinct classes: while the disorder-FM and disorder-not-$A$ transitions conform to the Potts CFT with free boundary conditions (b.c.)~\cite{CARDY1986186}, the RSPT-FM and RSPT-not-$A$ transitions exhibit different critical exponents. Notably, these exponents can be derived from the Potts CFT with the ``new'' conformal boundary conditions, which is $S_{3}$ symmetric and dual to the mixed b.c. and referred to as the dual-mixed b.c. These distinct conformal b.c. can be derived from the bulk effective field theory~\cite{Yu2022prl}. Consequently, although the topologically distinct quantum critical points in the Potts model do not exhibit degenerate edge modes, the numerical study unambiguously demonstrates that conformal b.c. can also classify these types of critical points. This further confirms that conformal b.c. can be regarded as a general topological invariant in quantum critical systems, at least in (1+1) dimensions.


Essentially, the conformal b.c. is fully determined by the Affleck-Ludwig boundary $g$-function~\cite{Affleck1991PRL}, which can be experimentally measured as follows: The thermal entropy of a critical chain of length $L$ at temperature $T$ scales as : $
S(L, T)=\frac{\pi c}{3v}LT+\ln g$, where $c$ is the central charge, $v$ is the velocity in the low-energy limit, and $\ln g$ is the Affleck-Ludwig boundary entropy, a universal constant that determine the conformal b.c. The excess boundary entropy associated with a given conformal b.c., relative to the fixed b.c., can be extracted from the entropy change induced by applying a magnetic field at the boundary. If degenerate edge states exist, they contribute an integer factor of degeneracy to $g$. However, for a generic conformal b.c., $g$ is not necessarily an integer but can be interpreted as an ``effective edge degeneracy.'' Consequently, the conformal b.c. generalizes the concept of edge degeneracy in characterizing quantum critical states and can serve as a topological invariant at criticality, at least in (1+1) dimensions~\cite{Yu2022prl}.

\subsubsection{Universal topological bulk-boundary correspondence at criticality}
\label{section3.5.3}

The second key aspect in characterizing topological phases of matter is the universal bulk-boundary correspondence. The most commonly used physical quantity for describing this correspondence is the topological invariant, which states that a bulk topological invariant—such as the $\mathbb{Z}_{2}$ topological index in a topological insulator under periodic boundary conditions—implies that the boundary of a topological phase hosts symmetry-protected gapless edge modes~\cite{Bernevig2006prl,Qi2011rmp,Hasan2010rmp}. These topological invariants are typically defined in parameter space, such as the winding number in the AIII or BDI symmetry class for topology band theory~\cite{Altland1997prb}. Another important quantity that reflects this correspondence is the entanglement spectrum. According to the well-known Li-Haldane conjecture~\cite{Li2008PRL}, in the gapped topological phases, the bulk entanglement spectrum encodes information about the boundary Hamiltonian. Specifically, this conjecture asserts that the low-lying entanglement spectrum in the bulk is in one-to-one correspondence with the universal part of the many-body energy spectrum under an open boundary. This implies that the bulk ground state wavefunction contains universal boundary information, such as the edge mode degeneracy in a gapped topological phase.  However, in the case of gSPT, the conventional topological invariants mentioned above no longer apply, as the presence of singularities in parameter space renders them ill-defined~\cite{verresen2020topologyedgestatessurvive}. This raises a fundamental question: Can bulk-boundary correspondence be established through the entanglement spectrum in gSPT states? In other words, is it possible to extract topological and universal boundary CFT information solely from the bulk wavefunction, without requiring open boundary conditions?

To address these questions, reference~\cite{Yu2022prl} first established a universal bulk-boundary correspondence at criticality through the bulk entanglement spectrum, thereby generalizing the Li-Haldane bulk-boundary correspondence to critical systems~\cite{Zhang2024pra,zhong2025quantumentanglementfermionicgapless}.  Specifically, we consider different families of one-dimensional quantum spin chains that host various gSPT phases. Each family features symmetry-protected topological edge modes, which are described by the corresponding boundary CFT. By analyzing the entanglement spectrum and energy spectrum of these systems, we demonstrate a one-to-one correspondence between the bulk entanglement spectrum and the energy spectrum under an open boundary. This finding implies that the entanglement spectrum encodes information not only about the topological edge states but also about the operator content of the boundary CFT. Furthermore, due to the conformal symmetry inherent in gSPTs, this universal spectral correspondence can be understood theoretically, establishing a robust bulk-boundary correspondence in 1+1D gSPT phases, as reviewed in the following.

\begin{figure}[tb]
    \includegraphics[width=1\textwidth]{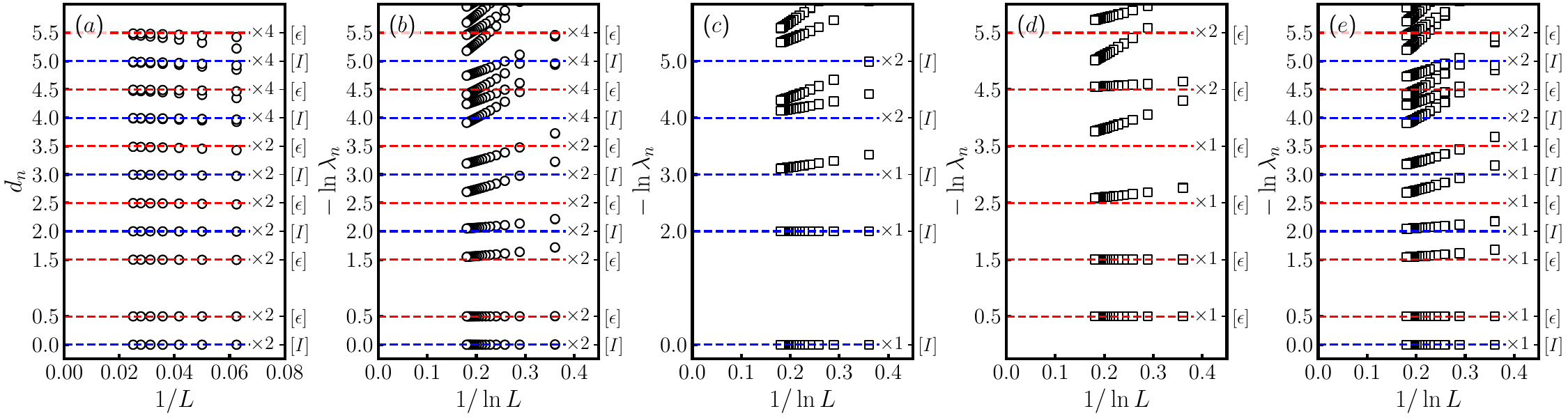}
    \caption{(a) Open boundary energy spectrum and (b) The bulk entanglement spectrum of the generalized cluster Ising chain $H''$ at the quantum critical point $h=1.0$ for several $L$. 
    The results of the bulk entanglement spectrum with additional projections on the boundary are shown in (c) for $(1+\sigma_{1}^{z})(1+\sigma_{L/2}^{z})$, in (d) for $(1+\sigma_{1}^{z})(1-\sigma_{L/2}^{z})$, and in (e) for $(1+\sigma_{1}^{z})$. All the spectra have been rescaled separately such that the first two levels are fixed to the corresponding values. For example, $d_{n} = 0.5\times(E_{n}-E_{1})/(E_{2}-E_{1})$ in (a). Open circles represent a two-fold degeneracy while open squares indicate a single degeneracy. The figures are adapted from Ref.~\cite{Yu2024prl}.}
    \label{fig22}
\end{figure}

As a first example of gSPT states corresponding to symmetry-enriched quantum critical points, we consider the cluster Ising chain, as discussed previously in Sec.~\ref{section3.2} and references~\cite{Verresen2021prx,Yu2022prl}. This model exhibits distinct topological behavior compared to a standard (1+1)D Ising CFT. The bulk entanglement spectrum and the many-body energy spectrum under open boundary conditions are shown in Fig.~\ref{fig22} (a) and (b), respectively.  After proper rescaling, we observe the following key features~\cite{Yu2024prl}:  
i) The entanglement spectrum exhibits the same doubly degenerate structure as the open-boundary energy spectrum, indicating the presence of nontrivial edge states.  ii) The bulk entanglement spectrum contains the same operator content as the corresponding boundary CFT. iii) The algebraic splitting of edge modes in this example can be identified through finite-size scaling of the bulk entanglement spectrum.  These results demonstrate that the bulk wave function effectively encodes both the topological features and the operator content information associated with open boundary conditions.

\begin{figure}[tb]
    \centering
    \includegraphics[width=0.45\textwidth]{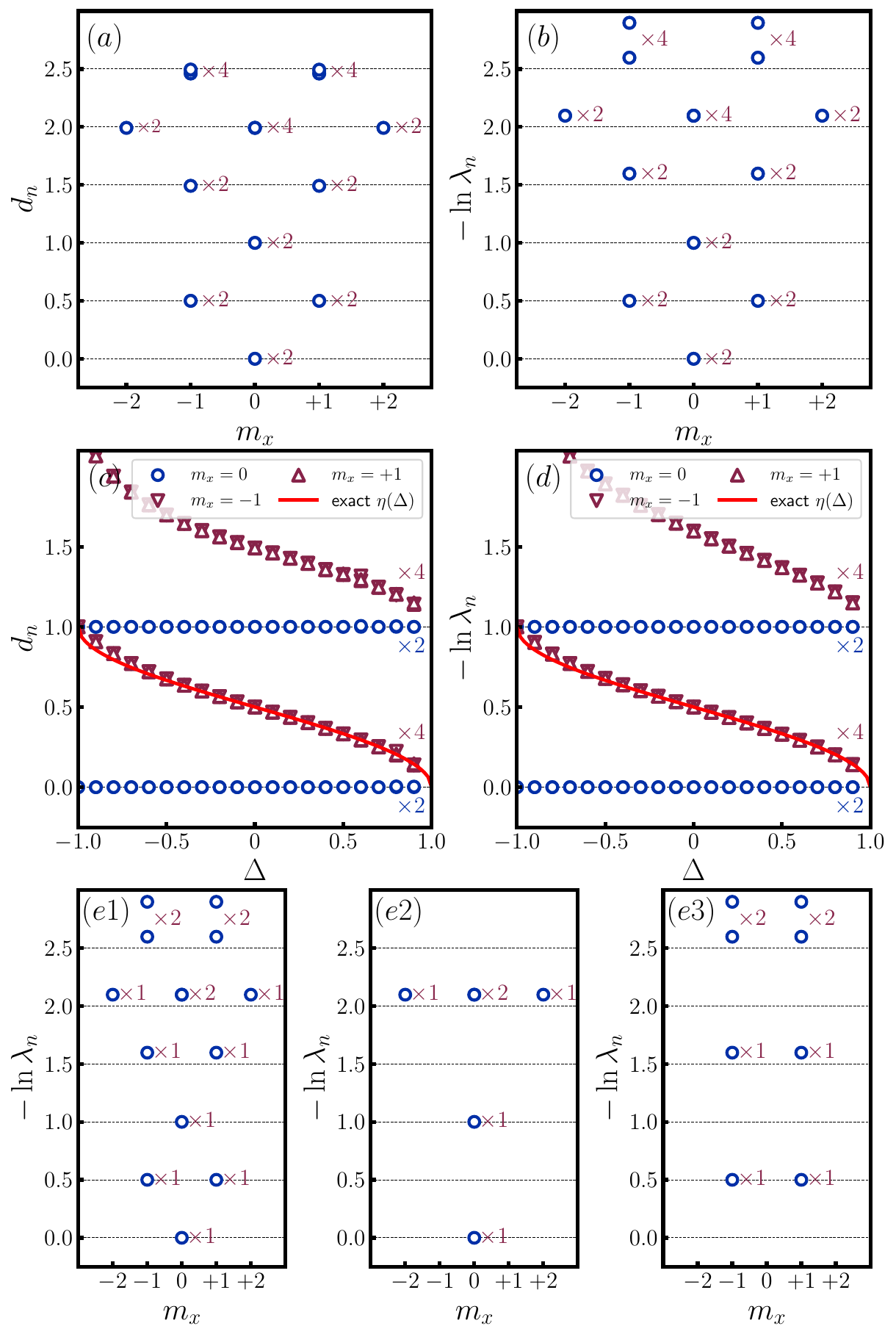}
    \caption{(a) Open boundary energy spectrum and (b) The bulk entanglement spectrum labeled by the quantum number $m_{x}$ for the intrinsically gSPT at $\Delta=0$. 
    The spectra are rescaled separately such that the first two levels within the $m_{x}=0$ sector are fixed to $0$ and $1$, respectively. 
    (c-d) The rescaled spectrum within the $m_{x}=0$ and $\pm1$ sectors as a function of $\Delta$. 
    The rescaled value of the first level in the $m_{x}=\pm1$ sector is related to the Luttinger parameter and is compared with the exact solution, $\eta(\Delta)=1-\arccos{(-\Delta)}/\pi$ (red solid line). 
    (e1)-(e3) display the resulting entanglement spectrum for $\Delta=0$ after the projection $(1+\sigma_{2L}^{z})$, $(1+\sigma_{L}^{z})(1+\sigma_{2L}^{z})$, and $(1-\sigma_{L}^{z})(1+\sigma_{2L}^{z})$ from left to right. (e1)-(e3) are separately rescaled to be directly compared with (b). The colored numbers indicate the degeneracy of each level. The figures are adapted from Ref.~\cite{Yu2024prl}}
    \label{fig23}
\end{figure}

Next, we examine bulk-boundary correspondence in stable topological critical phases, focusing on an intrinsically gSPT phase described by Eq.~(\ref{eq:igSPTHam}) with an XXZ-type perturbation in the $\tau$ degree of freedom, $\Delta \tau^{x}_{2i-1}\tau^{x}_{2i+1}$. We consider the regime where $|\Delta|<1$, in which the ground state remains an intrinsically gSPT phase, and the perturbation is exactly marginal and symmetric. This Hamiltonian can be derived by stacking an Ising SSB Hamiltonian with an XXZ chain through the KT transformation~\cite{Li2024scipost,li2023intrinsicallypurelygaplesssptnoninvertibleduality}. The low-energy effective theory is described by a $c=1$ free boson CFT, which shares the same quantum anomaly as the 2+1D Levin-Gu SPT phase~\cite{Levin2012prb}, ensuring a two-fold ground-state degeneracy. Notably, the sublattice magnetization, defined as $ m_{x} = \frac{1}{2} \sum_{i} \langle\tau_{2i-1}^{x}\rangle$, remains a good quantum number for any value of $\Delta$. Consequently, the full spectrum can be classified into different sectors labeled by $ m_{x}$. The energy and entanglement spectrum for $\Delta = 0$ is shown in Fig.~\ref{fig23} (a) and (b), respectively. We observe that the bulk entanglement spectrum not only exhibits the same degeneracy structure as the open-boundary energy spectrum but also shares the same operator content of underlying boundary CFT. Specifically, both spectra correspond to the operator content of the free boson boundary CFT~\cite{lauchli2013operatorcontentrealspaceentanglement}, indicating that topological and boundary CFT information can be extracted directly from the entanglement spectrum of the bulk wavefunction in this stable critical phase.

\begin{figure}[tb]
    \includegraphics[width=1\textwidth]{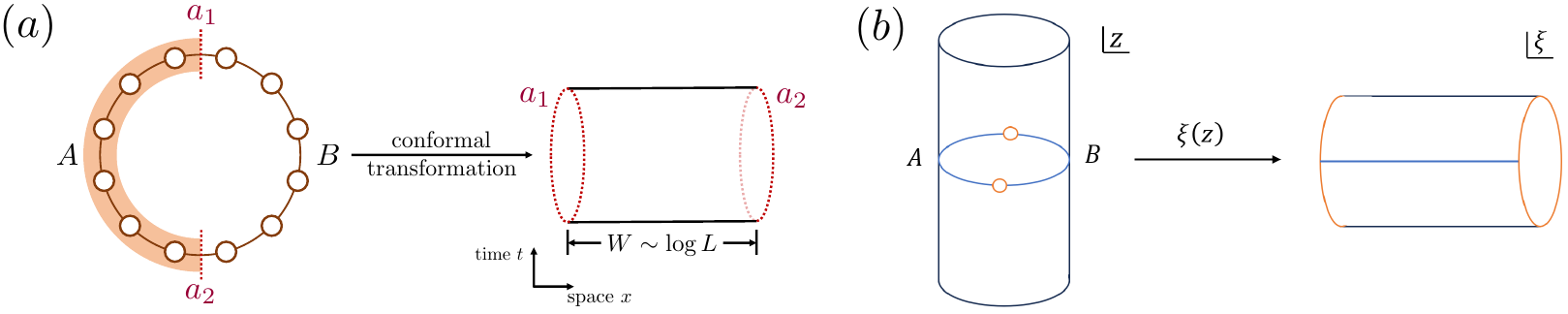}
    \caption{ (a) The setup involves a bipartition of one-dimensional periodic spin models. 
    The orange-shaded region denotes the subsystem $A$, and $B$ represents its complement. 
    The red dotted line represents the entanglement cut, and $a_{1,2}$ labels the boundary condition. 
    After conformal transformation, the reduced density matrix maps to a cylinder (annulus) with width $W \sim \text{log} L$.
    (b) The Euclidean theory is defined in an infinite cylinder with two entanglement cuts with separating subsystems $A$ and $B$.
    After a conformal transformation $\xi(z)$, the infinite cylinder with two cuts is mapped to an annulus. The figures are adapted from Ref.~\cite{Yu2024prl}.}
    \label{fig24}
\end{figure}

From the perspective of boundary CFT, there is a direct connection between the entanglement Hamiltonian and the Hamiltonian of an open-boundary chain~\cite{Cardy_2016,Ohmori_2015}. In the continuum limit, the entanglement cut is modeled as a small spatial region of thickness $\epsilon$ at the boundary between subsystems $A$ and $B$. In our case, we consider the ground state of a one-dimensional periodic chain of length $L$ with a bipartition $A: (-L/4, L/4)$ and its complement $B$, as illustrated in Fig.~\ref{fig24} (a). For simplicity and without loss of generality, we shift the entanglement cut from $(0, L/2)$ to $(-L/4, L/4)$. The Euclidean path integral corresponding to this setup is defined on an infinite cylinder with two entanglement cuts of radius $\epsilon$, as shown in Fig.~\ref{fig24} (b). The infinite length of the cylinder ensures that the Euclidean path integral projects onto the ground state. A complex coordinate $z$ is used to parametrize the cylinder, with imaginary time running along the $\text{Im}(z)$ direction, and the entanglement cuts located at $z = \pm L/4$. By applying the conformal transformation $\xi(z) = \log \left( \frac{e^{i2\pi z / L} - e^{- i\pi/2}}{e^{i\pi/2 }- e^{i 2\pi z/L}} \right)$, the cylinder is mapped onto an annulus, where $\xi = \xi + 2\pi i$ represents the annulus coordinate, with $\xi = x + i t$, as depicted in Fig.~\ref{fig24}(b). The conformal images of the entanglement cuts correspond to the annulus boundaries at $\xi \approx \pm \log \frac{2L}{\pi \epsilon}$, leading to an annulus width of $W = 2\log \frac{2L}{\pi \epsilon}$. After this transformation, the entanglement Hamiltonian is mapped to the conformal generator of translations in the $\text{Im}(\xi)$ direction. Consequently, the entanglement spectrum corresponds to the energy spectrum on the annulus with boundary conditions imposed at the entanglement cuts. In the low-energy regime, these boundary conditions flow to conformal boundary conditions, which we denote as $a_1$ and $a_2$. Given these boundary conditions and the annulus width $W$, the entanglement spectrum takes the form  $E^{(a_1, a_2)}_j = \frac{\pi}{W} \left(- \frac{c}{24} + \Delta_j^{(a_1, a_2)} \right)$, where $\Delta_j^{(a_1, a_2)}$ represents the scaling dimension of allowed operators consistent with the boundary conditions $a_1$ and $a_2$, and $c$ is the central charge of the underlying CFT. Notably, the energy levels are inversely proportional to the annulus width $W$, and through the conformal transformation, the entanglement spectrum scales as $\sim 1 / \log L$~\cite{lauchli2013operatorcontentrealspaceentanglement,Yu2022prl}.

Now, we apply these boundary CFT techniques to illustrate the topological degeneracy in the bulk entanglement spectrum for various types of gSPT states. We first consider symmetry-enriched Ising criticality, whose boundary CFT is characterized by a "superposition" state, $\tilde {\mathbb I} \oplus \tilde \epsilon$, where $\tilde {\mathbb I}$ and $\tilde \epsilon$ correspond to two fixed boundary conditions in the framework of boundary CFT~\cite{francesco2012conformal}. Physically, these two states represent boundary spins pointing in opposite directions, and the "superposition" indicates that the boundary exhibits spontaneous magnetization, with the two opposing magnetizations being equivalent. This results in the operator content $(\tilde {\mathbb I} \oplus \tilde \epsilon) \times (\tilde {\mathbb I} \oplus \tilde \epsilon) = 2 \times ( [\mathbb I] \oplus [\epsilon] )$, as illustrated in Fig.~\ref{fig22}. Here, $[\mathbb I], [\epsilon], [\sigma]$ denote the operator content of the three primary fields in the Ising CFT~\cite{ginsparg1988appliedconformalfieldtheory}. This behavior stands in stark contrast to the standard Ising CFT, where the boundary state is typically $\tilde \sigma$, corresponding to a free boundary condition without double degeneracy. The boundary condition beyond the "clear-cut" at the entanglement cut provides an additional means of manipulating the entanglement spectrum~\cite{Yu2024prl}. To explore this effect, we introduce projection operators at the entanglement cut, defined as $ P_{L/R} \propto (1 \pm \sigma^{z}_{1,L/2})$, and analyze the entanglement spectrum of the renormalized state $P_{L/R} \ket{\psi}$. The projection effectively fixes the boundary condition to either $\tilde{\mathbb{I}}$ or $\tilde{\epsilon}$. Consequently, the entanglement spectrum can be modified according to the following fusion rules:  
$\tilde{\mathbb{I}} \times \tilde{\mathbb{I}} = [{\mathbb{I}}],
\quad \tilde{\mathbb{I}} \times \tilde{\epsilon} = [\epsilon],
\quad \tilde{\mathbb{I}} \times (\tilde {\mathbb I} \oplus \tilde \epsilon) = [\mathbb I] \oplus [\epsilon],$  
as shown in Fig.~\ref{fig22} (c)-(f).

However, in the context of intrinsically gSPT phases described by a free boson $c=1$ CFT, the boundary condition in an open chain extends beyond the conventional Dirichlet boundary condition~\cite{Li2024scipost}. Recall that in a free boson CFT, the Dirichlet boundary condition supports states with energy $E_{m_x,n} \sim \frac{1}{W} (\eta(\Delta) m_x^2 + n)$, where $ m_x $ labels the topological sector, $ n $ denotes the descendant state, and the parameter $\eta(\Delta) = 1 - \frac{\arccos(-\Delta)}{\pi}$. However, in this case, the boundary state is enriched by symmetry fractionalization, characterized by the relation $ U^2 = \tau^x_{1/2} \sigma_1^z \sigma^z_{L} $ at both edges. In other words, in addition to the Dirichlet boundary condition, the boundary state acquires an extra label corresponding to spontaneous magnetization, $ \sigma_{1,2} $, at each edge. Consequently, the boundary state is enriched as $ |m_x , n, \sigma_1, \sigma_2 \rangle $, and it cannot be obtained simply as a superposition of conventional boundary conditions in the free boson boundary CFT. With parity symmetry, these states are classified into distinct parity sectors: $ |2k, n, \sigma, \sigma \rangle $ and $ |2k+1, n, \sigma, -\sigma \rangle $, where each state exhibits a double degeneracy associated with $ \sigma = \pm 1 $. This accounts for the double degeneracy observed in the entanglement spectrum, as shown in Fig.~\ref{fig23}, which is referred to in the literature as a \emph{symmetry-enriched boundary condition}~\cite{Yu2024prl}.  We can further manipulate the entanglement spectrum by modifying the entanglement cut through the introduction of projection operators, such as $(1 \pm \sigma_{L}^{z}) $ or $ (1 \pm \sigma_{2L}^{z})$. A single projection, e.g., $ (1 + \sigma_{2L}^{z}) $, lifts the double degeneracy and results in the selection of states $ | 2k, n, 1, 1 \rangle $ and $ |2k-1, n, 1, -1 \rangle $. Meanwhile, a joint projection, $ (1+\sigma_{L}^{z})(1+\sigma_{2L}^{z}) $ [or $ (1-\sigma_{L}^{z})(1+\sigma_{2L}^{z}) $], restricts the system to states with $ m_x \in 2 \mathbb Z $ (even sector) or $ m_x \in 2 \mathbb Z + 1 $ (odd sector), respectively, as illustrated in Fig.~\ref{fig23} (e1,e2,e3).

\subsubsection{Deconfined quantum criticality as an intrinsically gSPT state in one dimension}
\label{section3.5.4}
Unifying different types of quantum criticality beyond the traditional Landau-Ginzburg-Wilson symmetry-breaking paradigm plays a crucial role in exploring fundamental theories and intriguing phenomena in quantum many-body and statistical physics. One of the most prominent examples of a Landau-forbidden transition is deconfined criticality, which has been briefly reviewed in Sec.~\ref{section1}.  On a different front, another class of quantum critical states beyond the Landau paradigm is the gSPT state, which is the primary focus of this review. In particular, intrinsically gSPT phases represent an exotic class of gapless systems that exhibit emergent anomalies at low energy. This phenomenon is reminiscent of the mixed anomalies observed in DQCPs~\cite{Wang2017prx,Metlitski2018PRB}. Both critical systems exhibit quantum anomalies, raising several fundamental and intriguing questions: What is the deep connection between intrinsically gSPT phases and deconfined criticality? Moreover, can developing a theoretical framework for gSPT phases deepen our understanding of unconventional phase transitions?  Despite intensive research efforts~\cite{Ma2022scipost,myersonjain2024pristinepseudogappedboundariesdeconfined}, the deep connection between these two types of critical states, particularly in lattice realizations, remains elusive.

\begin{figure}[!h]
    \centering
    \includegraphics[width=0.45\linewidth]{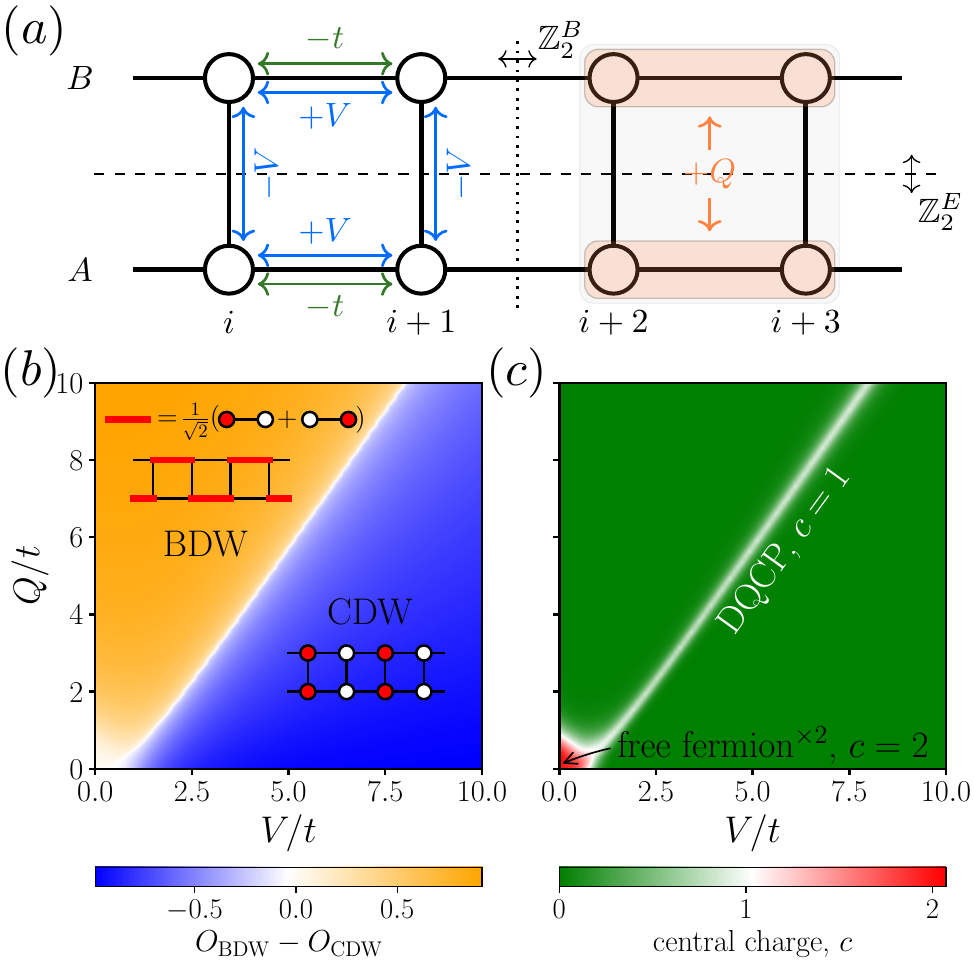}
    \caption{(a) Schematic plot of the two-leg fermion ladder, where the green, blue, and orange lines represent the hopping ($t$), density-density ($V$), and bond-bond ($Q$) interaction terms involving in the Hamiltonian Eq.~\eqref{eq:hamiltonian}. $\mathbb{Z}_{2}^{E(B)}$ denotes the leg-exchange (bond centered reflection) symmetry. The ground-state phase diagram is mapped out, respectively, by (b) the order parameters and (c) the estimated central charge, exhibiting four-fold degenerate BDW and two-fold degenerate CDW phases, along with a DQCP line ($c=1$). (b) is obtained from iDMRG calculations with MPS bond dimension $\chi=200$, while the central charge shown in (c) is estimated by $6\times(S_{2}-S_{1})/(\ln\xi_{2}-\ln\xi_{1})$, where labels ``$1$'' and ``$2$'' indicate the entanglement entropy ($S$) and MPS correlation length ($\xi$) from $\chi=100$ and $200$ iDMRG simulations, respectively. The figures are adapted from Ref.~\cite{yang2025deconfinedcriticalityintrinsicallygapless}.}
    \label{fig25}
\end{figure}

To address the above issues, Ref.~\cite{yang2025deconfinedcriticalityintrinsicallygapless} constructs a one-dimensional lattice model in which a DQCP coincides with an intrinsically gSPT state. Specifically, the authors consider an interacting spinless fermion model on a two-leg ladder of length $L$ at half-filling [see Fig.~\ref{fig25}(a)]:
\begin{eqnarray}
\label{eq:hamiltonian}
    H &=& -\, t \sum_{i=1}^{L} \sum_{\alpha=A,B} D_{i,\alpha} + Q \sum_{i=1}^{L}  (D_{i,A}-1) (D_{i,B}-1) \nonumber  \\
    && +\, V \sum_{i=1}^{L} \sum_{\alpha=A,B} Z_{i,\alpha}Z_{i+1,\alpha} - V \sum_{i=1}^{L} Z_{i,A}Z_{i,B} \,,
\end{eqnarray}
where $ Z_{i,\alpha} \equiv c^{\dagger}_{i,\alpha}c_{i,\alpha}-1/2 $ and $ D_{i,\alpha} \equiv c^{\dagger}_{i,\alpha}c_{i+1,\alpha} + h.c. $. The operator $ c^{\dagger}_{i,\alpha} $ ($ c_{i,\alpha} $) creates (annihilates) a spinless fermion at rung $ i $ on leg $ \alpha $. These fermions are \emph{not} physical degrees of freedom but are instead used as partons to construct a model where a DQCP emerges~\cite{chen2024emergentconformalsymmetrymulticritical,Zhou2024prx,Ippoliti2018prb}. \textcolor{red}{The ultraviolet (UV) lattice model has the anomaly-free symmetry $G_{UV}$ with a $D_4 \times \mathbb{Z}_2$ group structure, where the fermion parity $\mathbb{Z}_2^F$ (under which $c_{i,\alpha}\to-c_{i,\alpha}$) acts as a normal subgroup. In the low-energy theory, the fermion parity $\mathbb{Z}_2^F$ only acts on the gapped degrees of freedom, and the quotient symmetry $G_{UV}/\mathbb{Z}_2^F$ acts on the low-energy bosonic degrees of freedom.} In the Hamiltonian $ H $, $ t $ is the fermion hopping amplitude, while $ Q $ and $ V $ represent the strengths of the bond-bond and density-density interactions, respectively. The model preserves several symmetries: bond-centered reflection ($\mathbb{Z}_{2}^{B}$), layer exchange ($\mathbb{Z}_{2}^{E}$), and layer fermion parity ($\mathbb{Z}_{2}^{P}$). The ground-state phase diagram can be determined through DMRG simulations [see Fig.~\ref{fig25} (b)], exhibiting two ordered phases: a charge-density wave (CDW) phase, which spontaneously breaks the bond-centered reflection symmetry $\mathbb{Z}_{2}^{B}$, and a bond-density wave (BDW) phase, which breaks the layer-exchange symmetry $\mathbb{Z}_{2}^{E}$. Notably, the BDW is special long-range order phase in the sense that it hosts symmetry-protected topological edge modes near the boundary, which can be further revealed by a fourfold degeneracy in the bulk entanglement spectrum~\cite{Li2008PRL,Pollmann2010prb}. This phase is referred to as a spontaneous SPT phase~\cite{Ma2022scipost,liu2019superconductivity}, as it exhibits both spontaneous symmetry breaking and symmetry-protected topology—fundamentally different from conventional SPT phases that preserve all global symmetries. Now turning to the phase transitions, large-scale numerical simulations and field-theoretic arguments~\cite{yang2025deconfinedcriticalityintrinsicallygapless} unambiguously demonstrate that the transition line between the two distinct ordered phases is a (1+1)D DQCP, which is described by a compact boson CFT at central charge $c = 1$ with $U(1) \times U(1)$ symmetry. More importantly, the DQCP line also hosts topological edge states under open boundary condition,  which are protected by a mixed anomaly and therefore have no gapped counterparts. This conclusion is supported by numerical simulations and field-theoretic analyses~\cite{Huang2019prb,yang2025deconfinedcriticalityintrinsicallygapless}.

\textcolor{red}{We emphasize that our model possesses three $\mathbb{Z}_{2}$ symmetries: bond-centered reflection $\mathbb{Z}_{2}^{B}$, layer exchange $\mathbb{Z}_{2}^{E}$, and layer fermion parity $\mathbb{Z}_{2}^{P}$. We note the difference between the protecting symmetry responsible for the SPT order in the gapped BDW phase and the anomalous symmetry governing criticality at the phase transition. In the spontaneous SPT phase, the SPT order is protected by the symmetry $\mathbb{Z}_{2}^{B} \times \mathbb{Z}_{2}^{P}$. Crucially, the $\mathbb{Z}_{2}^{E}$ symmetry is spontaneously broken in this phase. In contrast, at the critical point, there exists an anomalous symmetry $\mathbb{Z}_{2}^{E} \times \mathbb{Z}_{2}^{B} \times \mathbb{Z}_{2}^{P}$, and no global symmetry is broken.} 

\textcolor{red}{In the BDW phase, the topological edge states are specifically protected by the $\mathbb{Z}_{2}^{B} \times \mathbb{Z}_{2}^{P}$ symmetry, while the $\mathbb{Z}_{2}^{E}$ symmetry can be spontaneously broken without trivializing the SPT order. 
The spontaneous breaking of $\mathbb{Z}_{2}^{E}$ naturally separates the four-fold degenerate ground states into two pairs: $\{ |b_{1}\rangle, |b_{2}\rangle \}$ and $\{ |b_{3}\rangle, |b_{4}\rangle \}$, related to each other through the action of $\mathbb{Z}_{2}^{E}$. 
 Crucially, within either pair, the protecting symmetry $\mathbb{Z}_{2}^{B} \times \mathbb{Z}_{2}^{P}$ acts projectively on the boundary. 
Specifically, within the pair $\{ |b_{3}\rangle, |b_{4}\rangle \}$, for example, 
\begin{center}
\includegraphics[width=0.4\linewidth]{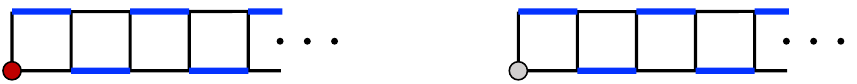}
\end{center}
$\mathbb{Z}_{2}^{B}$ exchanges the two edge states realized as a Pauli $\sigma_{x}$ operator, while $\mathbb{Z}_{2}^{P}$ assigns opposite charges for these two states realized as a Pauli $\sigma_{z}$ operator. 
The anti-commutation $\sigma_{x} \sigma_{z} = - \sigma_{z} \sigma_{x}$ implies the fact that the boundary hosts a projective representation of $\mathbb{Z}_{2}^{B} \times \mathbb{Z}_{2}^{P}$, which fundamentally underpins the stability of the topological edge modes.} 

\textcolor{red}{At the critical point, the quotient symmetry of the system, $G_{UV}/\mathbb{Z}_2^F \cong \mathbb{Z}_{2}^{B} \times \mathbb{Z}_{2}^{E} \times \mathbb{Z}_{2}^{P}$, exhibits an emergent mixed anomaly in the low-energy limit. This anomaly ensures the presence of topological degeneracies near the boundaries even at criticality. One can also recover this anomaly as follows, the compact boson CFT or the Luttinger liquid theory has $(U(1)_V\times U(1)_A)\rtimes\mathbb{Z}_2^C$ symmetry and suffers from the mixed anomaly between these groups. Here $U(1)_{V}$ and $U(1)_A$ correspond to the continuous symmetries generated by rotations of the bosonic fields $\phi$ and $\theta$ respectively—associated with U(1) momentum and winding symmetry in the string theory literature. The $\mathbb{Z}_{2}^{C}$ symmetry denotes the charge conjugation symmetry ($(\phi, \theta) \rightarrow (-\phi,-\theta)$). Once breaking $U(1)_V\times U(1)_A$ down to their $\mathbb{Z}_2$ subgroups, the mixed anomaly becomes the type-III anomaly among $\mathbb{Z}_{2}^{B} \times \mathbb{Z}_{2}^{E} \times \mathbb{Z}_{2}^{P}$ that is studied in our current model and also recent works~\cite{sal2024tdual,omer2025dqcp}.}

So far, we have recognized a DQCP that coincides with an intrinsically gSPT state at the lattice level. To uncover the deep connection between DQCPs and intrinsically gSPT phases, we address two fundamental questions: i) How can we demonstrate the existence of gapless fermionic boundary modes at the DQCP? ii) How can we show that these fermionic boundary modes are unique to gapless systems and have no gapped counterparts?

\textcolor{red}{For the first question, we examine the quantum anomalies between the $\mathbb{Z}_{2}$ symmetries associated with the distinct ordered phases. 
Specifically, we deform the Hamiltonian~\eqref{eq:hamiltonian} with additional pinning fields on the boundary, $-h \sum_{\alpha=A,B} (n_{1,\alpha} + n_{L,\alpha})$ where $h/t=10$ the same order as $Q/t$, to construct a CDW domain wall at criticality. 
The ground state of the deformed Hamiltonian exhibits a twofold degeneracy (more details can be found in Ref.~\cite{yang2025deconfinedcriticalityintrinsicallygapless}). 
In the subspace of the two-fold degenerate states at the critical point, the layer fermion parity $\mathbb{Z}_{2}^{P}$ and the layer-exchange $\mathbb{Z}_{2}^{E}$ symmetry operations can be represented as the effective $\sigma_{z}$ and $\sigma_{x}$ operators, respectively, which anticommute with each other. This indicates that the symmetry acts as the projective representation of the $\mathbb{Z}_{2}^{E} \times \mathbb{Z}_{2}^{P}$. 
In other words, the CDW domain wall carries the fractional charge of the BDW order, consistent with the emergent anomalies between the $\mathbb{Z}_{2}^B$ and $\mathbb{Z}_{2}^P \times \mathbb{Z}_2^E$ symmetries at the DQCP~\cite{Wang2017prx,Huang2019prb,Ma2022scipost}. More precisely, this corresponds to a type-III anomaly among $\mathbb{Z}_{2}^{E}\times \mathbb{Z}_{2}^{B}\times\mathbb{Z}_{2}^{P}$, which is analyzed in our current model~\cite{yang2025deconfinedcriticalityintrinsicallygapless} as well as in recent works~\cite{sal2024tdual,omer2025dqcp}.
This projective representation can only be realized by gapless fermions trapped at the CDW domain wall, which must originate from the boundary because the bulk fermions are gapped. 
Therefore, the emergent anomalies enforce the presence of gapless fermions near the boundary at the DQCP.}

For the first question, we examine the mixed anomalies between the $\mathbb{Z}_{2}$ symmetries associated with distinct ordered phases. Specifically, we deform the Hamiltonian in Eq.~(\ref{eq:hamiltonian}) by introducing pinning fields to create a CDW domain wall at criticality. The ground state of the deformed Hamiltonian exhibits a twofold degeneracy (more details can be found in Ref.~\cite{yang2025deconfinedcriticalityintrinsicallygapless}). Within this degenerate subspace, the layer fermion parity $\mathbb{Z}_{2}^{P}$ and the layer-exchange symmetry $\mathbb{Z}_{2}^{E}$ act as effective anticommuting $\sigma^{z}$ and $\sigma^{x}$ operators, respectively. This indicates that the symmetry group $\mathbb{Z}_{2}^{E} \rtimes \mathbb{Z}_{2}^{P}$ forms a projective representation of the dihedral group $ D_{4} $. In other words, the CDW domain wall carries the charge of the BDW order, consistent with the presence of mixed anomalies between the two $\mathbb{Z}_{2}$ symmetries at the DQCP~\cite{Wang2017prx,Huang2019prb,Ma2022scipost}. This fractional representation can only be realized by gapless fermions trapped at the CDW domain wall, as bulk fermions are gapped~\cite{yang2025deconfinedcriticalityintrinsicallygapless}. Therefore, the presence of mixed anomalies necessarily implies the existence of gapless boundary fermions at the DQCP.

To address the second question, it is well known that in a one-dimensional gapped system, gapless boundary fermions can only arise in a fermionic SPT phase, such as the Su-Schrieffer-Heeger model. In this case, realizing nontrivial topological edge states requires breaking at least one $\mathbb{Z}_{2}$ symmetry, regardless of the dimerization configuration. However, at the DQCP, all global symmetries remain unbroken. Consequently, it is fundamentally impossible to realize such topological edge states in any gapped system. This demonstrates that the gapless boundary fermions observed at the DQCP are inherently tied to gapless systems and cannot have gapped counterparts.

Consequently, we reveal that the mixed anomaly inherent in deconfined criticality enforces the presence of nontrivial topological edge modes near the boundary. This insight suggests that quantum anomalies serve as a general mechanism through which deconfined criticality manifests as a gapless topological state. Moreover, this progress not only provides a new perspective on deconfined criticality but also deepens our understanding of gapless topological phases of matter.

\subsection{Other generalization}
\label{section3.6}
So far, we have only mentioned the topological physics of short-range, many-body quantum critical systems in equilibrium settings, where the critical theory is well established~\cite{landau2013statistical,sachdev1999quantum}. With the rapid development of modern quantum simulation techniques, complex many-body quantum critical states that are challenging to realize in solid-state materials can now be engineered in tunable quantum simulators. However, these simulators often introduce factors beyond the traditional paradigms of condensed matter and statistical physics, such as non-equilibrium dynamics and long-range interactions. In this context, a universal theory of non-equilibrium or long-range quantum critical systems remains incomplete. Therefore, exploring topological physics beyond short-range, equilibrium quantum critical systems is of importance both theoretically and experimentally. On a different front, quantum critical systems exhibit distinct topological properties, leading to the possibility of phase transitions between topologically distinct critical systems. Studying these transitions and developing a universal theory that applies across arbitrary dimensions are crucial for advancing the theory of phase transitions in statistical physics. In this section, we briefly highlight recent progress on these fronts, including extensions topological physics into non-equilibrium quantum many-body systems~\cite{yu2025gaplesssymmetryprotectedtopologicalstates, Chang2020prr,ando2024gaugetheorymixedstate}, long-range interacting systems~\cite{Yang2025cp, Zhong2024pra, Jones2023prl}, phase transitions between topologically distinct critical points or phases~\cite{Yu2024prb, Zhang2024pra}, and a new classification framework based on topological holography~\cite{huang2023topologicalholographyquantumcriticality,huang2024fermionicquantumcriticalitylens,wen2024topologicalholographyfermions,wen2023classification11dgaplesssymmetry,wen2025stringcondensationtopologicalholography}.

\subsubsection{Generalized to non equilibrium setting}
\label{section3.6.1}
\begin{figure}[htp]
    \centering
    \includegraphics[width=0.55\linewidth]{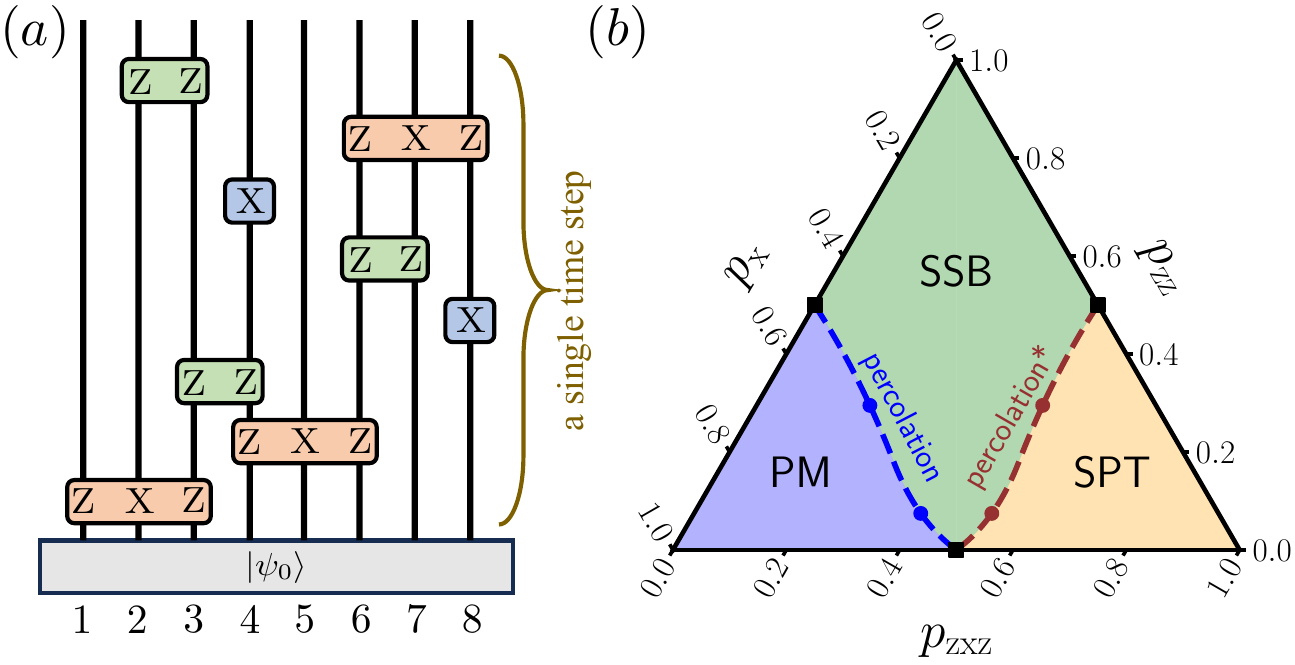}
    \caption{(a) Circuit diagram of the Ising cluster circuit model with $ L=8 $ qubits and 8 measurements (i.e., a single time step): blue, green, and orange rectangles represent the projective measurements $ X_{i} $, $ Z_{i}Z_{i+1} $, and $ Z_{i-1}X_{i}Z_{i+1} $, respectively.  
    (b) The steady-state phase diagram of the cluster circuit as a function of the probabilities $ p_{\text{x}} $, $ p_{\text{zz}} $, and $ p_{\text{zxz}} $.  
    The blue, green, and orange regions correspond to PM, SSB, and SPT orders, respectively.  
    Red circles indicate critical points obtained from data collapse of the generalized topological entanglement entropy [see Fig.2(a) in Ref.~\cite{yu2025gaplesssymmetryprotectedtopologicalstates}], while the red dashed line serves as a guide to the eye.  
    The label "Percolation*" denotes symmetry-enriched percolation universality.  
    Blue circles and the corresponding dashed line are obtained under the transformation $ p_{\rm x} \leftrightarrow p_{\rm zxz} $.  
    The figures are adapted from Ref.~\cite{yu2025gaplesssymmetryprotectedtopologicalstates}.}
    \label{fig26}
\end{figure}

To explore and reveal the mechanisms of topological physics in non-equilibrium quantum critical systems, a prominent platform for studying such phenomena is measurement-only quantum circuits incorporating non-commutative measurements~\cite{Xiang2013rmp,fisher2023random}. The interplay of competing measurements introduces a novel form of frustration, enabling the realization of quantum steady states characterized by distinct orders~\cite{Sang2021prr,BAO2021168618} and entanglement patterns~\cite{lavasani2021measurement,Lavasani2021prl,Morral2023prb}, as well as phase transitions between them~\cite{Liyaodong2018prb,Liyaodong2019prb,Jian2020prb,Skinner2019prx}. However, realizing gSPT phases in solid-state materials remains a significant challenge, highlighting the potential of quantum simulators in achieving these exotic quantum critical states. This naturally raises the question: Can gSPT phases be extended to non-equilibrium settings, such as measurement-only circuits? If so, how can the underlying mechanisms governing these phenomena be theoretically understood?  To address these questions, Ref.~\cite{yu2025gaplesssymmetryprotectedtopologicalstates} investigates various families of measurement-only quantum circuits designed to extend the notion of gSPT phases to non-equilibrium settings. The 1+1D measurement-only circuits are schematically illustrated in Fig.~\ref{fig26} (a). The circuit architecture consists of randomly applied measurements with certain probabilities, uniformly distributed along a one-dimensional qubit chain of length $ L $ under open boundary conditions.

The measurement protocol is structured as follows:  

$\bullet$ A single time step is defined as the application of $ L $ random measurement operations during the time evolution.  

$\bullet$ Each measurement operator is randomly selected from a predefined set with a specified probability.  

$\bullet$ Starting from an initial state $ \ket{\psi_0} $, the system evolves over a large number of time steps (set to $ 5L $ unless otherwise specified) to reach a steady state.

$\bullet$ Physical observables are then computed and averaged over different circuit realizations.  

We first identify phase transitions between different dynamical regimes and identify a new type of universality, termed \emph{symmetry-enriched percolation}~\cite{Verresen2021prx}, which is characterized by nontrivial boundary states, as shown in Fig.~\ref{fig26} (b). By generalizing and analyzing string operators with nontrivial symmetry flux~\cite{yu2025gaplesssymmetryprotectedtopologicalstates}, the symmetry-enriched percolation cannot be continuously connected to conventional percolation without passing through another fixed point, which corresponds to a double-copy percolation universality class. Beyond critical points, we further extend the concept of gSPT phases to non-equilibrium settings. Specifically, we outline the steady-state phase diagram of a $ \mathbb{Z}_4 $ circuit (for details, see Ref.~\cite{yu2025gaplesssymmetryprotectedtopologicalstates}), where a gSPT phase featuring nontrivial edge states and critical fluctuations emerges over a significant portion of the phase diagram. Additionally, the steady-state phase diagram can be understood by mapping the circuit model onto a Majorana loop model, providing a unified framework for studying steady-state gSPT phases in 1+1D measurement-only circuits. This approach uncovers novel quantum phases and critical points in measurement-only circuits, including the first example of a symmetry-enriched non-unitary CFT and a steady-state gSPT phase with robust edge modes. These findings broaden the concept of gSPT phases to non-equilibrium settings, making them more accessible to modern quantum simulation experiments.

Beyond the progress mentioned above, Ref.~\cite{Chang2020prr} explores the entanglement spectrum in topological non-Hermitian free fermion systems, revealing that the non-unitary CFT emerging at the critical point separating the topological parity-time (PT) symmetric phase and the spontaneously PT-broken phase can be viewed as a free-fermion analog of symmetry-enriched quantum criticality in non-Hermitian systems. A more detailed and systematic investigation of topological physics in non-Hermitian quantum critical systems remains an intriguing open question for future work.

\subsubsection{Gapless topological behavior in long range critical spin chain}
\label{section3.6.2}
\begin{figure}[tb]
    \centering
    \includegraphics[width=0.45\linewidth]{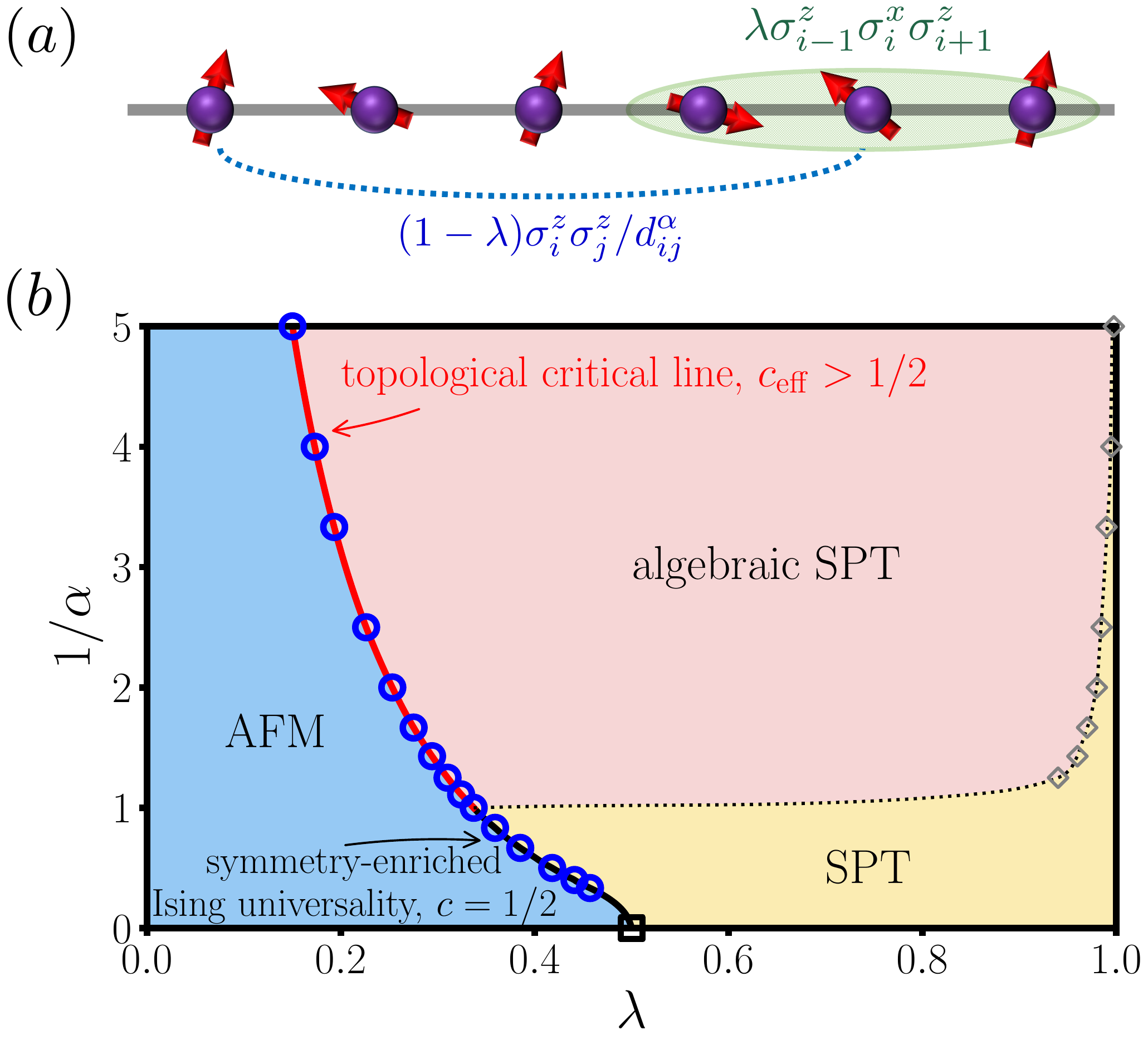}
    \caption{(a) A schematic of the LR cluster Ising spin chain, where the blue dashed line (green-filled ellipsoid) represents the LR AFM interaction (cluster interaction). Here, $ \lambda $ is the strength of the cluster term, $ \alpha $ is the LR exponent, and $ d_{ij} $ is the distance between two sites. (b) The global phase diagram of Hamiltonian~\eqref{E1}, featuring AFM (blue region), SPT (yellow region), and algebraic SPT (pink region) phases. The symmetry-enriched Ising universality class ($ c=1/2 $) is marked by the black line, and a topologically nontrivial critical line ($ c_{\rm eff} > 1/2 $) by the red line. Markers represent numerical results obtained from the peak positions of $ c_{\rm eff} $. The black square denotes the exact critical point $ \lambda=0.5 $ in the SR limit $ 1/\alpha \to 0 $. The black dashed line indicates a crossover between the SPT and algebraic SPT phases. The figures are adapted from Ref.~\cite{Yang2025cp}.}
    \label{fig27}
\end{figure}

LR power-law interactions ($1/r^{\alpha}$) represent a fundamental and practically significant form of nonlocal interactions, appearing ubiquitously in nature and various experimentally relevant systems~\cite{Kyprianidis2021Science,schirhagl2014nitrogen,Labuhn2016nature}. On the theoretical side, LR interactions are known to qualitatively alter physical phenomena, including modifications to the Mermin-Wagner theorem~\cite{Bruno2001prl,Maghrebi2017prl} and the Lieb-Schultz-Mattis theorem~\cite{Ma2024prb,liu2024liebschultzmattistheoremsgeneralizationslongrange}, the breakdown of the entanglement area law in gapped phases or of conformal symmetry at criticality~\cite{Koffel2012prl,Vodola2014prl,Vodola_2016}, deviations from conventional Lieb-Robinson bounds~\cite{Hauke2013prl,Gong2023prl}, and modify critical behavior~\cite{Defenu2023rmp}. For gapped, topological phases, numerous studies have demonstrated qualitative changes induced by LR interactions, including the emergence of massive edge modes~\cite{Vodola2014prl,Viyuela2016prb} and modifications to bulk-boundary correspondence~\cite{Jones2023prl}. However, whether LR interactions can significantly impact gapless topological phases remains an open question. This uncertainty arises because, while the gapless bulk is generally sensitive to LR interactions, topological properties often exhibit robustness against them. Therefore, it is crucial to explore the following key questions:  
i) How do gapless topological phases respond to LR interactions?  
ii) Can LR interactions give rise to novel quantum phases exhibiting previously unreported nontrivial gapless topological behavior?
To address the above questions, reference~\cite{Yang2025cp} investigates the LR interacting spin Hamiltonian on a lattice of length $L$, as illustrated in Fig.~\ref{fig27} (a):  

\begin{equation}
    \begin{split}
    \label{E1}
    H_{\text{LRCI}} = \sum_{i<j}\frac{(1-\lambda)}{d_{ij}^{\alpha}}\sigma^{z}_{i}\sigma^{z}_{j}+\lambda \sum_{j}\sigma^{z}_{j-1}\sigma^{x}_{j}\sigma^{z}_{j+1} \, ,
    \end{split}
\end{equation}  
where $ \sigma_{i}^{x,y,z} $ are Pauli matrices at site $ i $, $ \alpha $ is the power-law exponent of the LR interaction, and $ d_{ij} $ represents the distance between sites $ i $ and $ j $. This model preserves a global $ \mathbb{Z}_{2} \times \mathbb{Z}_{2}^{T} $ symmetry, generated by $ P = \prod_{i=1}^{L} \sigma^{x}_{i} $ and $ T=\mathcal{K} $ (complex conjugation). Notably, in the limit $ \alpha \to \infty $, Eq.~\eqref{E1} reduces to the short-range (SR) cluster Ising model (black square in Fig.~\ref{fig27} (b)), which is exactly solvable via the Jordan-Wigner transformation~\cite{Verresen2017prb, Smacchia2011pra, Guo2022pra}. This SR model can be experimentally realized in triangular optical lattices of ultracold atoms~\cite{Pachos2004prl,Becker_2010}. The parameter $ \lambda $ controls the competition between Ising and cluster interactions, driving the system into different phases, including antiferromagnetic (AFM) and cluster SPT phases. The transition between these phases falls into the symmetry-enriched Ising universality class, characterized by algebraically localized edge modes with two-fold degeneracy~\cite{Verresen2021prx, Yu2022prl}. This is in stark contrast to gapped topological phases with only SR interactions, where edge modes typically exhibit exponential localization.

Interestingly, this model is related to the LR transverse-field Ising model via an SPT entangler~\cite{Verresen2017prb}, suggesting that the bulk universality class of the LR cluster Ising model coincides with that of the LR transverse-field Ising model, sharing the same critical exponents and effective central charge $ c_{\text{eff}} $. Specifically, along the critical line, the critical exponents and $ c_{\text{eff}} $ vary monotonically with $ \alpha $ for $\alpha < \alpha_\text{c}$~\cite{Yang2023prb,Yuxuejia2022prb,Yuxuejia2023pre}. However, for $ \alpha > \alpha_\text{c} $, the critical exponents remain unchanged and align with those of the SR Ising universality class. Despite these bulk similarities, reference~\cite{Yang2025cp} highlights that the boundary topological properties of the LR cluster Ising model exhibit entirely different behavior from those of the LR transverse-field Ising model under open boundary conditions. The ground-state phase diagram, shown in Fig.~\ref{fig27} (b), is mapped as a function of $ \lambda $ and $ \alpha $. For large $ \alpha $, the system hosts an AFM (cluster SPT) phase when $ \lambda $ is below (above) a critical value $ \lambda_{\rm c} $, with a symmetry-enriched Ising quantum critical point separating them. However, for small $ \alpha < \alpha_{\rm c} \sim 1.0 $, the cluster SPT phase gives way to a distinct algebraic SPT phase featuring unique LR-induced edge modes~\cite{Yang2025cp}. Meanwhile, the symmetry-enriched Ising criticality extends into a topologically nontrivial critical line for $ \alpha < \alpha_\text{c} $.

\subsubsection{The phase transition between topological distinct quantum critical point or phases}
\label{section3.6.3}

The discovery of topological phases of matter has revealed that phase transitions can occur between topologically distinct quantum states without any symmetry breaking—these are known as topological "phase" transitions. Following this logic, quantum critical points or critical phases can also exhibit topological classifications. This naturally raises the question: can phase transitions occur between topologically distinct quantum critical points or critical phases, i.e., a "phase transition of phase transitions" induced by nontrivial topology?  

To this end, we first focus on topologically distinct quantum critical points, reference~\cite{Verresen2021prx,Yu2024prb,zhong2025quantumentanglementfermionicgapless} constructs an exactly solvable model that is a linear combination of the TFI model and the CI model discussed in Sec.~\ref{section3.2}:  
\begin{equation}  
\begin{split}  
\label{pspt}  
& H = \lambda H_{TFI} + (1-\lambda)H_{CI}, \\  
& H_{TFI} = -\sum_{j=1}^{N-1}\sigma^{x}_{j}\sigma^{x}_{j+1} - h\sum_{j=1}^{N}\sigma^{z}_{j}, \\  
& H_{CI} = -\sum_{j=1}^{N-1}\sigma^{x}_{j}\sigma^{x}_{j+1} + h\sum_{j=1}^{N-2}\sigma^{x}_{j}\sigma^{z}_{j+1}\sigma^{x}_{j+2}.  
\end{split}  
\end{equation}  

\begin{figure}[tb]
    \centering
    \includegraphics[width=0.7\linewidth]{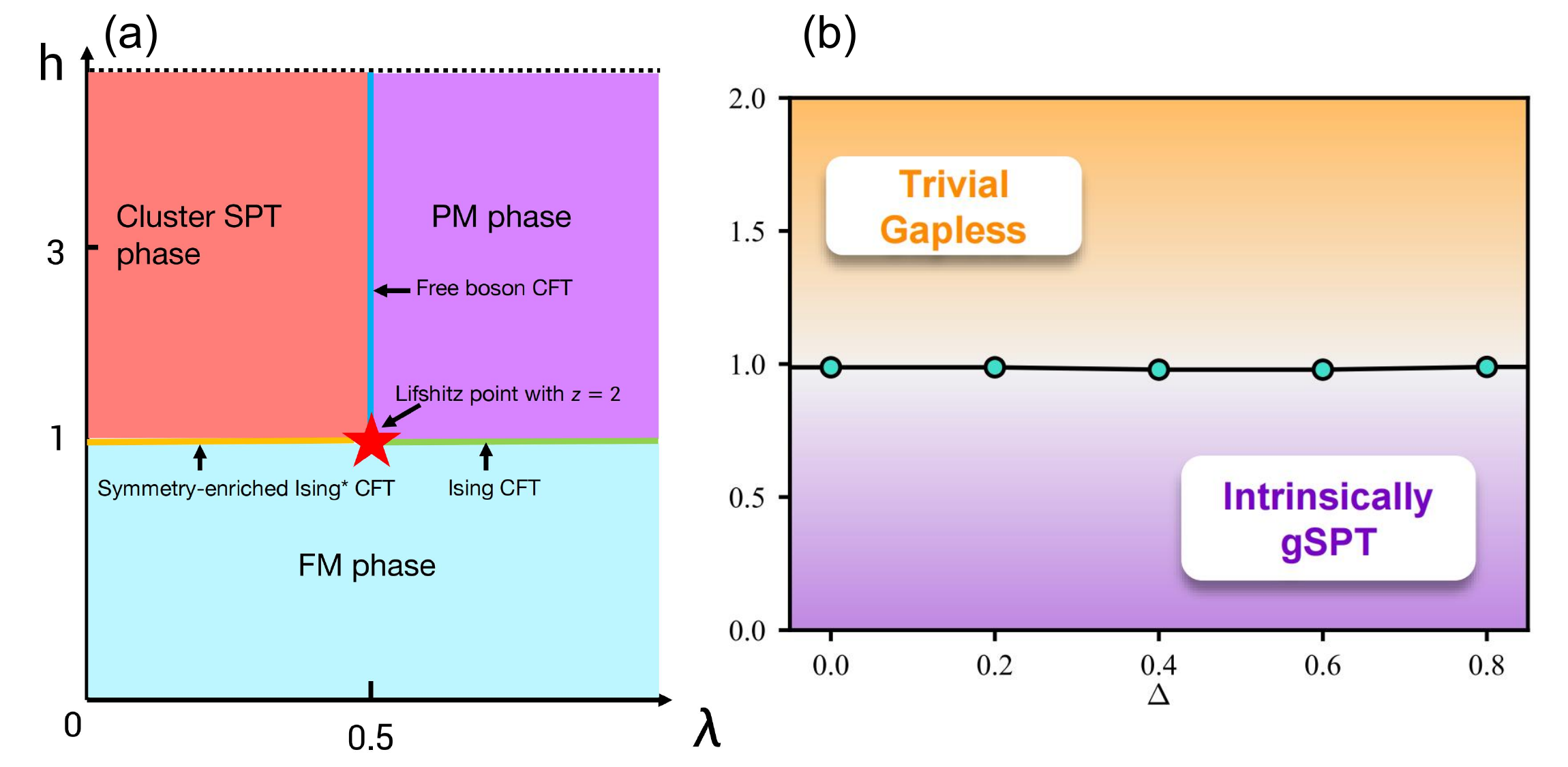}
    \caption{(a)  Schematic phase diagram of quantum Ising model interpolated with cluster interaction in terms of tuning parameters ($\lambda,h$). The phase diagram comprises three distinct regions: the $\mathbb{Z}_{2} \times \mathbb{Z}^{T}_{2}$ cluster SPT phase (light red area), the PM phase (purple area), and the FM order phase (light blue area). When $h<1.0$, the ground state belongs to the FM order phase. When $h=1.0$, the orange (green) solid critical line represents the topological (nontrivial) trivial Ising universality class between the  FM to (cluster SPT) PM phases. For $h>1.0$, the transition from cluster SPT to PM phase (blue solid line) is described by the free boson CFT with $c=1$. The red star denotes the multicritical Lifshitz point with dynamical exponent $z=2$. (b) Phase diagram  of the extended quantum XXZ spin chain with anisotropy parameter $\Delta$ and transverse field $h$. Symbols denote the numerical results of the critical values $h^*_c$. The figures are adapted from Ref.~\cite{Yu2024prb,Zhang2024pra}.}
    \label{fig28}
\end{figure}

Using fidelity susceptibility as a diagnostic tool, the global phase diagram can be unambiguously determined, as shown in Fig.~\ref{fig28} (a). For $ h = 1.0 $, tuning the parameter $ \lambda $ reveals a non-conformal multicritical Lifshitz point with a dynamical exponent $ z = 2 $, which separates topologically distinct Ising critical lines (green and orange lines in Fig.\ref{fig28} (a)). At this multicritical point, the scaling behavior of entanglement entropy and the entanglement spectrum exhibits anomalies, as discussed in recent studies~\cite{Wangke2022scipost,zhong2025quantumentanglementfermionicgapless}. Furthermore, for $ h > 1.0 $, fidelity susceptibility also detects the phase transition between the cluster SPT and PM phases, which is described by a (1+1)D free boson CFT with a central charge of $ c = 1 $. However, for $ h < 1.0 $, no phase transition occurs at all, and the ground state retains a ferromagnetic long-range order.

Beyond the critical points, reference~\cite{Zhang2024pra} takes the first step in addressing above question by investigating the quantum phase transition between topologically distinct stable critical phases. This is achieved by formulating a one-dimensional extended quantum XXZ spin model via the KT transformation (Sec.~\ref{section3.5.1} for details):

\begin{equation}  
\label{HigSPT}  
H = - \sum_{i=1}^{L} \Big( \tau_{2i-1}^{z}\sigma_{2i}^{x}\tau_{2i+1}^{z} + \tau_{2i-1}^{y}\sigma_{2i}^{x}\tau_{2i+1}^{y} + \Delta\tau_{2i-1}^{x}\tau_{2i+1}^{x} + \sigma_{2i}^{z}\tau_{2i+1}^{x}\sigma_{2i+2}^{z} + h\sigma_{2i}^{x} \Big).  
\end{equation}  

Similarly, by combining fidelity susceptibility with the string order parameter and entanglement spectrum, a comprehensive global phase diagram is obtained, capturing both intrinsically gSPT and trivial gapless phases, as shown in Fig.~\ref{fig28} (b). Moreover, as the XXZ-type anisotropy parameter $ \Delta $ varies, these topologically distinct gapless phases undergo a continuous phase transition. The critical points $ h_c $ and the correlation length exponent $ \nu $ remain unchanged from the $ \Delta = 0 $ case, and the transition is described by a CFT with central charge $ c = 3/2 $~\cite{francesco2012conformal,ginsparg1988appliedconformalfieldtheory, Mondal2023prb}. This can be understood as a combination of an Ising CFT and a free boson CFT~\cite{li2023intrinsicallypurelygaplesssptnoninvertibleduality}. These findings indicate that the unconventional critical point between topologically distinct gapless phases at $ \Delta = 0 $ extends into a critical line for general $ \Delta $.

\subsubsection{New perspective: topological holography in gSPT states }
\label{section3.6.4}
A new framework, known as topological holography, has recently emerged for classifying quantum matter and offering a holographic perspective on symmetry~\cite{moradi2022topologicalholographyunificationlandau,freed2024topologicalsymmetryquantumfield,Ji2020prr_a,thorngren2019fusioncategorysymmetryi,freed2022topological,KONG2018140}. In condensed matter physics, this framework is often referred to as symmetry topological order (SymTO) or categorical symmetry~\cite{Kong2020prr,Ji2020prr_a,Chatterjee2023prb_a,Chatterjee2023prb_b,bhardwaj2023categoricallandauparadigmgapped,bhardwaj2024gappedphases21dnoninvertible,bhardwaj2025gappedphases21dnoninvertible,bhardwaj2025gaplessphases21dnoninvertible}, while in high-energy physics, it is known as Symmetry Topological Field Theory (SymTFT)~\cite{Apruzzi_2023,kaidi2023symmetrytftsnoninvertibledefects,kaidi2023symmetrytftsanomaliesnoninvertible}. This approach provides a non-perturbative and unified framework for describing both gapped and gapless phases, effectively decoupling the dynamics of a quantum system from the underlying symmetry.

SymTO has proven to be a powerful framework for classifying gapped SPT phases protected by generalized symmetries, identifying anomalies in non-invertible symmetries, revealing dualities between different phases, and beyond~\cite{moradi2022topologicalholographyunificationlandau,Apruzzi_2023,freed2023introductiontopologicalsymmetryqft,zhang2023anomalies11dcategoricalsymmetries,Bhardwaj_2025}. Here, following Ref.~\cite{wen2023classification11dgaplesssymmetry}, we provide a brief overview of the application of SymTO for classifying gSPT states, with further details available in recent literature~\cite{wen2023classification11dgaplesssymmetry,huang2023topologicalholographyquantumcriticality}.



\textcolor{red}{For a system with a symmetry group $\Gamma$, the SymTO corresponds to a $\Gamma$-gauge theory in one higher dimension.  A complementary perspective involves recovering a $\Gamma$-symmetric system in $(d+1)$ dimensions through dimensional reduction. This is achieved by considering a thin slab of $(d+1)+1$ dimensional (twisted) $\Gamma$-gauge theory with appropriate open boundary conditions in the thin direction and periodic boundary conditions in the other directions.}
More concretely, we consider the simplest case of a (2+1)D $ \mathbb{Z}_2 $ gauge theory, which hosts anyon types $ e $, $ m $, and $ f = e \times m $. Suppose the system is defined on a slab, where the top boundary is fixed in a reference gapped state, corresponding to an anyon condensation that fully confines the topological order~\cite{KONG2014436,Kitaev2012CMP}. Here, we take this reference top boundary as an $ e $-condensed boundary. A global $ \mathbb{Z}_2 $ symmetry can be defined by nucleating a pair of $ m $ particles and dragging them around the periodic cycle. The gapped states of this quasi-1D system correspond to different possible gapped boundaries at the bottom of the slab. The $ m $-condensed boundary is symmetric under the defined $\mathbb{Z}_2$ symmetry, as it can absorb the $ m $-line responsible for generating the symmetry. In contrast, if the bottom boundary is $ e $-condensed, the system undergoes SSB, which results in a dual symmetry under pulling an $e$ particle from the top boundary condensed and moving it to the bottom boundary condensed. This dual symmetry anti-commutes with the original $\mathbb{Z}_2$ $m$-symmetry due to the mutual semionic statistics of $ e $ and $ m $ particles, leading to a twofold degenerate SSB ground state. The interplay between the global $ \mathbb{Z}_2 $ $ m $-symmetry and the dual $ \mathbb{Z}_2 $ $ e $-symmetry matches the structure of the spin-1/2 Ising model on a lattice. In the Ising model, the global $ \mathbb{Z}_2 $ m-symmetry, $ g_m = \prod_i \sigma^x_i $, is preserved in symmetric paramagnetic phases, while the dual domain-wall conservation symmetry, $ g_e = \prod_i \sigma^z_i \sigma^z_{i+1} $, is preserved in symmetry-broken phases. These two symmetries are anti-commute when restricted to overlapping finite intervals. The Ising critical point, which separates the two phases, corresponds to the phase transition between the $ e $- and $ m $-condensed bottom boundaries~\cite{Lichtman2021prb,Chatterjee2023prb_b}.  

The above procedure generalizes to any finite internal unitary symmetry in (1+1)D. In the SSB phase, a dual symmetry emerges, encoding the conservation of domain-wall excitations. Together, the original and dual symmetries form the categorical symmetry of the system, which can be uniformly described by a (2+1)D topological order. As demonstrated in the $\mathbb{Z}_2$ gauge theory example, the symmetry and dual symmetry correspond to anyon string operators associated with gauge flux and gauge charge, respectively. The statistics of anyons in the (2+1)D topological order encode the commutation relations between the symmetry and its dual. The thin-slab construction makes the categorical symmetry explicit: \textcolor{red}{if a (1+1)D system has a symmetry $ \Gamma $ with an anomaly classified by $ \omega \in H^3[\Gamma, U(1)] $, the system can be realized as the boundary of a twisted gauge theory (twisted quantum double) $ D_\omega(\Gamma) $.} The original and dual symmetries manifest through the process of dragging fluxes and charges around the periodic direction. \emph{Different phases of the (1+1)D system correspond to different boundary conditions imposed at the bottom boundary.}

As a first application of the SymTO framework to classify gapped SPT phases, we consider the quantum $\mathbb{Z}_2 \times \mathbb{Z}_{2}$ symmetric spin chain as an illustrative example: the dual toric code has different boundary conditions, corresponding to the distinct gapped phases of the $\mathbb{Z}_2$ symmetry. The SymTO in this case corresponds to the quantum double of $\mathbb{Z}_2 \times \mathbb{Z}_2$, which can be understood as two copies of the toric code. The anyons in each copy are labeled $e_i$, $m_i$, and $f_i = e_i \times m_i$ ($i = 1,2$).  The cluster SPT phase is dual to a symmetric, fully confining condensation in $D(\mathbb{Z}_2 \times \mathbb{Z}_2)$ with option $\mathcal{A}^{SPT} = \langle e_1 m_2, e_2 m_1 \rangle$ that satisfies all condition~\cite{wen2023classification11dgaplesssymmetry,huang2023topologicalholographyquantumcriticality}. The topological edge modes of the cluster SPT can
be seen in the thin-slab construction as follows. The top boundary condenses all charges $e_1$ and $e_2$, while the bottom boundary features an interface between the $\mathcal{A}^{SPT}$ condensation and the $\mathcal{A}^m = \langle m_1, m_2 \rangle$ condensation. The global symmetry of the system, represented by horizontal $m_1$ or $m_2$ strings, localizes to the edge of the SPT phase, forming a projective representation of $\mathbb{Z}_2 \times \mathbb{Z}_2$. Consequently, the condensation $\mathcal{A}^{SPT}$ corresponds to a gapped $\mathbb{Z}_2 \times \mathbb{Z}_2$-symmetric cluster SPT system in (1+1)D with nontrivial edge modes, as illustrated in Fig. 2 of Ref.~\cite{wen2023classification11dgaplesssymmetry}.

\color{red}
To illustrate the physics of gSPT within the framework of SymTO, we use a concrete $\mathbb{Z}_4$-symmetric intrinsic gSPT as an example to demonstrate its power. Specifically, the string order parameter and topological edge modes can be reconstructed from the perspective of SymTO. This phase is dual to a symmetric, partially confining condensation in $D(\mathbb{Z}_4)$. A suitable condensation choice is $\mathcal{A} = \langle e^2 m^2 \rangle$. Upon condensing $e^2 m^2$, the deconfined anyons are generated by $e^2$ and $em$. Both $e^2$ and $em$ acquire order 2 under condensation: trivially, $(e^2)^2 = 1$, while $(em)^2$ is identified with the vacuum. This results in four inequivalent anyons: $1$, $e^2$, $em$, and $e^3 m$. Notably, $em$ and $e^3 m$ are not bosons: $em$ is a semion with topological spin $\theta_{em} = i$, while $e^3 m$ is an anti-semion with $\theta_{e^3 m} = -i$. The resulting post-condensation topological order is the twisted quantum double $D_\omega(\mathbb{Z}_2)$ (twist $\omega \in \mathbb{Z}^3[\mathbb{Z}_2, U(1)]$ corresponds to the Levin-Gu anomaly~\cite{Levin2012prb}), which is known as the double semion theory. According to the low-energy equivalence principle~\cite{wen2023classification11dgaplesssymmetry}, the $e^2 m^2$ condensation describes a (1+1)D $\mathbb{Z}_4$-symmetric gapless system whose low-energy properties match those of a system with $\mathbb{Z}_2$ symmetry and the Levin-Gu anomaly. This suggests that the dual (1+1)D system hosts a low-energy sector with an effective anomalous \(\mathbb{Z}_2\) symmetry, consistent with the structure of $\mathbb{Z}_4$-intrinsically gSPT phases. Thus, the condensation $\mathcal{A} = \{1, e^2 m^2\}$ can be identified with the $\mathbb{Z}_4$-intrinsically gSPT phase.

\textbf{Strong order parameter from SymTO}: An important feature of a nontrivial gSPT is the existence of nontrivial string order parameters. To illustrate this, we employ the thin-slab construction with $e^2m^2$-condensation to reproduce the result. Following the standard prescription, we place the condensation $\mathcal{A}^e=\{1,e,e^2,e^3\}$ on the top boundary and $\mathcal{A}^{igSPT}=\{1,e^2m^2\}$ on the bottom boundary. Within the thin slab, a nonlocal $H$-shaped operator (see Fig. 3 of Ref.~\cite{wen2023classification11dgaplesssymmetry}) can be constructed. The middle segment of this operator is an $m^2$-string, while near its ends the $m^2$-string joins with an $e^2$-string from the top boundary to form an $e^2m^2$-string, which is then absorbed by the bottom boundary. The structure of this operator mimics that of the $\mathbb{Z}_4$ string operator~\cite{li2023intrinsicallypurelygaplesssptnoninvertibleduality,wen2023classification11dgaplesssymmetry}. In particular, the end of this operator anti-commutes with an intersecting $m$-string. Since the operator does not create any excitations, it acts as the identity on the physical state represented by the thin slab. Consequently, it serves as a nonlocal string order parameter with a nonzero vacuum expectation value, whose ends are charged under the $U_s$ symmetry. Thus, by employing the thin-slab construction, one can reconstruct the string order parameters of a gSPT from its dual anyon condensation.

\begin{figure*}
    \centering
    \includegraphics[width=0.55\columnwidth]{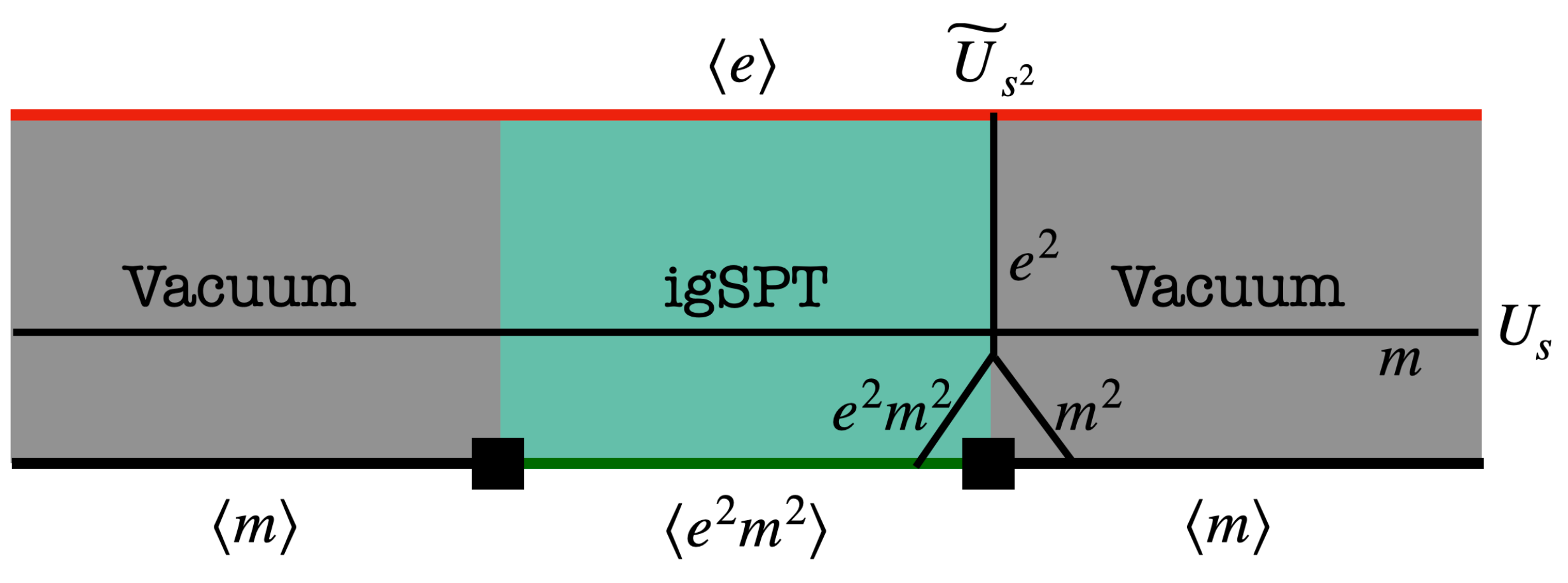}
    \caption{Edge modes of the $\mathbb{Z}_4$-intrinsically gSPT in the thin slab construction: The slab consists of a segment of an intrinsically gSPT phase with two interfaces to a trivial symmetric gapped phase (vacuum). The $\mathbb{Z}_2^A$ action (represented by the horizontal $m^2$-string, $\Tilde{U}_{s^2}$) can be localized at these interfaces, analogous to the localization of symmetry actions in conventional gapped SPT phases. The localized $A$-symmetry action intersects with the $G$-symmetry action (represented by the horizontal $m$-string, $U_s$) at a single point. Due to the mutual $\pi$ statistics between $e^2$ and $m$, the two operators anti-commute, leading to a two-fold ground state degeneracy. The figures are adapted from Ref.~\cite{wen2023classification11dgaplesssymmetry}.}
    \label{fig29}
    \end{figure*}

\color{red}
\textbf{Topological edge modes from SymTO: } We consider a setup where the intrinsically gSPT thin-slab is adjacent to slabs representing the vacuum, as shown in Fig.~\ref{fig29}. The vacuum is defined as an all-flux-condensed boundary ($m$-condensed boundary), since condensing all fluxes on the bottom boundary results in a symmetric, gapped (1+1)D system without a nontrivial string order parameter. In this setup, the bottom boundary consists of a segment of $e^2 m^2$ condensation adjacent to $m$ condensations. As shown in Fig.~\ref{fig29}, the symmetry action of $\mathbb{Z}_2^A$, represented by a horizontal $m^2$ string, localizes at the interfaces between the intrinsically gSPT and the vacuum and localized $\mathbb{Z}_2^A$ symmetry action anti-commutes with the $\mathbb{Z}_2$ symmetry action $U_s$, which is represented by a horizontal $m$ string. This anti-commutation relation aligns with the algebra of symmetries in the $\mathbb{Z}_4$ symmetric intrinsically gSPT phase, giving rise to a two-fold ground state degeneracy. 
\color{black}

More generally, \textcolor{red}{it is plausible that every $\Gamma$ symmetric gSPT phase is dual to a symmetric, partially confining condensation of the quantum double $D(\Gamma)$.} However, to fully establish this duality, it is crucial to understand the classification of two types of objects in two categories: (1) (1+1)D gSPTs and (2) symmetric, partially confining condensations of quantum doubles~\cite{wen2023classification11dgaplesssymmetry}. A careful examination of the structure of (1+1)D gSPTs provides a classification of (1+1)D gSPTs protected by internal unitary finite symmetries. Furthermore, the SymTO framework has been extended to higher dimensions~\cite{wen2025stringcondensationtopologicalholography,wen2025topologicalholography21dgapped} and fermionic gSPT states~\cite{huang2024fermionicquantumcriticalitylens,wen2024topologicalholographyfermions}, though we will not discuss these developments in this review.

\color{black}

\section{Summary and outlook}
\label{section4}
Topological physics in quantum critical systems is an emerging field that explores the possibility of nontrivial topology beyond gapped phases. 
The discovery of nontrivial topology in critical systems not only broadens the scope of topological physics to more challenging gapless quantum critical systems but also opens new avenues for classifying phase transitions based on topology rather than symmetry breaking. 
This fundamentally enriches the textbook understanding of phase transitions. 
In this subfield, gapless critical systems---including continuous phase transition points and critical phases---serve as essential platforms for realizing nontrivial topological states, even with gapless bulk fluctuations. 
This review explores topological physics in both non-interacting and interacting quantum critical systems, covering topics such as critical Majorana fermion chains~\cite{Verresen2018prl,Jones2019JSP}, transition points between distinct topological insulators or superconductors~\cite{verresen2020topologyedgestatessurvive,shen2024newboundarycriticalitytopological}, symmetry-enriched quantum criticality~\cite{Verresen2021prx,Yu2022prl}, gapless symmetry-protected topological phases~\cite{Scaffidi2017prx,Yu2024prl}, intrinsically gapless SPT states~\cite{Thorngren2022prb,li2023intrinsicallypurelygaplesssptnoninvertibleduality,Li2024scipost}, topological edge states at non-equilibrium transition points~\cite{zhou2024topologicaledgestatesfloquet,yu2025gaplesssymmetryprotectedtopologicalstates}, phase transitions between topologically distinct critical points or phases~\cite{Yu2024prb,Zhang2024pra,zhong2025quantumentanglementfermionicgapless}, deep connections to deconfined quantum criticality~\cite{yang2025deconfinedcriticalityintrinsicallygapless}, the impact of long-range interactions on topological critical systems, and the general properties of topological critical systems~\cite{Jones2023prl,Yang2025cp,Zhong2024pra}. 

Furthermore, we would like to make two remarks: 

i) Topological semimetals, which have attracted significant attention over the past decade~\cite{Armitage2018rmp}, can be regarded as a special class of non-interacting gSPT states, where topology is protected by translational symmetry.

ii) The nontrivial topology discussed here is an intrinsic property of the critical system itself. This fundamentally differs from a topological phase transition, which is either a transition driven by a topological defect (e.g., the Kosterlitz-Thouless transition) or a transition between distinct topological phases and is not inherently tied to criticality.

While significant progress has been made in establishing and advancing topological physics in quantum critical systems, numerous open questions remain, presenting exciting opportunities for further theoretical and experimental exploration.

\subsection{Higher dimension }
\label{section4.1}
In recent years, research on topological physics in many-body quantum critical systems has primarily focused on one-dimensional systems, both analytically and numerically. In contrast, studying higher-dimensional counterparts remains challenging due to the absence of well-established theoretical frameworks and efficient computational methods. In particular, numerical investigations of phase transitions in high-dimensional quantum many-body systems typically rely on unbiased quantum Monte Carlo simulations, but the presence of the sign problem severely limits their efficiency.  

Despite these challenges, recent efforts have begun to explore topological physics in higher-dimensional quantum critical systems. Notable developments include the charged twist membrane operator in non-intrinsic gSPT phases~\cite{Verresen2021prx}, the SymTO framework in higher dimensions~\cite{wen2025stringcondensationtopologicalholography,huang2024fermionicquantumcriticalitylens,wen2024topologicalholographyfermions,Andrea2025SciPost}, and multiplicative constructions to fermionic gSPT states in non-interacting systems~\cite{florescalderon2023topologicalquantumcriticalitymultiplicative}. However, these studies remain in their early stages, leaving many open questions.  

Dimensionality plays a key role in the classification of topological physics. For instance, in gapped systems, higher dimensions can host richer topological phases, such as topological spin liquids, whereas one-dimensional systems only exhibit short-range entangled SPT phases. Similarly, the topological physics of high-dimensional quantum critical systems is expected to be even richer. Developing a universal theoretical framework for these systems is crucial for advancing the understanding of topological physics in quantum critical systems. 

From a numerical perspective, insights from previous studies on topological physics at one-dimensional critical systems suggest that a quantum critical point between a gapped topological phase and an SSB phase is likely to exhibit nontrivial topological edge states. Therefore, constructing a high-dimensional lattice model that hosts both SSB phases and gapped topological phases (such as topological spin liquids) may provide a promising avenue for exploring topological properties in high-dimensional quantum critical systems.  To avoid the challenges associated with many-body numerical algorithms for higher dimensional critical systems, a key cue is to construct a modified exactly solvable model, such as the Kitaev toric code or honeycomb model~\cite{Kitaev_2001,KITAEV20062}, which can exhibit both topological spin-liquid phase and SSB long-range order. This enables a direct analytical study of the continuous phase transition between the two phases, as well as the potential emergence of nontrivial topological edge states at criticality.

To advance this direction, we propose the following two approaches:  

\emph{Construct high-dimensional lattice models without sign problems:} Within the framework of current quantum Monte Carlo algorithms, design high-dimensional lattice models that exhibit a continuous phase transition between SSB phases and gapped topological phases~\cite{Scaffidi2016prb,Dupont2021prb_a,Dupont2021prb_b,Nathanan2023scipost_b,Yu2024prl_a}.  

\emph{Develop new computational algorithms:} Enhancing the simulation efficiency of high-dimensional quantum critical points through the development of new quantum many-body computation algorithms, such as quantum Monte Carlo method~\cite{Sandvik2010review,assaad2008computational,berg2019monte,Yu2024prl_b}.

\subsection{The (intrinsically) purely gSPT states}
\label{section4.2}
As mentioned in Sec.~\ref{section3}, intrinsically and purely gSPT phases are among the most intriguing types of gSPT phases, exhibiting nontrivial topological properties beyond those of gapped counterparts. The former can exist only in gapless systems due to emergent anomalies, while the latter, which lacks a gapped sector, features algebraically decaying energy splitting of edge modes and therefore also exists exclusively in gapless systems.  Systematic constructions of gSPT and intrinsically gSPT phases with gapped sectors have been proposed in both field-theoretic frameworks and lattice simulations in one dimension (see Sec.~\ref{section3.5}). However, the only known lattice realization of purely gSPT phases in the literature is the critical cluster Ising spin chain with time-reversal symmetry~\cite{Verresen2018prl, Verresen2021prx, Yu2022prl}. 
It has been demonstrated that this model lacks a gapped sector, as evidenced by the algebratically energy splitting under open boundary conditions---an aspect we have repeatedly emphasized in previous sections.  
More importantly, this raises a fundamental question:
\emph{Does the critical ground state of a lattice model exhibit both nontrivial topological features that prohibit a gapped counterpart and algebraically localized edge modes, thereby realizing an intrinsically purely gSPT phase~\cite{wen2023classification11dgaplesssymmetry,Wen2023prb}?}
Although recent progress~\cite{li2023intrinsicallypurelygaplesssptnoninvertibleduality} has proposed methods for constructing both field-theoretic and lattice realizations of gSPT phases with additional gapped degrees of freedom, a systematic study and lattice realization of purely gSPT phases (both intrinsic and non-intrinsic) face significant challenges and remain in their early stages, particularly in higher dimensions. This research direction places gaplessness at the forefront, potentially leading to the discovery of novel topological phases unique to gapless systems.

Additionally, the strongly interacting many-body systems discussed in this review mainly focus on bosonic systems, such as quantum spin chains. The general theory of interacting fermionic gSPT states and their classification in various dimensions remains less explored~\cite{Keselman2015prb,Verresen2021prx,huang2024fermionicquantumcriticalitylens,wen2024topologicalholographyfermions}, and it is worth addressing in future studies.

\subsection{Gapless intrinsic topological order }
\label{section4.3}

Throughout this review, we have introduced SPT physics in both free-fermion and many-body quantum critical systems. These classifications are analogous to short-range entangled gapped topological phases, corresponding to topological insulators (band topology) in free-fermion systems and gapped SPT phases in many-body systems. 
The nontrivial topology of both gapped and gapless phases discussed so far is protected by underlying global symmetries.  Moreover, in gapped quantum many-body systems, there exist even richer topological phases that do not rely on symmetry protection, now known as intrinsic topological orders~\cite{wen1990topological, Wen2015nsr}. These phases exhibit ground-state degeneracy on nontrivial manifolds (e.g., under \emph{periodic boundary conditions}) and host fractionalized anyon excitations, which have potential applications in universal fault-tolerant topological quantum computing~\cite{Nayak2008rmp}. Unlike SPT phases, intrinsically topologically ordered states are considered highly entangled states of matter and cannot be transformed into a trivial product state through a finite-depth local unitary circuit~\cite{zeng2019quantum}. Representative examples of topological order include the fractional quantum Hall liquid~\cite{Tsui1982prl} and the gapped $\mathbb{Z}_{2}$ quantum spin liquid~\cite{Savary_2017,Zhou2017rmp}. The search for topologically ordered systems has garnered significant interest in both theory and experiment, as reviewed in comprehensive articles~\cite{Wen2015nsr,zeng2019quantum}.

Based on previous research experience of gapped topological phases, it is natural to extend these questions to gapless systems: Is there a topologically nontrivial gapless phase that hosts anyons/ground state degeneracy on non-trivial manifolds? More importantly, can we systematically construct lattice models and develop a complete theoretical framework for quantum critical systems with anyonic excitations in various dimensions? These novel critical systems can be regarded as gapless intrinsic topological orders, following the naming convention of their gapped counterparts. It is worth emphasizing that the fractionalized quantum critical points discussed in earlier literature---such as the Ising* and XY* critical points~\cite{xu2012unconventional}, as well as gapless/symmetry-enriched U(1) quantum spin liquids~\cite{Savary_2017, Zhou2017rmp,WangChong2013PRB,WangChong2016PRX,Zou2018PRB}---all fall within the category of gapless intrinsic topological orders in a broad sense. To systematically realize these novel phases, a promising clue comes from the analogy with non-intrinsic gSPT phases, which can emerge at the critical point between gapped SPT and SSB phases. By extension, we conjecture that gapless intrinsic topological order may appear at the transition point between gapped topological order and long-range ordered phases.  Since topological order can only exist in systems of two or more dimensions, numerical studies of topological order and its phase transitions face significant challenges, except in certain exactly solvable models~\cite{KITAEV20062}. Although there has been recent progress in this area~\cite{Lee2019PRX,Haller2023prr,Kaur2024NC,Singh2024NP,boesl2025quantumphasetransitionssymmetryenriched,xu2024criticalbehaviorfredenhagenmarcustring, Yu2024prl_a}, many fundamental questions remain open. This represents an important but highly challenging research direction, with deep connections to several modern hot topics in condensed matter physics.

\subsection{Experiment realizations}
\label{section4.4}

Although topological physics in quantum critical systems has attracted increasing interest in recent years, research has primarily focused on theoretical developments and numerical simulations. 
Experimentally probing these exotic critical systems remains highly challenging due to the intricate entanglement induced by quantum fluctuations across all length scales and the need for a large number of particles to reach criticality. 
Additionally, preparing the critical ground state in experiments is difficult, and the most commonly used methods for extracting physical observables at criticality are often impractical for large-scale implementation in real experimental settings. Consequently, unlike gapped SPT phases, which have been successfully realized in various experimental platforms~\cite{Zhang2022Nature,Sylvain2019,Xu2024NP,Sompet2022Nature,Iqbal2024Nature,semeghini2021probing,satzinger2021realizing}, the realization of nontrivial topology in gapless many-body systems remains elusive.  

Recent advances in quantum simulation experiments~\cite{tan2025exploringnontrivialtopologyquantum} have demonstrated how low-lying quantum states can be leveraged to probe key properties of gSPT states. In particular, superconducting quantum processors with up to 100 qubits have been used to simulate the critical cluster Ising model, providing the first experimental investigation of its nontrivial topological properties. By designing variational quantum circuits that respect the symmetries of the target Hamiltonian, the authors efficiently generated low-lying states at quantum critical points with a dominant ground-state component. Using these prepared states, they addressed fundamental challenges in extracting topological properties at criticality, including the identification of topological invariants and the verification of bulk-boundary correspondence, as reviewed in Sec.~\ref{section3.5}.  A key experimental advance was the development of an efficient method for measuring the boundary $g$-function, based on overlaps between low-lying critical wavefunctions under different boundary conditions. This method is scalable to systems with large qubit numbers. Furthermore, the bulk-boundary correspondence at criticality was explored by detecting the entanglement spectrum, which encodes universal information beyond entanglement entropy in characterizing gapless topological phases. Using entanglement Hamiltonian tomography (EHT)~\cite{Kokail2021NP}, the authors probed the entanglement spectrum of low-lying critical states, revealing a twofold degeneracy indicative of topological edge modes localized at the boundary. These results provide the first experimental evidence of nontrivial topological properties at quantum critical points and suggest that widely accessible low-lying many-body states could serve as valuable quantum resources for quantum simulation.

In addition to quantum simulators, free-fermion gSPT states in non-interacting systems, analogous to band topology, can be readily implemented in various classical simulation platforms, such as electric circuits~\cite{YANG20241, Lee2018CP, Wang2020NC}, acoustic systems~\cite{Cheng2022PRL, Yang2019NC, Gao2018NP, Yang2019Nature}, and others~\cite{kumar2025topologicaltransitiongaplessphases}. 
An interesting direction for future research is the exploration of topological physics in \emph{classical critical systems}, which may share connections with their quantum counterparts due to the quantum-classical correspondence discussed in Sec.~\ref{section1}. Extending the study of gSPT states to classical systems could offer a more convenient platform for their realization and allow for large-scale experimental studies, as mentioned at the beginning of this paragraph.

Regarding experimental realizations in condensed matter systems, the Haldane phase has been experimentally observed in one-dimensional quantum spin chains~\cite{Buyers1986PRL}, suggesting the possibility of realizing a symmetry-enriched quantum critical point at the transition between a gapped SPT and an SSB phase. 
In two dimensions, the deconfined quantum phase transition between a quantum spin Hall insulator and an s-wave superconducting phase has been shown to host an intrinsically gSPT phase~\cite{Grover2008prl,liu2019superconductivity,myersonjain2024pristinepseudogappedboundariesdeconfined, Ma2022scipost}. Notably, this deconfined transition could potentially be realized in WTe$_2$~\cite{song2024unconventional}.  

More broadly, weakly interacting topological semimetals, which exhibit gapless boundary states—such as three---dimensional Weyl semimetals with surface Fermi arcs~\cite{Wan2011prb, Armitage2018rmp}---can be viewed as a form of non-interacting gSPT phase, as discussed in Sec.~\ref{section2.3}. This perspective opens a promising avenue for realizing gSPT phases in condensed matter experiments, where entanglement spectrum measurements could serve as a key diagnostic for uncovering their nontrivial topology at criticality.

\addcontentsline{toc}{section}{Declaration of competing interest}
\section*{Declaration of competing interest}
The authors declare no competing financial interests that could have appeared to influence the work reported in this paper.
\addcontentsline{toc}{section}{Acknowledgments}
\section*{Acknowledgments}
We would like to thank Shao-Kai Jian, Sheng Yang, Yi-Zhuang You, Zi-Xiang Li, Zhiming Pan, Shuo Liu, Chao Song, Baile Zhang, Da-Chuan Lu, Long Zhang, Shou-Shu Gong, Rui-Zhen Huang, Chengxiang Ding, Yijian Zou, Linhao Li, Rui Wen, Ruochen Ma, Yuchen Guo, Weicheng Ye, Xueda Wen, Hongzheng Zhao, Yu-An Chen, Po-Yao Chang, Meng Zeng, Yunqing Zheng, Youjin Deng, Wei Zhu, Longwen Zhou, Chinghua Lee, and Ruben Verresen for stimulating discussions and collaborations. This work was supported by the National Natural Science Foundation of China (NSFC) (Grants No. 12405034) and and a start-up grant from Fuzhou University. The work of L.X. are supported by the National Natural Science Foundation of China under Grant No. 11935002, No. 12535001, No. 11525520 National Key R\&D Program under grant 2021YFA1400500. H.Q. Lin acknowledges the support from NSFC12088101 and 2022YFA1402701. 

\bibliography{ref}
\bibliographystyle{apsrev4-2}

\end{document}